\documentclass{amsart}
\usepackage{graphicx}
\usepackage{latexsym}
\usepackage{amssymb}
\usepackage{amsmath}
\usepackage{amsmath}
\numberwithin{equation}{section}

\newtheorem{theorem}{Theorem}[section]
\newtheorem{lemma}{Lemma}[section]
\newtheorem{proposition}{Proposition}[section]
\newtheorem{definition}{Definition}[section]

\begin{document}

\title[Hamiltonian structure for dispersive and dissipative dynamics]
{Hamiltonian structure for dispersive and dissipative dynamical
systems}
\author{Alex Figotin and Jeffrey H.\ Schenker}
\date{July 29, 2006, revised January 15, 2007}

\begin{abstract}
We develop a Hamiltonian theory of a time dispersive and dissipative
inhomogeneous medium, as described by a linear response equation
respecting causality and power dissipation. The proposed Hamiltonian
couples the given system to auxiliary fields, in the universal form
of a so-called canonical heat bath. After integrating out the heat
bath the original dissipative evolution is exactly reproduced.
Furthermore, we show that the dynamics associated to a minimal
Hamiltonian are essentially unique, up to a natural class of
isomorphisms. Using this formalism, we obtain closed form
expressions for the energy density, energy flux, momentum density,
and stress tensor involving the auxiliary fields, from which we
derive an approximate, ``Brillouin-type,'' formula for the time
averaged energy density and stress tensor associated to an almost
mono-chromatic wave.
\end{abstract}

 \maketitle

\section{Introduction}

The need for a Hamiltonian description of dissipative systems has
long been known. Forty years ago Morse and Feshbach gave an example
of an artificial Hamiltonian for a damped oscillator based on a
\textquotedblleft mirror-image\textquotedblright\ trick,
incorporating a second oscillator with negative friction \cite[Ch
3.2]{MorseFeshbach}. The resulting Hamiltonian is un-physical: it is
unbounded from below and under time reversal the oscillator is
transformed into its \textquotedblleft
mirror-image.\textquotedblright\ The artificial nature of this
construction was described in  \cite[Ch. 3.2]{MorseFeshbach}: ``By
this arbitrary trick we are able to handle dissipative systems as
though they were conservative. This is not very satisfactory if an
alternate method of solution is known...''

We propose here a quite general \textquotedblleft satisfactory
solution\textquotedblright\ to the general problem posed in
\cite{MorseFeshbach} by constructing a Hamiltonian for a time
dispersive and dissipative (TDD) dynamical system \emph{without
introducing negative friction} and, in particular, without
\textquotedblleft mirror-images.\textquotedblright\ Developing a
Hamiltonian structure for a TDD system  might seem  a paradoxical
goal \textemdash \ after all, neither dissipation nor time
dispersion occur in Hamiltonian systems. However, we will see that
if dissipation is introduced via a friction function, or
\emph{susceptibility}, obeying a power dissipation condition
\textemdash\ as it is for a linear dielectric medium described by
the Maxwell equations with frequency dependent material relations
\textemdash \ then the dynamics are \emph{exactly} reproduced by a
particular coupling of the TDD system to an effective model for the
normal modes of the underlying medium as independent oscillating
strings. For the combined system we give a non-negative Hamiltonian
with a transparent interpretation as the system energy.

An important motivation behind this effort is the clarification of
the definition of the radiation energy density and stress tensor for
a dissipative medium in the linear response theory, e.g., a
dielectric medium with complex valued frequency dependent material
relations. An intrinsic ambiguity in this definition has
led to problems interpreting the energy balance equation \cite[%
Sect. 77]{LandauLif}, \cite[Sect. 1.5a]{FM}, \cite[Sect. 6.8]{Jackson}, \cite%
{Loudon}. These difficulties do not persist if a fundamental
microscopic theory is considered. Consequently a number of efforts
\cite{Loudon,LaxNelson,NelsonChen} have been made to construct a
macroscopic theory of dielectric media, accounting for dispersion
and dissipation, based on a more fundamental microscopic theory. It
might seem that the introduction of an explicit realistic material
medium is the only way to model a TDD system. However, the
construction of this paper shows this is not so and provides a
consistent macroscopic approach \textit{within linear response
theory}.

As an example of our general construction, we analyze here TDD
dielectric media, including  a detailed analysis of the
electromagnetic energy and momentum densities. Part of that analysis
is the derivation of an approximate formula for the time averaged
Maxwell stress tensor similar to the Brillouin formula for the time
averaged energy density  \cite[Section 80]{LandauLif}.

Another important benefit of the approach developed here \textemdash
\ and in our previous work \cite{FS} \textemdash \ is that the
present formulation allows to apply the well developed scattering
theory for conservative systems \cite{Newton} to the long long
standing problem of scattering from a lossy non spherical scatter
---\ analyzed by other methods with limited success
\cite{Mishchenko}. This will be discussed in detail in forthcoming
work \cite{FS2}.

\subsection{Dissipative systems}\label{sec:dis}
We consider a system to be dissipative if energy tends to decrease
under its evolution.  It is common, taking energy conservation  as a
fundamental principle, to view a dissipative system as coupled to a
heat bath so that energy lost to dissipation is viewed as having
been absorbed by the heat bath.

We have shown in \cite{FS} that, indeed, a general linear causal TDD
system can be represented as a subsystem of a conservative system,
with the minimal such extension unique up to isomorphism. Let us
summarize the main ideas of that work here. One begins with an
evolution equation accounting for dispersion and dissipation, of the
form
\begin{equation}
\partial _{t}f\left( t\right) =  L f\left( t\right)
-\int_{0}^{\infty }a\left( \tau \right)f\left( t-\tau \right) \,\mathrm{d}%
\tau  \ + \ r(t),  \label{vA1}
\end{equation}%
where $f\left( t\right) $ describes the state of the system at time
$t$, specified by a point in a complex Hilbert space $H_{0}$, and
\begin{enumerate}
\item $L=-\mathrm{i} A$ with $A$ a self-adjoint operator on $H_{0}$,
\item $a\left(\tau \right) $ is an operator valued function,
called the \emph{friction function} and assumed to be of the form
\begin{equation}
a\left( \tau\right) =\alpha _{\infty }\delta \left( \tau \right)
+\alpha \left( \tau \right) \text{, }  \label{ax3a}
\end{equation}%
wtih $\alpha \left( \tau \right) $ strongly continuous for $\tau \geq 0$ and $%
\alpha _{\infty }$ self-adjoint.
\item
$r(t)$ is an external driving force.
\end{enumerate}
For $a(t)$ satisfying a \emph{power dissipation condition}
[\eqref{ava4}, below] one then constructs a complex Hilbert space
$\mathcal{H}$, an isometric injection $I: H_{0}\rightarrow \mathcal
H_0$ and a self adjoint operator $\mathcal A$ on $\mathcal H$ such
that the solution $f(t)$ to (\ref{vA1}) equals the projection
$I^\dagger F(t)$ onto $H_0$ of the solution $F(t)$ to
\begin{equation}
\partial _{t}F\left( t\right) =\mathcal{L}F\left( t\right) \ + \ I r(t) ,\
  \label{intro1}
\end{equation}%
with $\mathcal{L}=- \mathrm{i} \mathcal A$. That is
\begin{equation}\label{solvevA1}
  f(t) \ = \ I^\dagger F(t) \ = \ \int_{-\infty}^t I^{\dagger} \mathrm e^{(t-t')\mathcal
  L} I r(t') \mathrm d t'.
\end{equation}

The main object of this work is to extend \cite{FS} by considering
dissipative systems with a given Hamiltonian structure. Here and
below we call a system \emph{Hamiltonian} if its phase space is
endowed with a symplectic structure and a Hamiltonian function such
that: i.) the system evolves by Hamilton's equations, and ii.) the
physical energy of the system in a configuration associated to a
phase space point $u$ is equal to the value of the Hamiltonian
function at $u$. Accordingly, a dissipative system is by definition
not Hamiltonian. Nonetheless, almost every dissipative system of
interest to physics is a perturbation of a Hamiltonian system, with
the perturbation accounting for dispersion and
dissipation. 

The property of being Hamiltonian, as defined above, is more than a
formal property of the evolution equations, as it also involves a
physical restriction equating the Hamilton function and the system
energy. For a linear system such as \eqref{intro1} there are many
ways to represent the evolution equations as Hamilton's equations.
We circumvent this ambiguity by supposing given a Hamiltonian
structure on the given TDD system, whose evolution is a suitable
perturbation of the Hamilton equations. We then ask, and answer
affirmatively, the question, ``Is there a natural way to extend the
given Hamiltonian structure to the unique minimal conservative
extension of \cite{FS} so that the extended system is Hamiltonian?"
This will be achieved in a self contained way below, with reference
to the extension of \cite{FS}, by constructing a Hamiltonian
extension with additional degrees of freedom in the universal form
of a canonical heat bath as defined in \cite[Section 2]{JP2},
\cite[Section 2]{RB}.

\subsection{Hamiltonian systems}
We suppose given a dynamical system described by a coordinate $u$
taking values in \emph{phase space}, a \underline{real} Hilbert
space $V$. On $V$ there is defined a \emph{symplectic form}
$\mathfrak J(u,v) = \langle u, J v \rangle$, with $J:V\rightarrow V$ a linear map
such that
\begin{equation}
J^{\mathrm{T}}J=\mathbf{1}\;,\quad J^{2}=-\mathbf{1}%
\;.  \label{symplectic}
\end{equation}
We call a map $J$ satisfying \eqref{symplectic} a \emph{symplectic operator}.
Throughout we work with real Hilbert spaces and use $M^{\mathrm{T}}$
to denote the transpose of an operator $M$, i.e., the adjoint with
respect to a real inner product. Additional notation, used without
comment below, is summarized in Appendix \ref{Real} along with the
spectral theory for operators in real Hilbert spaces.

The evolution equation (in the limit of zero dissipation) is
to be Hamiltonian with respect to $\mathfrak J$. Thus, we
suppose given a Hamiltonian function $\mathrm{h}(u)$ such that when
dissipation is negligible $u$ evolves according to the symplectic gradient of $\mathrm{h}$
\begin{equation}
\partial _{t}u\ =\ J\frac{\delta \mathrm{h}(u)}{\delta u}\;.  \label{hamilton}
\end{equation}
For most applications the Hamiltonian $\mathrm{h}\left( u\right) $
is the system energy and is nonnegative (or at least bounded from
below). Mostly, we consider a \emph{quadratic}, non-negative
Hamiltonian
\begin{equation}
\mathrm{h}\left( u\right) \ =\ \frac{1}{2}\left\langle
Ku\,,\,Ku\right\rangle \label{hpq1}
\end{equation}
leading to a linear evolution equation. However, there is a natural
extension of the results presented here to a nonlinear system with
\begin{equation}
  \mathrm{h}\left ( u \right) \ = \ \frac{1}{2} \left\langle
Ku\,,\,Ku\right\rangle + \mathrm{h}'\left ( u \right ),
\end{equation}
where $\mathrm{h}'\left ( u \right )$ is an arbitrary function of
$u$. The construction carries over, provided the dissipation enters
linearly as in \eqref{hpq4} below. To keep the discussion as simple
as possible \textemdash \ and to avoid the difficult questions of
existence and uniqueness for non-linear systems \textemdash \  we
consider the Hamiltonian \eqref{hpq1} throughout the main text. A
few examples illustrating this point are discussed in Appendix
\ref{nonlinear}.

We call the operator $K$ the \emph{internal impedance operator} (see %
\eqref{hpq3} below), and suppose it to be a closed, densely defined map
\begin{equation}  \label{kVtoH}
K: \mathcal D (K) \ \rightarrow \ H, \quad  \mathcal D (K) \subset V
\end{equation}
with $H$ the \emph{stress space}. The (real) Hilbert spaces $V$ and
$H$ are respectively the system phase space and the state-space of
internal ``stresses.''\footnote{Abstractly, it is not
strictly necessary to distinguish $V$ and $H$.  We could
replace $H$ by $V$ and $K$ by $|K|$ (see Appendix \ref{Real}).
However, that might be physically unnatural,
and we find that the distinction clarifies the role of dissipation and
dispersion in applications. In particular, the
impedance operator is dimensionful (making it necessary
to distinguish domain and range) unless we parametrize phase space
by quantities with units $\sqrt{\text{energy}}$.

From a mathematical standpoint, using $|K|$ may introduce
complications. For instance, with $V = L^2(\mathbb{R}^3;
\mathbb{C})$,  $ H  = L^2(\mathbb{R}^3;\mathbb{C} ^3) $  and $K u
(\vec{r})  = \nabla u (\vec{r})$, the associated Hamiltonian, $
\mathrm{h}(u)  =  \int_{\mathbb{R}^3} \mathrm{d}^3 \vec{r} \ | \nabla u (%
\vec{r}) |^2  , $ produces the evolution $\partial_t
u_t(\vec{r})  =  - \mathrm{i} \Delta u_t(\vec{r}) $,
taking  $J=$ multiplication by $%
\mathrm{i}$. Of course we might take  $H = V$ and $K =
\sqrt{-\Delta} = |\nabla|$, instead. But it is more elegant (and more
natural) to work with the differential operator $\nabla$.}
The space of finite energy states is the operator domain
$\mathcal{D}(K)$. Physical examples and further discussion of the
operator $K$ are given in Section \ref{local}. Technical
assumptions and a discussion of the dynamics on $\mathcal{D}(K)$ are
given in Section \ref{result}

The equation of motion, in the absence of dissipation, is obtained from (%
\ref{hamilton}, \ref{hpq1}) by \emph{formal} differentiation.  It is
convenient split the equation in two:
\begin{subequations}
\begin{equation}
\partial _{t}u(t)\ =\ JK^{\mathrm{T}}f(t)\quad \text{(evolution equation),}
\label{hpq2}
\end{equation}%
with
\begin{equation}
f(t)\ =\ Ku(t) \quad (\text{material relation without dispersion or
dissipation}). \label{hpq3}
\end{equation}%
When dissipation is included, we replace \eqref{hpq3} with a
\emph{generalized material relation},
\begin{equation}\tag{\theequation$^\prime$}
f(t)+\int_{0}^{\infty }\mathrm{d}\tau \,\chi (\tau )f(t-\tau
)=Ku(t), \label{hpq4}
\end{equation}
\end{subequations}
where $\chi $ is the \textit{operator valued generalized
susceptibility}, a function of $\tau>0$ with values in the bounded
operators on $H$. Note that the
integral in \eqref{hpq4} explicitly satisfies \textit{causality}:
the left hand side depends only on times $t-\tau \leq t$.

The structure of (\ref{hpq2}, \ref{hpq4}) mirrors
the Maxwell equations for the electro-magnetic (EM) field in
a TDD medium. For a static \emph{non-dispersive} medium \textemdash
\ see Section \ref{maxwell} \textemdash \ eq.\ \eqref{hpq2} and \eqref{hpq3}
correspond respectively to the dynamical Maxwell equations and the material relations.
(The static Maxwell equations amount to a choice of coordinates.)
Dispersion and dissipation are incorporated in (\ref{hpq2}, \ref{hpq4}) by modifying the material relations in the same fashion as in the phenomenological theory of
the EM field in a TDD medium.

The vectors $u$ and $f$ of the TDD system (\ref{hpq2}, \ref{hpq4})
may interpreted physically as follows: $u$ specifies
the state of the system and $f$ specifies internal forces
driving the dynamics. Thus, we refer to $f$ as the
\emph{kinematical stress}. Similarly, we refer to $Ku$ as the
\emph{mechanical} or \emph{internal stress}, as the square of its magnitude is
the energy of the system. In a
dispersionless system, these quantities are equal, but in a TDD system they
are \emph{not}, and are related by \eqref{hpq4},
incorporating time dispersion.\footnote{We could
consider a relation inverse to \eqref{hpq4}, expressing the
kinematical stress as a function of the mechanical stress, $
f(t) = K u(t) + \int_0^\infty \mathrm{d} \tau \, \widetilde
\chi(\tau) K u(t-\tau).$ Under the power dissipation condition,
\eqref{PDC} below, we may
invert \eqref{hpq4} to obtain this equation and vice versa. However
\eqref{hpq4} appears in the standard form of Maxwell's equations and
is most convenient for our analysis.}

Associated to the dispersionless system (\ref{hpq2}, \ref{hpq3}) is
the initial value problem (IVP), which asks for $u(t)$, $t > t_0$ given the
initial condition $u(t_0)=u_0$. Under suitable hypotheses on $K$ and
$J$ this problem is well-posed for $u_0 \in \mathcal D(K)$, with
existence and uniqueness of solutions provable by standard spectral theory
(see \S \ref{sec:hamilton}). However, for the TDD system (\ref{hpq2},
\ref{hpq4}), the initial value problem is not well defined, because
the integral on the l.h.s.\ of \eqref{hpq4} involves $f(t)$ for $t
\rightarrow -\infty$. This dependence on history forces us to ask,
``how were the initial conditions $f_0$ and $u_0$ produced?'' Thus a
more physically sound approach is to suppose the system is driven
by a time dependent \emph{external force}
$\rho(t)$ (which we controll).  This leads us to the driven system:
\begin{subequations}\label{driven}
\begin{align}  \label{driven1}
\partial_t u(t) \ &= \ J K^{\mathrm{T}} f(t) \ + \ \rho(t) \\
K u(t) \ &= \ f(t) + \int_{0}^{\infty} \mathrm{d} \tau\, \chi(\tau)
f(t-\tau) , \label{driven2}
\end{align}
with initial conditions
\begin{equation}  \label{t-infic}
\lim_{t \rightarrow -\infty} u(t) \in \ker K , \quad \lim_{t
\rightarrow -\infty} f(t) \ = \ 0 \; ,
\end{equation}
\end{subequations}
so at $t = -\infty$ the system was at rest with zero
energy.
In the absence of dispersion, when $\chi =0$, eqs.\ (\ref{driven1}, \ref{driven2}) reduce to
\begin{equation}
\partial _{t}u(t)\ =\ JK^{\mathrm{T}}Ku(t)\ +\ \rho (t).  \label{hamdriven}
\end{equation}%
It is useful to note that \eqref{hamdriven} is Hamilton's
equation for the time dependent Hamiltonian
$\mathrm{h}_{t}(u)=\mathrm{h}(u)-\langle J\rho (t),u\rangle $.

We shall generally take the external force to be a bounded compactly
supported function $\rho: \mathbb{R}
\rightarrow V$. More generally we might ask only that $\rho \in L^1(\mathbb{R%
}, V)$ or even allow $\rho$ to be a measure. The initial value
problem for \eqref{hamdriven} amounts to the idealization $\rho(t) =
u_0 \delta(t-t_0)$.

\subsection{Hamiltonian extensions}

The main question addressed here is: \emph{when does the
system described by} (\ref{driven}) \emph{admit a Hamiltonian
extension?} We restrict ourselves to looking for a
\emph{quadratic Hamiltonian extension} (QHE), defined below. Our
main result is the existence of a QHE under physically natural
conditions on the susceptibility:
\begin{theorem}\label{thm:main}
Under mild regularity assumptions for the system operators $K$ and
$\chi $ \emph{(spelled out in Section \ref{result})}, if $\chi $ is
\emph{symmetric},
\begin{equation}
\chi (t)^{\mathrm{T}}\ =\ \chi (t)\;,  \label{symchi}
\end{equation}%
then there exists a quadratic Hamiltonian extension of the system
\emph{(\ref{driven})} if and only if $\chi $ satisfies the
\emph{power dissipation condition (PDC)}
\begin{equation}
\mathrm{Im}\left\{ \zeta \widehat{\chi}(\zeta )\right\} =\frac{1}{2\mathrm{i}}%
\left\{ \zeta \widehat{\chi}(\zeta )-\zeta ^{\ast
}\widehat{\chi}(\zeta )^{\dagger }\right\} \geq 0\text{ for all
$\zeta =\omega +\mathrm{i}\eta $, $\eta \geq 0 $, }  \label{PDC}
\end{equation}%
with $\widehat{\chi}$ the Fourier-Laplace transform of ${\chi }$,
\begin{equation}
\widehat{\chi}(\zeta ) \ =\ \int_{0}^{\infty }\mathrm{d%
}t\,\mathrm{e}^{\mathrm{i}\zeta t}\chi (t).  \label{FourLap}
\end{equation}%
\end{theorem}
\noindent{\textit{Remarks}:} i.) The operator $\widehat{\chi}(\zeta
)$ is \emph{complex linear}, defined on the complexification $
\mathbb{C}H$ of the real Hilbert space $H$ (see Appendix
\ref{Real}).  As indicated, the imaginary part in \eqref{PDC} refers
to the imaginary part with respect to the Hermitian structure on
$\mathbb{C} H$. Due to the symmetry condition \eqref{symchi}, this
is also the imaginary part with respect to the complex structure,
i.e., $\mathrm{Im}_{\mathbb{C}}\zeta \widehat{\chi}(\zeta ) = \frac{1}{2\mathrm{i}}%
\left\{ \zeta \widehat{\chi}(\zeta )-\zeta ^{*}\widehat{\chi}(\zeta
)^{*}\right\}$, where $^{*}$ denotes complex conjugation,
$A^{*}v=(Av^{*})^{*}$. ii.) As mentioned above, the result extends
with no extra effort to a non-linear system, with a non-quadratic
Hamiltonian $h(u)$, provided the dissipation is introduced linearly
as in \eqref{hpq4}. In that case, the extended Hamiltonian is, of
course, not quadratic as it maintains the non-quadratic part of the
initial Hamiltonian $h(u)$.\vskip0.1in

We verify the theorem by constructing an explicit extension based on
the following operator valued \emph{coupling function}
\begin{equation}
\varsigma (s)\ =\ \frac{1}{2 \pi }\int_{\mathbb{R}}\mathrm{d}\omega
\,\mathrm{e}^{- \mathrm i \omega s} \sqrt{2\omega \mathrm{Im}\widehat{%
\chi }(\omega )} \ =\ \frac{1}{2 \pi
}\int_{\mathbb{R}}\mathrm{d}\omega
\,\cos (\omega s)\sqrt{2\omega \mathrm{Im}\widehat{%
\chi }(\omega )} , \label{Excoupling}
\end{equation}
and the associated map
\begin{equation}\label{Excoupling2}
T\varphi \ =\ \int_{-\infty }^{\infty }%
\mathrm{d}\sigma \varsigma (\sigma )\varphi (\sigma )\, , \quad
T:L^{2}(\mathbb{R} ,H)\rightarrow H.
\end{equation}
The extended Hamiltonian is
\begin{align}
\mathrm{H}(U)\ &=\ \frac{1}{2}\left\{ \left\Vert Ku-T\varphi
\right\Vert _{H}^{2}+\int_{-\infty}^{\infty }\left[ \left\Vert
\theta \left( s\right) \right\Vert _{H}^{2}+\left\Vert \partial
_{s}\varphi \left( s\right) \right\Vert
_{H}^{2}\right] \,\mathrm{d}s\right\}  \label{ExHamiltonian} \\
&= \ \frac{1}{2}\langle \mathcal{K}U,\mathcal{K}U\rangle ,\ \notag
\end{align}
with
\begin{equation}\label{eq:calK}
\mathcal{K}U=%
\begin{pmatrix}
K & 0 & -T \\
0 & \mathbf{1} & 0 \\
0 & 0 & \partial _{s}%
\end{pmatrix}%
\begin{pmatrix}
u \\
\theta (s) \\
\varphi (s)%
\end{pmatrix}.
\end{equation}
The \emph{extended impedance} $\mathcal K$ is a densely
defined closed map from \emph{extended phase space}
\begin{equation}\label{eq:calV}
\mathcal{V}\ =\ V\oplus L^{2}(\mathbb{R},H)\oplus
L^{2}(\mathbb{R},H)\;
\end{equation}
into  \emph{extended stress space}
\begin{equation}\label{eq:calH}
\mathcal{H}\ :=\ H\oplus L^{2}(\mathbb{R},H)\oplus
L^{2}(\mathbb{R},H).
\end{equation}
The symplectic structure on $\mathcal V$ is given by the following
extension of $J$:
\begin{equation}\label{eq:calJ}
\mathcal{J}\ =\
\begin{pmatrix}
J & 0 & 0 \\
0 & 0 & -\mathbf{1} \\
0 & \mathbf{1} & 0%
\end{pmatrix}%
:\mathcal{V}\rightarrow \mathcal{V}.
\end{equation}
We denote by $I_{V}$ and $I_{H}$ the isometric injections $
V \hookrightarrow \mathcal{V}$ and $H \hookrightarrow \mathcal H$
respectively:
\begin{equation}\label{isoinject}
  I_V u \ = \ \begin{pmatrix} u \\ 0 \\ 0 \end{pmatrix} \quad
  \text{and} \quad I_H f \ = \ \begin{pmatrix} f \\ 0 \\ 0
  \end{pmatrix}.
\end{equation}

The driven Hamilton equations for the extended system are
\begin{align}
\partial_t u(t) \ &= \ J K^{\mathrm{T}} f(t) + \rho(t)\label{ExDy1} \\ \partial_t
  \theta(s,t)  \ &= \ \partial_s^2
  \varphi(s,t) +
  \varsigma\left(  s\right ) f(t) ,
  \label{ExDy2} \\
\label{ExDy3}
  \partial_t \varphi(s,t)  \ &= \
  \theta(s,t)
\end{align}
where we have set the driving force $R(t) \ = \ I_V \rho(t)$
and  introduced the kinematical stress $f$ in terms
terms of the extended system:
\begin{equation}\label{ExStress}
  f(t) \ = \ K u(t) - \int_{-\infty}^\infty \mathrm{d}
  \sigma \varsigma(\sigma) \varphi(\sigma,t) \; .
\end{equation}
We think of $\varphi$ as the displacement of an infinite
``hidden string'' in $\mathbb R \times H$. The equilibrium
configuration of this string is $\mathbb R \times \{ 0\}$, and
displacements in the directions described by $H$ move
harmonically, driven by the time dependent force $\varsigma(s)f(t)$.

This explicit extension is an example of what we call a
\emph{Quadratic Hamiltonian extension} of \eqref{driven}. Namely, it
is a dynamical system described by a vector coordinate $U$, taking
values in an extended phase space $\mathcal V$, with the following
properties:
\begin{enumerate}
\item The system is a \emph{quadratic Hamiltonian system}.  That is, there are
an extended symplectic operator $\mathcal J : \mathcal V \rightarrow \mathcal V$ and
an extended impedance operator $\mathcal K : \mathcal V \rightarrow
\mathcal H$, taking values in extended stress space, such that the
evolution of  $U\in \mathcal V$ is governed by
\begin{subequations}\label{eq:QHE2}
\begin{align}\label{eq:QHE2a}
\partial_t U(t) \ &= \ \mathcal J \mathcal K^{\mathrm{T}} F(t) \ + \ R(t) , \\
F(t) \ &=  \ \mathcal K U(t) \label{eq:QHE2b}
\end{align}
\end{subequations}
with $R(t)$ the external force. In other words, the dynamics are
Hamiltonian with symplectic form
$\mathfrak J(U,U') = \langle U, \mathcal J U' \rangle$ and  Hamiltonian
\begin{equation}\label{eq:QHE1}
\mathrm{H}(U) \ = \ \frac{1}{2} \langle \mathcal K U, \mathcal K U
\rangle.
\end{equation}

\item The system \emph{extends} (\ref{driven}), as follows. There are
isometric injections \begin{equation}\label{eq:QHE2c} I_V :V
\rightarrow \mathcal V \quad \text{and} \quad I_H : H \rightarrow
\mathcal H \end{equation} such that
\begin{equation}\label{eq:QHE3}
I_H K \ = \ \mathcal K I_V \, , \quad \mathcal J I_V \ = \ I_V J ,
\end{equation}
and the solution $u(t)$ to \eqref{driven}, with given initial
condition $u_{-\infty} \in \ker K$ and driving force $\rho(t)$, is
$u(t) = I_V^{\mathrm{T}} U(t)$, where $U(t)$ is solves
\eqref{eq:QHE2} with
\begin{equation}
  \lim_{t \rightarrow -\infty} U(t) \ = \ I_V u_{-\infty} ,  \quad \text{and } R(t) = I_V \rho(t)
  .
\end{equation}
\end{enumerate}
\begin{figure}
\includegraphics{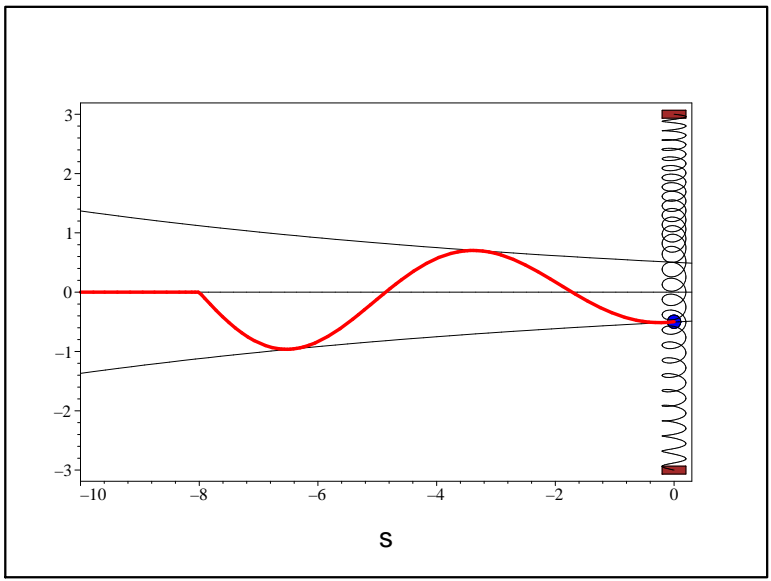}
\caption{The Lamb model introduced in \cite{Lamb} to describe
radiation damping, is a point mass attached to an infinite elastic
string and a Hook's law spring. The point mass evolves as a
classical linearly damped oscillator.}\label{Lamb}
\end{figure}

Thus the TDD dynamics of \eqref{driven} may be modeled by
describing $u(t)$ as one component of an extended vector.
The motion of the extended system is reversible, but an
irreversible motion of the underlying TDD system results. This is
demonstrated in its simplest form by the Lamb model \cite{Lamb}
\textemdash\  see Fig.\ \ref{Lamb} \textemdash \ in which the energy
of an oscillator escapes to infinity along an attached flexible
string. For a simple damped harmonic oscillator, the Hamiltonian
theory proposed here is \emph{precisely} the Lamb model, and is
otherwise a generalization of the Lamb model, obtained by coupling
an infinite classical elastic string to every degree of freedom of
the initial Hamiltonian system, illustrating that, from the
standpoint of thermodynamics, dissipation in classical linear
response is an idealization which assumes infinite heat capacity of
(hidden) degrees of freedom.

\subsection{Evolution in stress space and a minimal extension}\label{sec:minimal}
The extension just described is closely related to the extension
theory of \cite{FS}, summarized in \S\ref{sec:dis}. To understand
the relation between the present work and \cite{FS}, it is useful to
recast the evolution \eqref{driven} in stress space. If $\chi(\tau)$
is, say, continuous on $[0, \infty)$ and differentiable for $\tau >
0$ then, by \eqref{driven2},
\begin{align}  \label{cov3}
\partial_t Ku(t) \ &= \ \partial_t f(t) + \int_0^\infty \mathrm{d} \tau \,
\chi(\tau) \partial_t f(t - \tau ) \mathrm d \tau \\
&= \ \partial_t f(t) + \chi(0) f(t) + \int_0^\infty \mathrm{d} \tau
\, [\partial_\tau \chi](\tau) f(t - \tau ) \mathrm d \tau\; . \notag
\end{align}
Combining this with \eqref{driven1}, we obtain:
\begin{equation}  \label{cov4}
\partial_{t} f(t) \ = \ K J K^{\mathrm{T}} f(t) - \int_{[0,\infty)} \mathrm{d%
} \tau \, a(\tau) f(t - \tau) + K \rho(t) \; ,
\end{equation}
where $a$ is the operator valued distribution
\begin{equation}  \label{cov5}
a(t) \ = \ \chi(0) \delta(t) \ + \ \partial_t \chi(t) \; .
\end{equation}

The evolution \eqref{cov4} is essentially of the form \eqref{vA1},
with the minor difference that it is defined on a \emph{real}
Hilbert space with a skew-symmetric generator.  This is of no
consequence, as the main result of \cite{FS} holds in this context:
  \begin{theorem}\label{FSthm}
  Suppose given a linear dynamical system described by a point $f$ taking values
   in a real Hilbert space $H$ which evolves according to \eqref{vA1} with a skew-symmetric
   generator $L= -L^{\mathrm{T}}$.  If the friction function $a(\cdot) = \alpha_\infty \delta(\cdot)
   + \alpha(\cdot)$ satisfies the power
   dissipation condition
   \begin{equation}
    \int_{-\infty }^{\infty }\int_{-\infty }^{\infty }\left( f\left(
    t\right) ,a_{e}\left( t-\tau \right) f\left( \tau \right) \right)
    \,\mathrm{d}t\mathrm{d}\tau  \ge 0 \label{ava4} ,
    \end{equation}
    for compactly supported continuous functions $f:\mathbb{R} \rightarrow H_0$,
    where
    \begin{equation}
    a_{e}\left( t\right) =2\alpha _{\infty }\delta \left( t\right)
    +\left\{ \begin{array}{ccc}
    \alpha \left( t\right) & \text{if} & t>0 \\
    \alpha ^{\dagger }\left( -t\right) & \text{if} & t<0%
    \end{array}%
    \right. \text{ , }-\infty <t<\infty ,  \label{ava2}
     \end{equation}
     then there exist a real Hilbert space extension
   $\mathcal H \overset{I}{\hookleftarrow} H$ and a skew-symmetric operator $\mathcal L$ defined
   on $\mathcal H$ such that \eqref{solvevA1} holds.

    If, furthermore, the pair $(\mathcal H, \mathcal L)$ is \emph{minimal}, in the sense that
    $\mathcal  H$ is the smallest invariant subspace for $\mathcal L$ containing the range of $I$,
    then the pair $(\mathcal H, \mathcal L)$ is unique up to
    transformation by an orthogonal isometry.
\end{theorem}
\noindent \textit{Remark}: The existence of an extension
follows from the results of \cite{FS} applied to the complexification of
\eqref{vA1}(with $f$ a point in  $\mathbb C H$).
This extension will not be minimal in general, but we can restrict
the generator to a suitable real subspace to get
the minimal extension.  Uniqueness may be verified by the
arguments of \cite{FS}. For completeness we give a more detailed
sketch of the proof in Appendix \ref{Sketch}.

The power dissipation condition \eqref{PDC} of the present work
implies  the PDC \eqref{ava4} of \cite{FS}  for the friction
function $a$ defined in \eqref{cov5}, since
\begin{equation}
a_e(t) \ = \ \partial_t \chi^{\mathrm{o}}(t) ,
\end{equation}
with $\chi^{\mathrm{o}}$ the odd extension of the susceptibility
$\chi$,
\begin{equation}  \label{chio}
\chi^{\mathrm{o}}(\tau) \ : = \
\begin{cases}
\chi(\tau) \; , & \tau > 0 \\
-\chi(-\tau)^{\mathrm{T}} \; , & \tau < 0 \; .%
\end{cases}
\end{equation}
In the present work,  as in \cite{FS},
the energy of the dissipative system at time $t$ is
$\mathcal{E}(t) := \frac{1}{2} \| f(t)\|^2$. For a
trajectory $f(\cdot)$ which evolves according to \eqref{cov4} this
gives a total change in energy from $t=-\infty$ to $t=+\infty$ of
\begin{equation}  \label{ava3}
\int_{-\infty}^\infty \partial_t \mathcal{E }%
(t)   \mathrm{d} t \ = \ \mathcal{W}_{\mathrm{fr}}[f] +
\int_{-\infty}^\infty \mathrm{Re}  \left ( f(t), r(t) \right )
\mathrm{d} t ,
\end{equation}
where
\begin{equation}
\mathcal{W}_{\mathrm{fr}}[f]:= - \frac{1}{2} \int_{-\infty }^{\infty
}\int_{-\infty }^{\infty }\left( f\left( t\right) ,a_{e}\left(
t-\tau \right) f\left( \tau \right) \right)
\,\mathrm{d}t\mathrm{d}\tau ,  \label{ava4a}
\end{equation}
It is natural to interpret the two terms on the r.h.s.\ of
\eqref{ava3} as the total work done by the frictional and external
forces, respectively. Thus, the PDC \eqref{ava4} essentially requires that
\emph{the
total work done by frictional forces is always non-positive}. This
physically natural property also
provides a necessary and sufficient condition for an extension of the form (\ref%
{intro1}) to exist.

The theorem guarantees the existence of a unique minimal
extension of the form \eqref{intro1} to the evolution in stress
space \eqref{cov4}. However, for this to be a \emph{Hamiltonian}
extension we must impose a
Hamiltonian structure on the dynamical system \eqref{intro1}.  In
particular we must express the generator $\mathcal L$ as a product
\begin{equation}\label{productrepresentation}
\mathcal L \ = \ \mathcal K \mathcal J \mathcal K^{\mathrm{T}} ,
\end{equation}
with $\mathcal J $ a symplectic operator.
But, given a skew-adjoint
operator, there are in general many ways  to decompose it in this
fashion and thus many ways to impose a Hamiltonian structure on the
evolution \eqref{intro1}.  For the resulting structure to be
naturally related to the Hamiltonian structure of the original
dynamical system (\ref{hpq2},\ref{hpq3}) it is necessary that $\mathcal K$ and
$\mathcal J$ \emph{extend} the original operators $K$ and  $J$
respectively.  The
main point of this work is to exhibit an explicit Hamiltonian
extension with these properties, that may then be used in the
analysis of conservation laws for the dissipative system
\eqref{driven}.

So, by following the motion of the \emph{extended stress vector}
$F(t) =  \mathcal K U(t) $, we find one
extension of the type guaranteed by Theorem \ref{FSthm}. Indeed, \eqref{eq:QHE2} implies
\begin{equation}\label{extstressevol}
  \partial_t F(t) \ = \ \mathcal{K J K}^{\mathrm{T}} F(t) \ + \
  I_H K \rho(t) ,
\end{equation}
where the generator
$\mathcal{KJK}^{\mathrm{T}}$ is skew-symmetric and has the formal
expression
\begin{equation}\label{extstressgen}
  \mathcal{KJK}^{\mathrm{T}} \ = \ \begin{pmatrix}
    KJK^{\mathrm{T}} & - T & 0 \\
    T^{\mathrm{T}} &  0 & \partial_s \\
    0 & \partial_s & 0
  \end{pmatrix} .
\end{equation}
The solution to \eqref{extstressevol} is easily expressed
\begin{equation}\label{FSolve}
  F(t) \ = \ \int_{-\infty}^t \mathrm e^{(t-t')\mathcal{KJK}^{\mathrm{T}}} I_H K \rho(t')
  \mathrm d t',
\end{equation}
in terms
of the one parameter group $\mathrm e^{t\mathcal{KJK}^{\mathrm{T}}}$
of orthogonal transformations.  By the properties of the QHE, the solution to \eqref{cov4} is
therefore expressed as
\begin{equation}
  f(t) \ = \ I_H^{\mathrm{T}} F(t) \ = \ \int_{-\infty}^t I_H^{\mathrm{T}}
  \mathrm e^{(t-t')\mathcal{KJK}^{\mathrm{T}}} I_H K \rho(t')
  \mathrm d t'.
\end{equation}

It is natural to ask is the extension \eqref{extstressgen} is
the unique minimal extension of Theorem \ref{FSthm}.  In
fact, it is \emph{not} minimal.  Indeed, easily shows that
any configuration of the hidden string resulting from a physical
driving force $I_V \rho(t)$ is symmetric under $s \leftrightarrow
-s$.   That is, we still have a QHE if we replace
$\mathcal V$ and $\mathcal H$ respectively by
 \begin{equation}\label{eq:symmetricsubspaces}
 \mathcal V_s \ = \ V \oplus \mathcal S(\mathbb R, H) \oplus \mathcal S(\mathbb R, H)\,  \text{ and }
 \mathcal H_s  \ = \ H \oplus \mathcal S(\mathbb R, H) \oplus \mathcal A(\mathbb R, H) ,
 \end{equation}
 with
\begin{equation}\label{eq:symmetricsubspaces2}
\begin{aligned}
\mathcal S(\mathbb R, H) \ &= \ \{ \phi \in L^2(\mathbb R, H) \ : \ \phi(s) = \phi(-s) \} \\
\mathcal A (\mathbb R, H) \ &= \ \{ \phi \in L^2(\mathbb R, H) \ : \
\phi(s) = -\phi(-s) \}.
\end{aligned}
\end{equation}
Note that $\mathcal K: \mathcal V_s \rightarrow \mathcal H_s$ and
that $\mathcal J:\mathcal V_s \rightarrow \mathcal V_s$.  If the
kernel of the susceptibility $\ker \widehat \chi(\omega)$ is
non-trivial on a set of positive measure we will see in
\S\ref{StringFT} that further reductions are possible.

There is however no harm in working with an extension which is
non-minimal, which we do for convenience of notation.  Indeed, by
\eqref{FSolve}, the solution $F(t)$ remains in the subspace
$\mathcal H_0$ that is the smallest invariant subspace for $\mathcal
K \mathcal J \mathcal K^{\mathrm{T}}$ containing the range of $I_H$.
The restriction of $\mathcal K \mathcal J \mathcal K^{\mathrm{T}}$
to this subspace is the unique minimal extension of
Theorem \ref{FSthm}.  Thus even if we employ a non-minimal
extension, we effectively work with the unique minimal
extension anyway.  In \S\ref{StringFT} we give an explicit
description of $\mathcal H_0$ as well as the minimal subspace
$\mathcal V_0 \subset \mathcal V$ such that $\mathcal K : \mathcal
V_0 \rightarrow \mathcal H_0$ and $\mathcal J : \mathcal V_0
\rightarrow \mathcal V_0$.

Finally, we note that even in the minimal extension there is a
great deal of freedom to change variables  and thus alter the
explicit expressions for the extended impedance $\mathcal K$.
Indeed, given a symplectic operator $\mathcal J$ there is a natural
\emph{symplectic group} of symmetries of phase space $\mathcal V$
consisting of linear maps $\mathcal M$ such that $\mathcal
M^\mathrm{T} \mathcal J \mathcal M  =  \mathcal J $. Likewise the
Hamiltonian $\frac{1}{2} \| \mathcal K U\|^2$ does not change if we
replace the impedance $\mathcal K$ by $\mathcal O \mathcal K$ with
$\mathcal O$ any \emph{orthogonal map} of stress space, $\mathcal
O^\mathrm{T} \mathcal O  =  \mathbf{1} .$ Thus the impedance
$\mathcal K$ is essentially defined only up to re-parameterizations
of the form
\begin{equation}\label{reparameterization}
\mathcal K \ \mapsto \ \mathcal O \mathcal K \mathcal M^{-1}, \quad
\mathcal O^\mathrm{T} \mathcal O \ = \ \mathbf{1}  \text{ and }
\mathcal M^\mathrm{T} \mathcal J \mathcal M \ = \ \mathcal J .
\end{equation}
We refer to a combined mapping \eqref{reparameterization} of phase
and stress space as a \emph{symplectic/orthogonal} isomorphism. (A
symplectic map $\mathcal M^{\mathrm{T}} \mathcal J \mathcal M =
\mathcal J$ need not be bounded in infinite dimensions, making it
somewhat difficult to formulate the change of variables
\eqref{reparameterization} in complete generality.)

\subsection{Relation with the previous literature\label%
{relation}}

Analysis of a dispersive and dissipative medium based on the
construction of
its Lagrangian or Hamiltonian is a well established area, see \cite%
{LaxNelson, Loudon, NelsonChen, MH, MR} and references therein.
However, all of those works have relied on specifying an underlying
micro-structure for
the material medium, such as an infinite lattice of dipoles as in \cite%
{LaxNelson}. In contrast, our approach is phenomenological. Our
hidden variables are not ``real'' microscopic variables as in
\cite{LaxNelson}, but describe effective modes which exactly produce
a prescribed causal frequency dependent susceptibility. As regards
the underlying microscopic theory, our construction can be seen as
giving an \emph{effective Hamiltonian} for those modes well
approximated by linear response.

In this section, we compare the approach developed in
this paper and our previous work \cite{FS} with a number of
other efforts to describe dissipative and or dispersive media via
extensions (instead of microscopic variables).

\subsubsection{Dilation theory}
The dilation theory \textemdash \ beginning with the Sz.-Nagy--Foias
theory of contractions \cite{SzN,SzNF} and Naimark's theory of
positive operator valued measures \cite{Naimark} and subsequently
extended by a number of other authors \textemdash \ was the first
general method for constructing a spectral theory of dissipative
operators and has ultimately provided a complete treatment of
dissipative linear systems \emph{without dispersion}. A key
observation of our previous work \cite{FS} is that many of the
classical tools of dilation theory, in particular Naimark's theorem,
are useful for describing the generic case of dissipative and
dispersive systems.

Let us recall the basics of the dilation theory as presented by
Pavlov in his extensive review \cite{Pav1} and his more
recent work \cite{Pav2}. Although there are a number of approaches
to the subject, Pavlov uses Lax-Phillips scattering theory,
\cite{LaxPh}, which provides a conceptually useful picture of the
extended operators. That theory
assumes the existence of: (i) a dynamical unitary evolution group $U_{t}=\mathrm{e}^{%
\mathrm{i}\Omega t}$ in a Hilbert space $H$ where $\Omega $ is a
self-adjoint operator in $H$; (ii) an \emph{incoming} subspace
$D_{-}\subset
H$ invariant under the semi-group $U_{t}$, $t<0$, and an \emph{%
outgoing} subspace $D_{+\text{ }}\subset H$ invariant under to the
semi-group $U_{t}$, $t>0$. The invariant subspaces $D_{\pm \text{
}}$ (called \emph{scattering channels}) are assumed to be
orthogonal. Then one introduces the \emph{observation} subspace
$Q=H\ominus \left( D_{-}\oplus D_{+}\right) $, assumed to be
\emph{co-invariant with respect to the unitary group,} in the sense
that the restriction of $U_{t}$, $t>0$, to $Q$ is a semigroup, i.e.,
\begin{equation}
Z_{t}=\left. P_{Q}U_{t}\right\vert _{Q}=\mathrm{e}^{\mathrm{i}Bt},
\label{zp1}
\end{equation}%
where $P_Q$ is the orthogonal projection onto $Q$.

In many interesting cases the generator $B$ of the semigroup $Z_{t}$
is dissipative, i.e. $\operatorname{Im}B\geq 0 $, and the relation
(\ref{zp1}) provides a natural setting for dissipative operators
within the Lax-Phillips scattering theory. The dilation theory reverses
the Lax-Phillips construction by
constructing the Lax-Phillips spaces given $B$ and $Q$.  The
generator $\Omega$ of the constructed unitary group is called the
dilation of $B$ and has the property
\begin{equation}
f(B) \ = \ P_{Q} f(\Omega) \vert_Q
\end{equation}
for suitable analytic functions $f$. Thus the self-adjoint operator
$\Omega$ provides an effective spectral theory for the non-self
adjoint $B$.

Unfortunately, the dilation theory fails to describe many important
physical situations simply because the assumption that dissipation occurs
without dispersion, i.e. that $Z_t$ is a
semi-group, is too restrictive.  In  systems such as  (\ref{vA1})
dissipation comes with dispersion, and dilation
theory applies only in the very special case of instantaneous
(Markovian) friction $a\left( t\right) =\alpha_{\infty}\delta \left( t\right)
$. Many phenomenological models, such as Lorentz or Debeye
dielectric media, employ friction functions which are not
instantaneous. For such systems one must use a more general approach
as developed in \cite{FS} and here.

\subsubsection{The work of Tip} The recent papers of Tip  \cite{Tip, Tip2}
is more closely related to this work. For the special case of the EM field
in a linear absorptive dielectric, he has given a
Hamiltonian formalism involving auxiliary fields similar to our
``hidden string.''  His formalism made possible an analysis of
energy conservation, scattering, and quantization \cite{Tip} and led
to a clarification of the issue of boundary conditions in piecewise
constant dielectrics \cite{Tip2}.  Stallinga \cite{Stallinga} has
used this formalism to give formulas for the energy density and
stress tensor in dielectric media.

We do not rely on Tip's work, but the present work follows and
parallels it to some extent. In particular, the present paper gives
a general context in which some results of \cite{Tip, Tip2}
may be seen as special cases of results valid in a large class of
linear dispersive Hamiltonian systems.

\subsubsection{Heat bath and coupling\label{heatbath}}
The evolution equations (\ref{ExDy2}, \ref{ExDy3})
describing the hidden string are identical to those of a so-called
\emph{canonical heat bath} as defined in \cite[Section 2]{JP2},
\cite[Section 2]{RB}.
As the canonical heat bath as described in \cite[Section 2]{JP2}, \cite[%
Section 2]{RB} has naturally appeared in our construction, let us
look at it in more detail.

The Hamiltonian of our extended system \eqref{ExHamiltonian} can be
expressed  as a sum of two contributions
\begin{equation}\label{splittheenergy}
\mathrm{H}(U) \ = \ \mathrm{H}_{\mathrm{sys}}(U) +
\mathrm{H}_{\mathrm{str}}(U) ,
\end{equation}
the \emph{system energy}
\begin{equation}
\mathrm{H}_{\mathrm{sys}}(U) \ = \  \frac{1}{2}\left\Vert
Ku-T\varphi \right\Vert _{H}^{2} \ = \ \frac{1}{2} \left \Vert f
\right \Vert^2,
\end{equation}
where $f$ is the kinematical stress as defined by (\ref{ExStress}),
and the \emph{string energy}
\begin{equation}\label{Hhb1}
\mathrm{H}_{\mathrm{str}}(U) \ = \ \frac{1}{2}
\int_{-\infty}^{\infty }\left[ \left\Vert \theta \left( s\right)
\right\Vert _{H}^{2}+\left\Vert \partial _{s}\varphi \left( s\right)
\right\Vert _{H}^{2}\right] \,\mathrm{d}s.
\end{equation}
We conceive of $\mathrm{H}_{\mathrm{sys}}$ as the energy of an open
system  dynamically coupled to a ``heat bath'' with energy $\mathrm{H}_{\mathrm{str}}$,  described by the hidden string.

The physical concept of a heat bath originates in statstical
mechanics. There general considerations show that for a system
to behave according to thermodynamics it should be properly
coupled to a heat bath. Dynamical models at the
mathematical level of rigor were introduced, motivated and described
rather recently, see \cite[Section 1]{KKS}, \cite[Section 2]%
{JP2}, \cite[Section 2]{RB} and references therein. According to the
references, based on statistical mechanical arguments, the heat
bath must be governed by a self-adjoint operator with absolutely
continuous spectrum and no gaps, i.e. the spectrum must be the
entire real line $\mathbb{R}$, and the spectrum must be of a uniform
multiplicity. These requirements lead to a
system equivalent to a system with the Hamiltonian $%
\mathrm{H}_{\mathrm{str}}(\varphi ,\theta )$ as in (\ref{Hhb1}), \cite[%
Section 2]{JP2}.

Our construction of the unique extended Hamiltonian produces
an auxiliary system with Hamiltonian in the
form $\mathrm{H}_{\mathrm{%
str}}(\varphi ,\theta ) $ as in (\ref{Hhb1}) and gives another way to
obtain the canonical heat bath as a natural part of the conservative
system extending a dissipative and dispersive one under the
condition of its causality. In our Hamiltonian setting
(\ref{splittheenergy}) the coupling $\left( Ku,T\varphi \right) $
can be classified as the \emph{dipole approximation}, \cite[Section
 1,2]{RB}, associated with a bilinear form.

\subsection{Organization of the paper}
The main body of this paper has two parts. The first, comprising
Sections \ref{construction} -- \ref{maxwell}, is essentially the
physics part of the paper. It consists of a formal derivation of the
quadratic Hamiltonian extension (\S\ref{construction}), containing
all relevant physical details,  followed by an application of the
extension to TDD wave equations (\S\ref{local}) with Maxwell's
equations for the electro-magnetic field in a TDD medium considered
as a detailed example (\S\ref{maxwell}).  In particular, in
\S\ref{local} we write the extended Hamiltonian for a TDD wave
system as the integral of a local energy density and derive
expressions for the energy flux and stress tensor. We also derive
general approximations for the time average of these quantities in
the special case of an almost mono-chromatic wave. In
\S\ref{maxwell} we specialize these formulas to the Maxwell
equations.

The second part, consisting solely of Section \ref{result}, is a
more detailed mathematical examination of the quadratic Hamiltonian
extension. Here we give a precise formulation and proof of the main
results leading to Theorem \ref{thm:main}, with a rigorous analysis
of the unbounded operators involved.

The appendices contain supplementary material, including  a.)\ a
brief review of notation and spectral theory for operators on real
Hilbert spaces, b.) a sketch of the proof of Thm.\ \ref{FSthm}, c.)\
a few examples illustrating the application of our construction to
non-linear systems with linear friction and d.)\ a derivation of the
symmetric stress tensor for a system with a Lagrangian density, used
in Section \ref{local}.\maketitle

\section{Formal construction of a Hamiltonian}\label{construction}
We begin by analyzing extended systems of the type outlined in
(\ref{ExHamiltonian}-\ref{eq:calJ}), with an unspecified
symmetric operator valued coupling function $\varsigma(s)$.  It is a
simple matter to obtain, via a formal calculation,
evolution equations of the form (\ref{driven}) for the reduced
system. In this way, we get a relation between the susceptibility
and the coupling $\varsigma$ \textemdash \
\eqref{fb11} below. As it turns out, the symmetry \eqref{symchi} and
power dissipation \eqref{PDC} conditions are necessary and
sufficient for inverting \eqref{fb11} to write $\varsigma$ as a
function of $\chi$.

The extensions we consider are described by a vector coordinate $U$
taking values in the extended phase space $ \mathcal{V}  =  V \oplus
L^2(\mathbb{R}, H)
  \oplus L^2(\mathbb{R}, H),$ with $U \in \mathcal V$ denoted
\begin{equation}\label{HPQ5}
U= \begin{pmatrix} u \\ \theta(s) \\
\varphi(s)
\end{pmatrix} \; .
\end{equation}
Recall that we interpret $\phi(s)$ and $\theta(s)$ as the
displacement and momentum density of an $H$-valued string,
consistent with the equations of motion
(\ref{ExDy1}-\ref{ExStress}), namely
\begin{align}
\partial_t u(t) \ &= \ J K^{\mathrm{T}} f(t) + \rho(t)\label{St0} \\ \partial_t
  \theta(s,t)  \ &= \ \partial_s^2
  \varphi(s,t) +
  \varsigma\left(  s\right)  f(t) ,
  \label{St2} \\
\label{St1}
  \partial_t \varphi(s,t)  \ &= \
  \theta(s,t)
\end{align}
with kinematical stress $f$,
\begin{equation}\label{abstractf}
  f(t) \ = \ K u(t) - \int_{-\infty}^\infty \mathrm{d}
  \sigma \varsigma(\sigma) \varphi(\sigma,t) \; .
\end{equation}
Here we take $\varsigma$ to be an (as yet) unspecified operator
valued distribution.

Upon eliminating $\theta$ from (\ref{St2}, \ref{St1}), we find that
the string displacement $\varphi$ follows a driven wave equation
\begin{equation}
\label{drivenwave}\left\{  \partial_{t}^{2} - \partial_{s}^{2}
\right\} \varphi\left( s,t\right)  = \varsigma\left(s\right) f(t) .
\end{equation}
Taking $f$ as given, we solve \eqref{drivenwave} for $\varphi$
with initial values $\lim_{t \rightarrow -\infty} \varphi(t) = 0 $, $\lim_{t \rightarrow -\infty} \partial_t \varphi(t) = 0$,
corresponding to the string at rest in the distant past:
\begin{equation}
\label{solvedrivenwave} \varphi\left( s,t\right)=
 \frac{1}{2}\int_{0}^{\infty} \mathrm{d} \tau \int_{s-\tau}^{s+\tau} \mathrm{d}
\sigma\, \varsigma \left( \sigma\right) f(t-\tau),
\end{equation}
 where we have tacitly assumed that $f(t)$ is integrable. Recalling
that $f$ is related to $\varphi$ by \eqref{abstractf}, we obtain the
following equation relating $f$ and $u$
\begin{equation}\label{fb10}
f(t) + \frac{1}{2} \int_{-\infty}^\infty \mathrm{d}
  s \, \int_{0}^{\infty} \mathrm{d} \tau
  \int_{s-\tau}^{s+\tau} \mathrm{d}
\sigma\, \varsigma(s)  \varsigma  \left ( \sigma\right)f(t-\tau)  \
= \ K u(t) \; ,
\end{equation}
which is of the form of the generalized material relation
\eqref{driven2} with susceptibility
\begin{equation}
\label{fb11}\chi\left(  \tau \right)  = \frac{1}{2}\int_{-\infty
}^{\infty} \mathrm{d} s  \int_{s - \tau}^{s + \tau} \mathrm{d}
\sigma\, \varsigma\left( s \right) \varsigma\left( \sigma\right)  \;
.
\end{equation}

Thus, the reduced system described by $u$ is a TDD system of the
form (\ref{driven}), with susceptibility given by \eqref{fb11}. To
construct a quadratic Hamiltonian extension to \eqref{driven} it
essentially suffices to write the string coupling $\varsigma$ as a
function of the susceptibility $\chi$ by inverting \eqref{fb11}.
Note that the r.h.s.\ of \eqref{fb11} is a symmetric operator, so
the symmetry condition \eqref{symchi} is certainly necessary. As we
will see the power dissipation condition \eqref{PDC} is also
necessary, and together the two are sufficient.

Note that \eqref{fb11} holds also for $\tau < 0$, with $\int_b^a :=
 - \int_a^b$, provided we replace $\chi$ by its odd extension
$\chi^{\mathrm{o}}$, defined in \eqref{chio}. Differentiating with
respect to $\tau$ then gives
\begin{equation}\label{fb13}
  \partial_{\tau} \chi^{\mathrm{o}}(\tau) \ = \
\frac{1}{2}\int_{-\infty }^{\infty} \mathrm{d} s \, \varsigma\left(
s \right) \left \{ \varsigma\left( s + \tau\right)  +
\varsigma\left( s - \tau\right) \right \} \; .
\end{equation}
If $\chi(0+) \neq 0$ then $\chi^{\mathrm{o}}$ has a jump
discontinuity at $0$ and \eqref{fb13}, which holds in the sense of
distributions, implies that $\varsigma$ includes a Dirac delta
contribution at $s=0$.

To understand the role of the PDC
\eqref{PDC} here, let us suppose that
\begin{equation}\label{pwl} \widehat \chi(\omega) = \lim_{\eta
\downarrow 0} \widehat \chi(\omega + \mathrm{i} \eta)
\end{equation}
exists and is continuous for $\omega \in \mathbb R$, as holds  for
instance if $\chi \in L^1[0,\infty)$. Then the PDC \eqref{PDC}
implies that
\begin{equation}\label{PDCo1}
  \omega \mathrm{Im} \widehat \chi(\omega) \ \ge \  0 \; ,
\end{equation}
which may be expressed as
\begin{equation}\label{PDCo5}
  \omega \mathrm{Im} \widehat \chi(\omega) \ = \
  \frac{1}{2} \widehat { \partial_\tau \chi^{\mathrm{o}}}(\omega)  \;
  ,
\end{equation}
where
\begin{equation}\label{PDCo3}
\widehat {\chi^{\mathrm{o}}}(\omega) \ =  \ \int_{-\infty}^\infty
\mathrm{d} \tau \, \mathrm{e}^{\mathrm{i} \omega \tau}
\chi^{\mathrm{o}}(\tau) \ = \ 2 \mathrm{i} \int_{0}^\infty
\mathrm{d} \tau
 \sin(\omega \tau) \chi(\tau) \; .
\end{equation}

To solve for $\varsigma$, we take the Fourier transform of
\eqref{fb13}, which by \eqref{PDCo5} is
\begin{equation}\label{fb14}
    2 \omega \mathrm{Im} \widehat \chi(\omega)
    \ = \  \frac{1}{2}\left \{ \widehat \varsigma(-\omega)
    \widehat \varsigma(\omega) + \widehat \varsigma(\omega)
    \widehat \varsigma(-\omega) \right \} \; ,
\end{equation}
with  $\widehat \varsigma(\omega) =
  \int_{-\infty}^\infty \mathrm{d} s  \mathrm{e}^{\mathrm{i}\omega s}
  \varsigma(s). $
Note that
\begin{equation}
  \widehat \varsigma(\omega)^\dagger \ = \
  \widehat \varsigma(-\omega) \; ,
\end{equation}
where $\bullet^\dagger$ denotes the Hermitian conjugate. Therefore
\eqref{fb14} is the same as
\begin{equation}\label{fb14a}
    2 \omega \mathrm{Im} \widehat \chi(\omega)
    \ = \  \frac{1}{2}\left \{ \widehat \varsigma(-\omega)
    \widehat \varsigma(-\omega)^{\dagger} +
    \widehat \varsigma(\omega) \widehat \varsigma(\omega)^{\dagger}
    \right \} \; .
\end{equation}
Clearly the r.h.s.\ is non-negative and we see, in particular, that
\eqref{fb11} implies the power dissipation condition \eqref{PDC}.
(Once the inequality $\omega \mathrm{Im} \widehat \chi(\omega) \ge
0$ is known on the real axis, it extends to the entire upper half
plane because $ \mathrm{Im} \zeta\widehat \chi(\zeta)$ is a harmonic
function.  See (\ref{Udefn}, \ref{herglotz}) below.)

A solution to \eqref{fb11} is not unique.  However, there is a
unique solution with $\widehat \varsigma(\omega)$ a
\emph{non-negative real symmetric operator} for each $\omega$, i.e.,
\begin{equation}\label{fb12}
  \widehat \varsigma(\omega) \ = \ \ = \ \widehat
  \varsigma(\omega)^\mathrm{T} \ = \ \widehat \varsigma(\omega)^\dagger \ =
  \ \widehat \varsigma(-\omega) ,
\end{equation} and
\begin{equation}\label{fb15}
   \widehat \varsigma(\omega)  \ \ge \ 0 \; .
\end{equation}
Indeed, under the symmetry condition \eqref{fb12}, eq.\
\eqref{fb14a} simplifies to
\begin{equation}\label{fb16}
    2 \omega \mathrm{Im} \widehat \chi(\omega)
    \ = \  \widehat \varsigma(\omega)^2 \; .
\end{equation}
(This is consistent since $\mathrm{Im}\widehat  \chi(\omega)$ is a
real operator as we see from the formula $
  \mathrm{Im} \widehat \chi(\omega)  =
  \int_0^\infty \mathrm{d} t
  \sin(\omega t) \chi(t) $.)
The  unique non-negative solution to \eqref{fb16} is given by
the operator square root,
\begin{equation}\label{fb17}
  \widehat \varsigma(\omega) \ = \
  \sqrt{2 \omega
  \mathrm{Im} \widehat \chi(\omega)
  } \; .
\end{equation}
Thus, \emph{a quadratic Hamiltonian extension of the
system \emph{(\ref{driven})} is given by
\emph{(\ref{ExHamiltonian})} with the coupling function $\varsigma$
given by Fourier inversion of the r.h.s.\ of \eqref{fb17}, i.e.,
\begin{equation}\label{fb18}
  \varsigma(s) \ = \ \frac{1}{2 \pi} \int_{\mathbb{R}}
  \mathrm{d} \omega \, \cos(\omega s) \sqrt{
  2 \omega
  \mathrm{Im} \widehat \chi(\omega)} \; ,
\end{equation}
which is \eqref{Excoupling}.}

\subsection{A minimal extension}\label{StringFT}
The system with Hamiltonian \eqref{ExHamiltonian} has a mechanical
interpretation as strings coupled to the degrees of freedom of the
underlying TDD system and provides a conceptual picture of the TDD
dynamics in terms of absorption and emission of energy by those
``hidden'' strings. However, for calculations and to describe the
\emph{minimal} extension, it is easier to work with a system in
which the string displacement is replaced by its Fourier transform
\begin{equation}
\label{strft} \widetilde \varphi(\kappa) \ = \ \int_{-\infty}^\infty
\mathrm{e}^{\mathrm{i}\kappa s}
    \varphi(s) \mathrm{d} s .
\end{equation}
To make the change of variables symplectic, we replace the momentum
density $\theta$ by
\begin{equation}\label{strftmom}\widetilde \theta(\kappa)
    \ = \  \frac{1}{ 2\pi} \int_{-\infty}^\infty
    \mathrm{e}^{\mathrm{i}\kappa s} \theta(s)  \mathrm{d} s .
\end{equation}

The resulting transformation of phase space
 \begin{equation}\label{eq:transform}
 U \ \mapsto \ \mathcal M U \ = \ \begin{pmatrix}
1 & 0 & 0 \\
0 & \frac{1}{2 \pi} \mathcal F & 0\\
0 & 0 &  \mathcal F
              \end{pmatrix} U,
\end{equation}
with $\mathcal Ff(\kappa) \ = \ \int_{\mathbb R}\mathrm e^{\mathrm i
\kappa s} f(s) \mathrm d s$, is a symplectic map \textemdash \
$\mathcal M \mathcal J \mathcal M^{\mathrm{T}}= \mathcal J$, since
$\mathcal F^{-1} = \frac{1}{2\pi} \mathcal F^{\mathrm{T}}$. Note
that the Fourier transform $\mathcal F$ maps the symmetric space
$\mathcal S(\mathbb R, H)$, defined in
\eqref{eq:symmetricsubspaces2}, onto itself, since
$\phi(s)=\phi(-s)$ implies that $\mathcal F \phi(\kappa)$ is real.
Thus $\mathcal M$ is well defined as a symplectic map on the reduced
phase space $\mathcal V_s$ (see \eqref{eq:symmetricsubspaces}).
Correspondingly, we transform stress space by the orthogonal map
\begin{equation}\label{eq:Ftransform}
F \ \mapsto \ \mathcal O F \ = \ \begin{pmatrix} 1 & 0 & 0 \\
0 & \frac{1}{\sqrt{2\pi}} \mathcal F & 0 \\
0 & 0 &   \frac{1}{\sqrt{2\pi}} \mathrm{i}\mathcal F
              \end{pmatrix} F.
\end{equation}
The map $\mathrm{i} \mathcal F$ maps the anti-symmetric space
$\mathcal A(\mathbb R, H)$ (see \eqref{eq:symmetricsubspaces2}) onto
itself, so $\mathcal O$ is well defined as an orthogonal map of the
reduced stress space $\mathcal H_s$ (see
\eqref{eq:symmetricsubspaces}).  Together the two transformations
amount to an symplectic/orthogonal isomorphism of the form
(\ref{reparameterization}), and the impedance is transformed to
\begin{equation}\label{eq:OSFtransfrom}
\mathcal K \ \mapsto \ \widehat {\mathcal K} \ = \ \mathcal O
\mathcal K \mathcal M^{-1} ,
\end{equation}
where
\begin{equation}\label{ftimpedance}
\widehat{\mathcal K} \begin{pmatrix}
      u \\
      \widetilde{\theta}(\kappa) \\
      \widetilde{\varphi}(\kappa)
    \end{pmatrix}\ = \ \begin{pmatrix}
      K & 0 & -\frac{1}{2 \pi} \widehat{T} \\
      0 & \sqrt{2 \pi} \mathbf{1} & 0 \\
      0 & 0 &  \frac{1}{\sqrt{2 \pi}}\kappa
    \end{pmatrix} \begin{pmatrix}
      u \\
      \widetilde{\theta}(\kappa) \\
      \widetilde{\varphi}(\kappa)
    \end{pmatrix} .
\end{equation}
Here
\begin{equation}\label{widehatT}
\widehat T \widetilde \varphi \ = \ \int_{-\infty}^\infty \widehat
\varsigma(\kappa) \widetilde \varphi(\kappa) \mathrm d \kappa ,
\end{equation}
where  $\widehat \varsigma$ was defined in \eqref{fb17}.

The associated equations of motion, from the Fourier transform of
(\ref{St0}--\ref{abstractf}),  are
\begin{align}
\partial_t u(t) \ &= \ J K^{\mathrm{T}} f(t) \label{St0ft}  \\ \label{St2ft} \partial_t
   \widetilde \theta(\kappa,t)  \ &= \ - \frac{1}{2 \pi} \kappa^2
  \widetilde \varphi(\kappa,t) + \frac{1}{2 \pi}
  \widehat \varsigma (  \kappa )  f(t) \\
\label{St1ft}
  \partial_t \widetilde \varphi(\kappa,t)  \ &= \ 2 \pi
   \widetilde \theta(\kappa,t) ,
\end{align}
with
\begin{equation}\label{St3ft}
    f(t) \ = \ K u(t) - \frac{1}{ 2 \pi}\int_{-\infty}^\infty \widehat \varsigma(\kappa)
    \widetilde \varphi(\kappa,t) \mathrm{d} \kappa .\end{equation}
Combining \eqref{St2ft} and \eqref{St1ft} we obtain the Fourier
Transform of \eqref{drivenwave}
\begin{equation}
\partial_t^2 \widetilde \varphi(\kappa,t) \ = \ - \kappa^2 \widetilde \varphi(\kappa,t) +
\widehat \varsigma(\kappa) f(t) ,
\end{equation}
with solution
\begin{equation}
\widetilde \varphi(\kappa,t) \ = \ \widehat \varsigma(\kappa) \cdot
\left [ \int_{-\infty}^t \frac{\sin(\kappa (t-t'))}{\kappa} f(t')
\mathrm d t' \right ] .
\end{equation}
Clearly the resulting string displacement satisfies
\begin{equation}
\widetilde \varphi(\kappa,t) \ \in \ \operatorname{ran} \left (
\widehat \varsigma(\kappa) \right ) \text{ for every } \kappa \in
\mathbb R .
\end{equation}
The same holds for the momentum density $\widetilde \theta$, since
\begin{equation}
\widetilde \theta(\kappa,t) \ = \ \partial_t \widetilde
\varphi(\kappa,t) \ = \ 2 \pi \widehat \varsigma(\kappa) \cdot \left
[ \int_{-\infty}^t \cos(\kappa (t-t'))  f(t') \mathrm d t' \right ]
.
\end{equation}

Thus, we may restrict the phase space to the Hilbert space
\begin{equation}\label{calV0}
\mathcal V_0 \ = \ V \oplus \mathcal S_{\widehat \varsigma}(\mathbb
R, H) \oplus \mathcal S_{\widehat \varsigma}(\mathbb R, H) ,
\end{equation}
where
\begin{equation}
\mathcal S_{\widehat \varsigma}(\mathbb R, H) \ = \ \left \{ f \in
\mathcal S (\mathbb R, H) \ : \ f(\kappa) \in \operatorname{ran}
\left ( \widehat \varsigma(\kappa) \right ) \text{ for every }
\kappa \in \mathbb R \right \}.
\end{equation}
We denote by  $ \mathcal J_0$ and $\widehat{\mathcal K}_0$ the
restrictions of the symplectic operator $\mathcal J$ and impedance
$\widehat{\mathcal K}$ to $\mathcal V_0$. Thus $\mathcal J_0$ still
has the block matrix form (\ref{eq:calJ}) and $\widehat{\mathcal
K}_0$ is defined by the r.h.s.\ of (\ref{ftimpedance}) for vectors
$U=(u,\widetilde \theta, \widetilde \phi) \in \mathcal V_0$. We
consider the impedance $\widehat {\mathcal K}_0$ as a map from
$\mathcal V_0$ to the restricted stress space
\begin{equation}\label{calH0}
\mathcal H_0 \ = \ H \oplus \mathcal S_{\widehat \varsigma}(\mathbb
R, H) \oplus \mathcal A_{\widehat \varsigma}(\mathbb R, H)
\end{equation}
with (see \eqref{eq:symmetricsubspaces2})
\begin{equation}
\mathcal A_{\widehat \varsigma}(\mathbb R, H) \ = \ \left \{ f \in
\mathcal A (\mathbb R, H) \ : \ f(\kappa) \in  \operatorname{ran}
\left ( \widehat \varsigma(\kappa) \right ) \text{ for every }
\kappa \in \mathbb R \right \}.
\end{equation}
Clearly $\mathcal J_0, \widehat{\mathcal K}_0 $ give a quadratic
Hamiltonian extension to \eqref{driven}. We claim that the resulting
extension to \eqref{cov4} is the unique minimal extension guaranteed
by Theorem \ref{FSthm}.  Indeed the generator has the expression
\begin{equation}\label{eq:ftgenerator}
\widehat{\mathcal K}_0 \mathcal J_0\widehat{\mathcal
K}_0^{\mathrm{T}} \ = \ \begin{pmatrix}
K JK^{\mathrm{T}} & - \frac{1}{\sqrt{2 \pi}} \widehat T & 0 \\
\frac{1}{\sqrt{2 \pi}} \widehat T^{\mathrm T} & 0 & - \kappa \\
0 & \kappa & 0
                                                          \end{pmatrix} ,
\end{equation}
where $[\widehat T^{\mathrm T} f](\kappa) \ = \ \widehat
\varsigma(\kappa) f,$
 by \eqref{widehatT}.
One easily verifies there is no subspace of $\mathcal H_0$ invariant
under $ \widehat{\mathcal K}_0 \mathcal J_0\widehat{\mathcal
K}_0^{\mathrm{T}}$ and
 containing $H
\oplus 0 \oplus 0$  (the range of $I_H$).  Thus:
\begin{theorem}There exists a quadratic Hamiltonian extension with
$(\mathcal H, \mathcal K \mathcal J \mathcal K^{\mathrm{T}})$ the
unique minimal extension of Theorem \ref{FSthm}.
\end{theorem}


For the purpose of calculation it is sometimes useful to take the
Fourier-Laplace  transform \eqref{FourLap}  with respect to time,
setting
\begin{equation}
  \begin{pmatrix}
    \widehat u(\zeta)  \\
    \widehat \theta(\kappa, \zeta) \\
    \widehat \varphi(\kappa, \zeta)
  \end{pmatrix} \ = \ \int_{-\infty}^\infty \mathrm{e}^{\mathrm{i} \zeta t}
  \begin{pmatrix}
    u(t)  \\
    \widetilde \theta(\kappa, t) \\
    \widetilde \varphi(\kappa, t)
  \end{pmatrix} \mathrm{d} t , \quad \mathrm{Im} \zeta  > 0 .
\end{equation}
We obtain the system of equations
\begin{align}
-\mathrm{i} \zeta \widehat u(\zeta) \ &= \ J K^{\mathrm{T}} \widehat
f(\zeta) \label{St0fft}
\\ -\mathrm{i} \zeta
   \widehat \theta(\kappa, \zeta) \ &= \ -\frac{1}{2 \pi} \kappa^2
  \widehat \varphi(\kappa, \zeta) +
\frac{1}{ 2 \pi}  \widehat \varsigma (  \kappa )  \widehat f(\zeta)\label{St2fft} \\
  -\mathrm{i} \zeta \widehat \varphi(\kappa,\zeta)  \ &= \ 2 \pi
   \widehat \theta(\kappa, \zeta), \label{St1fft}
\end{align}
with
\begin{equation}\label{St3fft}
    \widehat f(\zeta) \ = \ K \widehat u(\zeta) - \frac{1}{ 2\pi}
    \int_{-\infty}^\infty \widehat \varsigma(\kappa)
    \widehat \varphi(\kappa,\zeta) \mathrm{d} \kappa .
\end{equation}
In particular (\ref{St2fft}, \ref{St1fft}) together imply
\begin{equation}\label{St4fft}
    \widehat \varphi(\kappa,\zeta) \ = \ \frac{1}{\kappa^2 - \zeta^2} \widehat
    \varsigma (\kappa) \widehat f(\zeta),
\end{equation}
which, with \eqref{St3fft}, yields
\begin{equation}\label{St5fft}
\widehat f(\zeta) + \frac{1}{2\pi}
    \int_{-\infty}^\infty \frac{1}{\kappa^2 - \zeta^2}
    \widehat \varsigma(\kappa) ^2 \mathrm{d} \kappa
     \widehat f(\zeta) \ = \ K \widehat u(\zeta) .
\end{equation}
This is suggestive of the identity
\begin{equation}\label{St6fft}
    \widehat \chi(\zeta ) \ = \ \frac{1}{2 \pi} \int_{-\infty}^\infty
    \frac{1}{\kappa^2 - \zeta^2}
    \widehat \varsigma(\kappa) ^2 \mathrm{d} \kappa ,
\end{equation}
which holds for $\widehat \varsigma$ as in \eqref{fb17} as we shall
see in the proof of Theorem \ref{thm:dissipation} below. In fact,
\eqref{St6fft} is a consequence of the Herglotz-Nevanlina
representation for an (operator valued) analytic function in the
upper half plane with non-negative imaginary part (see \cite[Section
59]{AkhGlaz} and \cite[Section 32.3]{Lax}) or, what is essentially
the same, the Kramers-Kronigs relations (see \cite[Sec.
62]{LandauLif}).

\subsection{TDD Lagrangian systems}\label{TDDLagrang} In many applications
the phase space $V$ decomposes naturally as $V_{\mathrm{p}} \oplus
V_{\mathrm{q}}$ ($V_{\mathrm{p}}= V_{\mathrm{q}}$), with the
symplectic operator in the canonical form:\footnote{Any symplectic
operator can be written in the form \eqref{jj2} by a suitable choice
of basis for $V$, abut the subspaces $V_{\mathrm{p,q}}$ are not
unique. See Lemma \ref{lJ2}.}
\begin{equation}\label{jj2}
  J = \begin{pmatrix}
    0 & -\mathbf{1} \\
  \mathbf{1} & 0
  \end{pmatrix} \; .
\end{equation}
The two components of $u =\begin{pmatrix} p \\ q \end{pmatrix} \in
V$ are ``momentum'' and ``coordinate'' respectively. If the linear
map $K$ is block diagonal
\begin{equation}
  K \ = \ \begin{pmatrix}
    K_{\mathrm{p}} & 0 \\
    0 & K_{\mathrm{q}}
  \end{pmatrix} \; ,
\end{equation}
then the linear maps $K_{\mathrm{p}}$ and $K_{\mathrm{q}}$ can be
thought of as follows:
\begin{eqnarray*}
&&K_{\mathrm{p}}^{\mathrm{T}}K_{\mathrm{p}}\text{ is the inverse
mass
(mobility) operator, and } \\
&&K_{\mathrm{q}}^{\mathrm{T}}K_{\mathrm{q}}\text{ is the stiffness
(inverse flexibility) operator.}
\end{eqnarray*}
Correspondingly, we suppose that $K_{\mathrm{p}}$ is boundedly
invertible, or at least invertible, as otherwise there are
``infinitely massive" modes. The equations of motion are
\begin{equation}
    \partial_t \begin{pmatrix}
      p(t) \\
      q(t)
    \end{pmatrix} \ = \ \begin{pmatrix}
      - K_{\mathrm{q}}^{\mathrm{T}} f_{\mathrm{q}}(t) \\
      K_{\mathrm{p}}^{\mathrm{T}} f_{\mathrm{p}}(t)
    \end{pmatrix} , \quad  \begin{pmatrix}
       f_{\mathrm{p}}(t) \\
       f_{\mathrm{q}}(t)
    \end{pmatrix} \ = \ \begin{pmatrix} K_{\mathrm{p}} p(t) \\
      K_{\mathrm{q}}q(t)
    \end{pmatrix}.
\end{equation}

A system in this form has an equivalent Lagrangian formulation, with
Lagrangian
\begin{equation}
\mathrm{L}(q,\partial_t q) \ = \ \langle p, \partial_t q \rangle -
\mathrm{h}(p,q) ,
\end{equation}
where we express $p$ as a function of $\partial_t q$ using the
equation of motion for $q$, i.e.,
\begin{equation}\label{ptoqdot}
p = \left [ K_{\mathrm{p}}^{\mathrm{T}} K_{\mathrm{p}} \right ]^{-1}
\partial_t q .
\end{equation}
As $K_{\mathrm{p}}$ is boundedly invertible, eq.\ \eqref{ptoqdot} is
unambiguous. Thus, the Lagrangian is
\begin{equation}\label{quadLang}
  \mathrm{L}(q,\partial_t q) \ = \ \frac{1}{2} \left \| \left [ K_{\mathrm{p}}^{-1}
  \right ]^{\mathrm{T}} \partial_t q
   \right \|_{H_{\mathrm{p}}}^2 -
   \frac{1}{2} \left \|  K_{\mathrm{q}}  q\right \|_{H_{\mathrm{q}}}^2 ,
\end{equation}
where we have $H = H_{\mathrm{p}} \oplus H_{\mathrm{q}}$ with
$K_{\mathrm{w}} \in \mathcal{L}(V_{\mathrm{w}}, H_{\mathrm{w}})$,
$\mathrm{w}=\mathrm{p,q}$. The trajectory $q(t)$ may be obtained
from the Lagrangian by noting that it is a stationary point for the
action $ \mathrm{A}([q(\cdot)];t_0,t_1) = \int_{t_0}^{t_1}
\mathrm{L}(q(t),\partial_t q(t)) \mathrm{d} t .$ The Euler-Lagrange
equation obtained by setting the variation of $\mathrm{A}$ equal to zero is
\begin{equation}\label{quadELang}
    \left [K_{\mathrm{p}}^{\mathrm{T}}
    K_{\mathrm{p}} \right ]^{-1} \partial_t^2 q \ = \
    -K_{\mathrm{q}}^{\mathrm{T}} K_{\mathrm{q}}q  ,
\end{equation}
which is formally equivalent to (\ref{hpq2}, \ref{hpq3}), with $u =
\begin{pmatrix} q \\ [K_{\mathrm p}^{\mathrm{T}} K_{\mathrm p }
]^{-1} \partial_t q \end{pmatrix}$.

For a Lagrangian system of this form, we generally make the
physically natural assumption that an external driving force
$\rho(t)$ couples through the r.h.s.\ of \eqref{quadELang}. That is
the equation of motion is
\begin{equation}\label{drivenquadELang}
     \left [K_{\mathrm{p}}^{\mathrm{T}}
    K_{\mathrm{p}} \right ]^{-1} \partial_t^2 q(t) \ = \
    -K_{\mathrm{q}}^{\mathrm{T}} K_{\mathrm{q}}q(t) + \rho(t) ,
\end{equation}
with $\rho(t) \in V_{\mathrm{p}}=V_{\mathrm{q}}$.  This amounts to
consider the time dependent Lagrangian $\mathrm{L}(q,\partial_t q,)
+ \langle q(t), \rho(t) \rangle_{V_{\mathrm{q}}}$, or Hamiltonian
$\mathrm{h}(q,\partial_t q) - \langle q(t), \rho(t)
\rangle_{V_{\mathrm{q}}}$.

For a TDD system \eqref{driven} with Hamiltonian $\mathrm{h}$ of
this form, the extended Hamiltonian (\ref{ExHamiltonian}),
$\mathrm{H} (U) = \frac{1}{2} \| \mathcal{K} U \|^2$, is of the form
\begin{equation}\label{extlagrK}
    \mathcal{K} \ = \ \begin{pmatrix}
      K_{\mathrm{p}} & 0  & 0 & -T_{\mathrm{p}} \\
      0 & K_{\mathrm{q}} & 0 & -T_{\mathrm{q}} \\
      0 & 0 & 1 & 0 \\
      0 & 0 & 0 & \partial_s
    \end{pmatrix} \; ,
\end{equation}
where $T_{\mathrm{p}}$, $T_{\mathrm{q}}$ are the $p$ and $q$
components of the coupling operator $T$ (see (\ref{Excoupling2})):
\begin{equation}\label{decomposeT}
    T \varphi \ = \ \begin{pmatrix}
      T_{\mathrm{p}} \varphi \\
      T_{\mathrm{q}} \varphi
    \end{pmatrix} \ = \ \int_{-\infty}^\infty \begin{pmatrix}
      \varsigma_{\mathrm{p}}(s) \\
      \varsigma_{\mathrm{q}}(s)
    \end{pmatrix} \varphi(s) \mathrm{d} s ,
\end{equation}
with momentum and coordinate string coupling functions $\varsigma _{\mathrm{p%
}}$ and $\varsigma _{\mathrm{q}}$ respectively. Notice that the
constitutive relation (\ref{abstractf}) turns into%
\begin{equation}
f_{\mathrm{q}}=K_{\mathrm{q}}q-T_{\mathrm{q}}\varphi ,\ f_{\mathrm{p}}=K_{%
\mathrm{p}}p-T_{\mathrm{p}}\varphi ,  \label{KHpq1}
\end{equation}%
readily implying the following representation for the Hamiltonian
\begin{multline}
\begin{aligned}
\mathrm{H}(U)\ & = \ \frac{%
1}{2}\left\{ \left\Vert f_{\mathrm{q}}\right\Vert _{H_{\mathrm{q}%
}}^{2}+\left\Vert f_{\mathrm{p}}\right\Vert _{H_{\mathrm{p}}}^{2}\right\} +%
\frac{1}{2}\int_{0}^{\infty }\left[ \left\Vert \theta \left(
s\right) \right\Vert _{H}^{2}+\left\Vert \partial _{s}\varphi \left(
s\right)
\right\Vert _{H}^{2}\right] \,\mathrm{d}s  \label{HKpq} \\
& = \  \frac{1}{2}\left\{ \left\Vert
K_{\mathrm{q}}q-T_{\mathrm{q}}\varphi
\right\Vert _{H_{\mathrm{q}}}^{2}+\left\Vert K_{\mathrm{p}}p-T_{\mathrm{p}%
}\varphi \right\Vert _{H_{\mathrm{p}}}^{2}\right\}  \\
& \quad +\frac{1}{2}\int_{0}^{\infty }\left[ \left\Vert \theta
\left( s\right) \right\Vert _{H}^{2}+\left\Vert \partial _{s}\varphi
\left( s\right) \right\Vert _{H}^{2}\right] \,\mathrm{d}s,
\end{aligned}
\end{multline}
where $H=H_{\mathrm{q}%
}\oplus H_{\mathrm{p}}.$

 We form a Lagrangian for the
extended system, taking $p$ and $\theta$ as momenta,
\begin{equation}\label{extLagr}
    \mathrm{L}(q, \varphi, \partial_t q, \partial_t \varphi) \ = \
    \langle p, \partial_t q \rangle + \langle \theta, \partial_t \varphi
    \rangle - \mathrm{H}(p,q,\theta,\varphi) .
\end{equation}
where we must write $p$ and $\theta$ as functions of $\partial_t q$,
$\partial_t \varphi$ and $\varphi$,
\begin{equation}
  \theta \ = \ \partial_t \varphi, \text{ and }
  p \ = \ \left [ K_{\mathrm{p}}^{\mathrm{T}} K_{\mathrm{p}} \right ]^{-1}
  \partial_t q + K_{\mathrm{p}}^{-1} T_{\mathrm{p}} \varphi ,
\end{equation}
using the equations of motion. Thus the above and (\ref{extLagr})
imply \begin{multline}\label{extLagr2}
    \mathrm{L}(q, \varphi, \partial_t q, \partial_t \varphi) \ = \
    \frac{1}{2} \left \| \left [ K_{\mathrm{p}}^{-1}
  \right ]^{\mathrm{T}}\partial_t q  \right
  \|_{H_{\mathrm{p}}}^2 + \left \langle \left [ K_{\mathrm{p}}^{-1}
  \right ]^{\mathrm{T}} \partial_t q,
  T_{\mathrm{p}}\varphi \right \rangle_{H_{\mathrm{p}}}
  \\ + \frac{1}{2} \int_{-\infty}^\infty
  \left \| \partial_t \varphi(s) \right \|_{H}^2 \mathrm{d} s
  - \frac{1}{2} \left \| K_{\mathrm{q}} q - T_{\mathrm{q}} \varphi \right
  \|^2_{H_{\mathrm{q}}} - \frac{1}{2} \int_{-\infty}^{\infty}
  \left \| \partial_s \varphi(s) \right \|_{H}^2 \mathrm{d} s.
\end{multline}
The Lagrangian form of the equations of motion, with driving force
$\rho$, is
\begin{equation}\label{lagr1}
  \partial_t \left \{ \left [ K_{\mathrm{p}}^{\mathrm{T}} K_{\mathrm{p}} \right ]^{-1}
  \partial_t q(t)  + K_{\mathrm{p}}^{-1} \int_{-\infty}^\infty
  \varsigma_{\mathrm{p}}(s)  \varphi(s,t) \mathrm{d} s  \right \}\ = \ -
  K_{\mathrm{q}}^{\mathrm{T}} f_{\mathrm q}(t) + \rho(t),
\end{equation}
\begin{equation}\label{lagr2}
  \partial_t^2 \varphi(s,t) \ = \ \partial_s^2 \varphi(s,t) +\varsigma_{\mathrm{q}}(s)  f_{\mathrm q}(t)
  - \varsigma_{\mathrm{p}}(s) \left [ K_{\mathrm{p}}^{-1}
  \right ]^{\mathrm{T}} \partial_t q(t) ,
\end{equation}
with
\begin{equation}\label{fqt}
 f_{\mathrm q}(t)  \ = \ K_{\mathrm{q}} q(t)  - \int_{-\infty}^\infty
  \varsigma_{\mathrm{q}}(s) \varphi(s,t)  \mathrm{d} s .
\end{equation}

\section{Local TDD Lagragians and conserved currents}\label{local}
Many physical systems of interest are described by wave motion of
vector valued fields, with the coordinate variable $u$ a function of
the position $\vec{r}\in \mathbb{R}^d$ (usu.\ $d=3$) valued in a
Hilbert space $\widehat V$. That is, phase space $V =
L^2(\mathbb{R}^d; \widehat V)$. Of particular interest are systems
with a Hamiltonian that is an integral over $\mathbb{R}^d$ of a
density, whose value at a point $\vec{r}$ is a function of the field
$u(\vec{r})$ and its derivatives at the point $\vec{r}$. In this
section we focus on extended TDD Lagrangian systems of this type
with $u(\vec{r}) = (p(\vec{r}), q(\vec{r}))$ and the symplectic
operator $J$ in the canonical representation \eqref{jj2}.

That is, we take a system of the type considered in
\S\ref{TDDLagrang} and suppose the spaces
$V_{\mathrm{q}}=V_{\mathrm{p}}$ and $H_{\mathrm{p,q}}$ are of the
form
\begin{equation}\label{localspaces}
V_{\mathrm{q}}=L^2(\mathbb{R}^d,V_0) \quad \text{and} \quad
H_{\mathrm{w}}=L^2(\mathbb{R}^d, H_{\mathrm{w}}^0),  \quad
\mathrm{w} = \mathrm{p,q},
\end{equation}
with $V_0$, $H_{\mathrm{p,q}}^0$ real Hilbert spaces. So the
coordinate $q$ is a vector field $q(\vec{r}) \in V_0$. We suppose
further that the impedance operator $K$ is of the form
\begin{equation}\label{impedancedensity}
    K \begin{pmatrix}
      p \\
      q
    \end{pmatrix}(\vec{r}) \ =\ \begin{pmatrix}
      K_{\mathrm{p}}(\vec{r}) & 0 \\
    0 & K_{\mathrm{q}}(\vec{r}) + \mathbf{Y}_i(\vec{r}) \cdot \partial_i
    \end{pmatrix} \begin{pmatrix}
      p \\
      q
    \end{pmatrix}(\vec{r})  ,
\end{equation}
where repeated indices are summed from $i=1,\ldots, d$. For each
$\vec{r}$, $K_{\mathrm{p}}(\vec{r})$ and $K_{\mathrm{q}}(\vec{r})$,
$\mathbf{Y}_i(\vec{r})$, $i=1,...,d$ are bounded operators from
$V_0$ to $H_{\mathrm{p}}^0$ and $H_{\mathrm{q}}^0$ respectively and
$\partial_i =
\partial/\partial_{\vec{r}_i}$. This form covers classical linear
elastic, acoustic and dielectric media.

The system, in the absence of dissipation, is governed by a
Lagrangian
\begin{equation}\label{localLagr}
    \mathrm{L}(q,\partial_t q) \ =\ \int_{\mathbb{R}^d} \mathsf{L}(q(\vec{r}),
    \nabla q(\vec{r}),
    \partial_t q(\vec{r})) \mathrm{d}^d \vec{r} ,
    \end{equation}
with a Lagrangian density second order in the coordinate
$q(\vec{r})$ and first derivatives:
\begin{multline}\label{localLagrdens}
    \mathsf{L}(q(\vec{r}),\nabla q(\vec{r}) , \partial_t q(\vec{r})) \\ = \ \frac{1}{2} \left \{
    \left \| \left [ K_{\mathrm{p}}(\vec{r})^{\mathrm{T}} \right ]^{-1}
    \partial_t q(\vec{r}) \right \|^2 - \left \| K_{\mathrm{q}}(\vec{r})
    q(\vec{r}) + \mathbf{Y}(\vec{r}) \cdot \nabla q(\vec{r}) \right \|^2 \right
    \} .
\end{multline}
By a suitable choice of $K_{\mathrm{p},\mathrm{q}}$ and $\mathbf{Y}$
we can obtain any Lagrangian density of the form $\mathsf{L} =
T(\partial_t q) - V(q, \nabla q)$ with $T$ and $V$ homogeneous of
degree two.

Now suppose there is time dispersion and dissipation in the system
so that the equations of motion and material relations according to
(\ref{driven}) and (\ref{impedancedensity}) are
\begin{gather}
\partial _{t}%
\begin{pmatrix}
p(\vec{r},t) \\
q(\vec{r},t)%
\end{pmatrix}%
\ =\
\begin{pmatrix}\
-K_{\mathrm{q}}(\vec{r})^{\mathrm{T}}f_{\mathrm{q}}(\vec{r},t)+\partial
_{i}\left\{ \mathbf{Y}_{i}(\vec{r})^{\mathrm{T}}f_{\mathrm{q}}(\vec{r}%
,t)\right\} \\
K_{\mathrm{p}}^{\mathrm{T}}f_{\mathrm{p}}(\vec{r},t)%
\end{pmatrix}
\\
\begin{pmatrix}
f_{\mathrm{p}}(\vec{r},t) \\
f_{\mathrm{q}}(\vec{r},t)%
\end{pmatrix}%
\ +\int_{0}^{\infty }\chi (\tau ;\vec{r})%
\begin{pmatrix}
f_{\mathrm{p}}(\vec{r},t-\tau ) \\
f_{\mathrm{q}}(\vec{r},t-\tau )%
\end{pmatrix}%
\,\mathrm{d}\tau \ =\
\begin{pmatrix}
K_{\mathrm{p}}p(\vec{r},t) \\
K_{\mathrm{q}}q(\vec{r},t)+\mathbf{Y}_{i}(\vec{r})\partial _{i}q(\vec{r},t)%
\end{pmatrix}%
,\label{pqdisp}
\end{gather}
with $\chi(\tau;\vec{r})$ a $\mathcal{B}(H_{\mathrm{p}}^0 \oplus
H_{\mathrm{q}}^0)$ valued susceptibility function.\footnote{This
form for the susceptibility precludes spatial dispersion, which
would involve integration over $\vec{r}$ on the l.h.s. of
\eqref{pqdisp}. Our general construction  works in the presence of
spatial dispersion, but the extended Lagrangian is non-local making
it difficult to give a meaningful definition of the energy density
and stress tensor.} The string coupling operators constructed above
then fiber over $\mathbb{R}^d$ in the same way
\begin{equation}\label{nospatialdisp}
    \left [\varsigma_{\mathrm{p}}(s) p \right ] (\vec{r}) \ =\
    \varsigma_{\mathrm{p}}(s, \vec{r}) p(\vec{r}), \quad
    \left [\varsigma_{\mathrm{q}}(s) q \right ] (\vec{r}) \ =\
    \varsigma_{\mathrm{q}}(s, \vec{r}) q(\vec{r}) ,
\end{equation}
and the extended Lagrangian \eqref{extLagr2} is the integral of a
Lagrangian density:
\begin{equation}
  \mathrm{L}(Q, \partial_t Q) \ = \ \int_{\mathrm{R}^d} \mathsf{L}(Q(\vec{r}),\nabla Q(\vec{r}),
  \partial_t Q(\vec{r})) \mathrm{d}^d \vec{r} ,
\end{equation}
where $Q(\vec{r})=(q(\vec{r}), \varphi(s,\vec{r}))$ and
\begin{multline}\label{lagrandens}
  \mathsf{L}(Q(\vec{r}),\nabla Q(\vec{r}),
  \partial_t Q(\vec{r}) ) \\ = \ \frac{1}{2} \left \|
  \left [K_{\mathrm{p}}^{\mathrm{T}}(\vec{r}) \right ]^{-1} \partial_t q(\vec{r})
  \right \|_{H_{\mathrm{p}}^0}^2 + \left \langle
  \left [K_{\mathrm{p}}^{\mathrm{T}}(\vec{r}) \right ]^{-1}\partial_t q(\vec{r}),
  \int_{-\infty}^\infty \varsigma_{\mathrm{p}}(s,\vec{r})\varphi(s,\vec{r})
  \mathrm{d} s\right \rangle_{H_{\mathrm{p}}^0}  \\ + \frac{1}{2}
  \int_{-\infty}^\infty \left \| \partial_t \varphi(s,\vec{r}) \right
  \|_{H_{\mathrm{p}}^0
  \oplus H_{\mathrm{q}}^0}^2
  \mathrm{d} s - \frac{1}{2} \int_{-\infty}^\infty
  \left \| \partial_s \varphi(s,\vec{r}) \right \|_{H_{\mathrm{p}}^0
  \oplus H_{\mathrm{q}}^0}^2
  \mathrm{d} s
  \\ - \left \| K_{\mathrm{q}}(\vec{r}) q(\vec{r}) + \mathbf{Y}(\vec{r}) \cdot \nabla
  q(\vec{r}) - \int_{-\infty}^\infty \varsigma_{\mathrm{q}}(s,\vec{r})\varphi(s,\vec{r})
  \mathrm{d} s \right \|_{ H_{\mathrm{q}}^0}^2 .
\end{multline}
\textit{Remark}: $\varphi$ is an element of $L^2(\mathbb{R}, H)$
with $H = H_{\mathrm{p}} \oplus H_\mathrm{q} = L^2(\mathbb{R}^d,
H_{\mathrm{p}}^0 \oplus H_\mathrm{q}^0)$. We identify
$L^2(\mathbb{R}, H)$ with $L^2( \mathbb{R} \times \mathbb{R}^d,
H_{\mathrm{p}}^0 \oplus H_\mathrm{q}^0)$, writing $\varphi$  as
$(s,\vec{r}) \mapsto \varphi(s,\vec{r})\in H_{\mathrm{p}}^0 \oplus
H_\mathrm{q}^0$.

In Appendix \ref{sec:stress} we recall some basic constructions for
a system with a Lagrangian density.  In particular, we obtain
suitable expressions for the energy flux vector and the stress
tensor of a homogeneous system.  In this section, we apply the
expressions derived there to the extended TDD Lagrangian
\eqref{lagrandens}

\subsection{Energy density and flux}\label{sec:energyflux}
Because $\mathsf{L}$ does not depend explicitly on time, the
\emph{total energy}, which  can be expressed as the integral of a
density (see \eqref{Hamdens2})
\begin{equation}
  \mathcal{E} \ = \ \int_{\mathrm{R}^d} \mathsf{H}(\vec{r},t) \mathrm{d}^d
  \vec{r},
\end{equation}
is conserved (in the absence of an external driving force). The
value of the total energy $\mathcal{E}$ is, of course, just the
extended Hamiltonian $\mathrm{H}$ evaluated ``on-shell,'' at a field
configuration evolving according to the equations of motion.

We can express the energy density $\mathsf{H}$  as a sum of two
contributions
\begin{equation}\label{hamdens}
    \mathsf{H}(\vec{r},t) \ = \ \mathsf{H}_{\mathrm{sys}}(\vec{r},t) +
    \mathsf{H}_{\mathrm{str}}(\vec{r},t) ,
\end{equation}
which we interpret as the energy density of the TDD system and the
heat bath, as described by the hidden  strings, respectively. Here
\begin{equation}\label{mediumenergy}
    \mathsf{H}_{\mathrm{str}}(\vec{r},t) \ = \ \frac{1}{2} \int_{-\infty}^\infty
  \left \| \partial_s \varphi(s,\vec{r},t)\right \|_{H_{\mathrm{p}}^0 \oplus H_{\mathrm{q}}^0}
  \mathrm{d} s
  + \frac{1}{2} \int_{-\infty}^\infty
  \left \| \theta(s,\vec{r},t)\right \|_{H_{\mathrm{p}}^0 \oplus H_{\mathrm{q}}^0}
  \mathrm{d} s ,
\end{equation}
with $\theta(s,\vec{r},t)  =  \partial_t \varphi(s,\vec{r},t)$, and
\begin{equation}\label{systemenergy}
\mathsf{H}_{\mathrm{sys}}(\vec{r},t) \ = \ \frac{1}{2} \left \|
f_{\mathrm{p}}(\vec{r},t) \right \|_{H_{\mathrm{p}}^0}^2
  + \frac{1}{2} \left \|f_{\mathrm{q}}(\vec{r},t) \right
  \|_{H_{\mathrm{q}}^0}^2 ,
\end{equation}
with $f_{\mathrm{q,p}}$ the coordinate and momentum parts of the
kinematical stress, \begin{align} \label{fq} f_{\mathrm{q}}(\vec{r}
) \ &= \ K_{\mathrm{q}}(\vec{r}) q(\vec{r}) +
    \mathbf{Y}(\vec{r})
    \cdot \nabla
  q(\vec{r}) - \int_{-\infty}^\infty
  \varsigma_{\mathrm{q}}(s,\vec{r})\varphi(s,\vec{r})
  \mathrm{d} s , \intertext{and}
  \label{fp}
    f_{\mathrm{p}}(\vec{r} ) \ &= \
    \left [K_{\mathrm{p}}^{\mathrm{T}}(\vec{r}) \right ]^{-1} \partial_t
    q(\vec{r},t) \ = \ K_{\mathrm{p}}(\vec{r}) p(\vec{r},t) -
  \int_{-\infty}^\infty \varsigma_{\mathrm{p}}(s,\vec{r})\varphi(s,\vec{r},t)
  \mathrm{d} s , \end{align}
  as follows from (\ref{abstractf}), (%
\ref{impedancedensity}) and (\ref{pqdisp}). By
(\ref{hamdens}--\ref{systemenergy}), $\mathsf{H}_{\mathrm{sys}}$
includes the interaction energy between the system and the hidden
strings.

The energy density, expressed in canonical coordinates,
$\mathsf{H}(\vec{r},t)  = $ $\mathsf{H}\bigl(q(\vec{r},t),$ $\nabla
q(\vec{r},t),$ $\varphi(\cdot,\vec{r},t),$ $p(\vec{r},t),$
$\theta(\cdot,\vec{r},t) \bigr)$, is also the Hamiltonian density.
The equations of motion can be recovered
 by variation
\begin{align}\label{qmotion}
  \partial_t q(\vec{r},t)   &=  \frac{\delta \mathsf{H}}{\delta p}
  (\vec{r},t)  =  K_{\mathrm{p}}^{\mathrm{T}}(\vec{r})
  f_{\mathrm{p}}(\vec{r},t) , \\ \label{pmotion}
  \partial_t p(\vec{r},t)  &=
  \left \{ -\frac{\delta \mathsf{H}}{\delta q}
   + \partial_i\frac{\delta \mathsf{H}}{\delta \partial_i q} \right
   \}
  (\vec{r},t)  =  -K_{\mathrm{q}}^{\mathrm{T}}(\vec{r})
   f_{\mathrm{q}}(\vec{r},t) + \partial_i \mathbf{Y}_i^{\mathrm{T}}(\vec{r})
   f_{\mathrm{q}}(\vec{r},t),\\ \label{thetamotion}
   \partial_t \theta(s,\vec{r},t)
    &=  - \frac{\delta \mathsf{H}}{\delta \varphi(s) }
  (\vec{r},t) \\ \notag &=  \partial_s^2 \varphi(s,\vec{r},t) +
   \varsigma_{\mathrm{p}}^{\mathrm{T}}(\vec{r},s)
   f_{\mathrm{p}}(\vec{r},t) +
   \varsigma_{\mathrm{q}}^{\mathrm{T}}(\vec{r},s)
   f_{\mathrm{q}}(\vec{r},t) \\ \label{varphimotion}
  \partial_t \varphi(s,\vec{r},t)  &=  \frac{\delta \mathsf{H}}{\delta \theta(s)}
  (\vec{r},t)  =
  \theta(s,\vec{r},t)  .
\end{align}
If the system is driven by an external force $\rho(\vec{r},t)\in
V_0$, we replace \eqref{pmotion} by
\begin{equation}\label{drivenp}
    \partial_t p(\vec{r},t) \ = \ - K_{\mathrm{q}}^{\mathrm{T}}(\vec{r})
    f_{\mathrm{q}}(\vec{r},t)
    + \partial_i \mathbf{Y}_i^{\mathrm{T}}(\vec{r}) f_{\mathrm{q}}(\vec{r},t)
    + \rho(\vec{r},t) .
\end{equation}

When the driving force vanishes, the total energy is conserved and
the energy density satisfies a local conservation law
\begin{equation}\label{encontinuity}
    \partial_t \mathsf{H}(\vec{r},t) + \partial_i \mathbf{S}_i(\vec{r},t) \ = \
    0 ,
\end{equation}
with $\mathbf{S}$ the energy flux vector, an expression for which is
derived Appendix \ref{sec:stress}. For the present case the energy
flux is  (see \eqref{Efluxdens})
\begin{equation}\label{energyflux}
    \mathbf{S}_i (\vec{r},t) = - \langle \partial_t q(\vec{r},t),
    \mathbf{Y}_i^{\mathrm{T}}(\vec{r}) f_{\mathrm{q}}(\vec{r},t)
    \rangle_{V_0}
    =  -  \langle
    K_{\mathrm{p}}^{\mathrm{T}}(\vec{r})
    f_{\mathrm{p}}(\vec{r},t), \mathbf{Y}_i^{\mathrm{T}}(\vec{r}) f_{\mathrm{q}}(\vec{r},t)
     \rangle_{V_0} .
\end{equation}

With a non-zero driving force, one has
\begin{equation}\label{endrivencontinuity}
    \partial_t \mathsf{H}(\vec{r},t) + \partial_i \mathbf{S}_i(\vec{r},t)
    \ = \ \langle \partial_t q(\vec{r},t), \rho(\vec{r},t) \rangle_{V_0}
\end{equation}
in place of \eqref{encontinuity} (see Theorem
\ref{thm:drivencontinuity}). Thus
\begin{equation}\label{energychange}
    \partial_t \mathcal E \ = \
    \int_{-\mathbb{R}^d} \langle \partial_t q(\vec{r},t), \rho(\vec{r},t) \rangle_{V_0}
    \mathrm{d}^d \vec{r} ,
\end{equation}
consistent with our interpretation of  $\langle
\partial_t q(\vec{r},t), \rho(\vec{r},t) \rangle_{V_0}$ as the power
density of the external force.

We conceive of the four vector fields $p(\vec{r},t)$,
$q(\vec{r},t)$, $f_{\mathrm{p}}(\vec{r},t)$, and
$f_{\mathrm{q}}(\vec{r},t)$ as specifying the ``state'' of the
reduced TDD system. By \eqref{energyflux}, the energy flux at time
$t$ is a function of these fields evaluated at time
$t$.\footnote{This is a consequence of the absence of spatial
dispersion. By adding terms involving $\nabla \varphi$ to the
Lagrangian, we could extend the above set up to systems with spatial
dispersion, resulting in an energy flux with non-trivial
contributions from $\varphi$.} The energy density depends in a more
essential way on the configuration of the hidden strings.
Nonetheless, we may use \eqref{endrivencontinuity} to give a
definition of  the energy density intrinsic to the TDD system by
writing it as integral over the history of the system, namely
\begin{equation}\label{energydensityhistoryintegral}
 \mathsf{H}(\vec{r},t) \ = \  \int_{-\infty}^t \left
 \{ -\partial_i \mathbf{S}_i(\vec{r},t')
 + \langle \partial_t q(\vec{r},t'), \rho(\vec{r},t') \rangle_{V_0}
 \right \} \mathrm{d} t',
\end{equation}
if the energy density was zero at $t = -\infty$, i.e. the system and
medium were \emph{at rest}.

\subsection{Homogeneity, isotropy, wave momentum and the stress tensor}\label{sec:enmom}
Suppose the extended TDD system has a Lagrangian density
\eqref{lagrandens} which is \emph{homogeneous} \textemdash \
invariant under spatial translations. Then \emph{the total wave
momentum} $\mathbf{P}$ is conserved, and there is a corresponding
conserved current, \emph{the wave momentum density}
$\mathbf{p}(\vec{r},t)$, analyzable using Noether's Theorem (see
\cite[Section 5.5]{MH}).\footnote{We follow \cite{MorseFeshbach} in
using the term ``wave momentum'' for the conserved quantity
associated to translation invariance. This avoids confusion with
``canonical momenta,'' the variables $p$ and $\theta$.} If the
Lagrangian density is also \emph{isotropic} \textemdash \ invariant
under spatial rotations \textemdash, then the anti-symmetric tensor
of angular momentum about the origin $\mathsf{M}_{i,j}$ is
conserved.\footnote{In dimension $d=3$, the usual angular momentum
pseudo-vector $\boldsymbol{\ell}$ is obtain from $\mathsf{M}_{i,j}$
as $\boldsymbol{\ell}_i = \epsilon_{i,j,k} \mathsf{M}_{j,k}$ with
$\epsilon_{i,j,k}$ the fully anti-symmetric symbol with
$\epsilon_{1,2,3}=1$.} In appendix \ref{sec:stress}, following
\cite{Barut}, we recall the formulation of the symmetric stress
tensor $\mathsf{T}$ and wave momentum density $\mathbf{p}$ for a
homogeneous and isotropic system.

To say that the Lagrangian in \eqref{lagrandens} is isotropic, we
must specify how $q$ and $\varphi$ transform under rotations. Thus
we suppose given representations $\Upsilon$ and
$\Upsilon_{\mathrm{w}}$, $\mathrm{w}=\mathrm{p,q}$, of the rotation
group $\mathsf{SO}(d)$ by orthogonal operators in $\mathcal B(V_0)$
and $\mathcal B(H_{\mathrm{w}}^0)$, $\mathrm{w}=\mathrm{p,q}$,
respectively, so that under a global rotation of the coordinate
system about the origin,
\begin{equation}
    \vec{r} \, ' \ = \ R \cdot \vec{r}
\end{equation}
with $R \in \mathsf{SO}(d)$ an orthogonal matrix, the fields $q$ and
$\varphi$ transform as
\begin{equation}\label{qtransform}
    q'(\vec{r}\, ') \ = \ \Upsilon(R) q(R^{-1} \vec{r} \, ')
\end{equation}
and
\begin{equation}
    \varphi'(s,\vec{r}\,') \ = \ \begin{pmatrix}
      \Upsilon_{\mathrm{p}}(R) & 0 \\
      0 & \Upsilon_{\mathrm{q}}(R)
    \end{pmatrix} \varphi(s,R^{-1} \vec{r}\,') .
\end{equation}
(Recall that $\varphi$ is $H_{\mathrm{p}}\oplus
H_{\mathrm{q}}$-valued.) The representations $\Upsilon$ and
$\Upsilon_{\mathrm{p},\mathrm{q}}$ are specified by families of
skew-adjoint operators $g_{i,j}$ and $g_{i,j}^{\mathrm{w}}$,
$\mathrm{w}=\mathrm{p,q}$, on $V_0$ and $H_{\mathrm{w}}^0 $,
$\mathrm{w}=\mathrm{p,q}$, respectively, $i,j=1,...,d$. A rotation
$R=\mathrm{e}^{\omega}$, with $\omega$ an anti-symmetric matrix, has
representatives (see appendix \ref{sec:stress}):
\begin{equation}\label{representatives}
    \Upsilon(\mathrm{e}^{\omega}) \ = \ \mathrm{e}^{\frac{1}{2} \omega_{i,j}
    g_{i,j}}, \quad  \Upsilon_{\mathrm{w}}(\mathrm{e}^{\omega}) \ = \
    \mathrm{e}^{\frac{1}{2} \omega_{i,j}
    g_{i,j}^{\mathrm{w}}}  \quad (\text{summation convention}).
\end{equation}
In addition to being skew-adjoint, the operators $g_{i,j}$ and
$g_{i,j}^{\mathrm{p},\mathrm{q}}$ satisfy
\begin{multline}
    g_{i,j}^\sharp  \ = \ - g_{j,i}^\sharp , \\ \text{and} \quad
    g_{i,j}^\sharp g_{i',j'}^\sharp -  g_{i',j'}^\sharp g_{i,j}^\sharp
    \ = \ -\delta_{i,i'} g_{j,j'}^\sharp + \delta_{i,j'} g_{j,i'}^\sharp
    +\delta_{j,i'} g_{i,j'}^\sharp - \delta_{j,j'} g_{i,i'}^\sharp .
\end{multline}
Since $V_0$ and $H_{\mathrm{p},\mathrm{q}}$ are finite dimensional
in our application to the Maxwell equations below, we assume
$g_{i,j}$ and $g_{i,j}^{\mathrm{w}}$ are bounded for simplicity.
\begin{definition}\label{def:isohomo} The Lagrangian density \eqref{lagrandens}
is \emph{homogeneous}  if $K_{\mathrm{p,q}}(\vec{r})$,
$\mathbf{Y}(\vec{r})$, and $\varsigma_{\mathrm{p,q}}(\vec{r},s)$ are
independent of $\vec{r}$, and is \emph{isotropic} if
\begin{gather}
K_{\mathrm{w}} g_{i,j} \ = \ g_{i,j}^{\mathrm{w}}
K_{\mathrm{w}} , \quad \mathrm{w}=\mathrm{p,q} \\
\label{Yiso} \mathbf{Y}_{k} g_{i,j} \ = \ g_{i,j}^{\mathrm{q}}
\mathbf{Y}_{k} + \delta_{j,k} \mathbf{Y}_i - \delta_{i,k}
\mathbf{Y}_{j} \intertext{and} \varsigma_{\mathrm{w}} (s)
\begin{pmatrix}
g_{i,j}^{\mathrm{p}} & 0 \\
      0 & g_{i,j}^{\mathrm{q}}
\end{pmatrix} \ = \  g_{i,j}^{\mathrm{w}} \varsigma_{\mathrm{w}}(s)
, \quad \mathrm{w}=\mathrm{p,q} .
\end{gather}
\end{definition}
\noindent \emph{Remarks:} i.) In Appendix \ref{sec:stress} we give
more general definitions \ref{def:homogeneity} and
\ref{def:isotropy}, which are consistent with \ref{def:isohomo}.
ii.) The last two terms on the r.h.s.\ of \eqref{Yiso} result from
the fact that $\mathbf{Y}_i$ appear coupled with a spatial
derivative in the Lagrangian \eqref{lagrandens}.  iii.) Recall that
$\varsigma_{\mathrm{w}}(s,\vec{r}) : H_{\mathrm{p}}\oplus
H_{\mathrm{q}} \rightarrow H_{\mathrm{w}}$.

Theorem \ref{thm:stress} below gives the following expressions for
the wave momentum density $\mathbf{p}$ and stress tensor
$\mathsf{T}$, expressed here in canonical coordinates:
\begin{multline}\label{momdens}
  \mathbf{p}_i(\vec{r},t)  \ = \ \left \langle \partial_i q(\vec{r},t),p(\vec{r},t)
  \right \rangle_{V_0} \\
  + \int_{-\infty}^\infty \left \langle\partial_i  \varphi(s,
  \vec{r},t),
  \theta(s, \vec{r},t) \right \rangle_{H_{\mathrm{p}}^0
  \oplus H_{\mathrm{q}}^0}
  \mathrm{d} s - \partial_j \Phi_{i,j}(\vec{r},t) ,
\end{multline}
\begin{multline}\label{momflux} \mathsf{T}_{i,j}(\vec{r},t)  \\ = \ -\left \langle
 \partial_{i} q(\vec{r},t) ,
 \mathbf{Y}_j^{\mathrm{T}}  f_{\mathrm{q}}(\vec{r},t) \right \rangle_{V_0} -
\delta_{i,j} \mathsf{L}(\vec{r},t) + \partial_t \Phi_{i,j} +
\partial_k \Psi_{i,j,k}(\vec{r},t),
\end{multline}
where $f_{\mathrm{p,q}}$ are defined in (\ref{fq}, \ref{fp}),
$\mathsf{L}$ is  the Lagrangian density
\begin{multline}\label{canonLagr} \mathsf{L}(\vec{r},t) \
= \ \left \langle \mathsf{K}_{\mathrm{p}}^{\mathrm{T}}
f_{\mathrm{p}}(\vec{r},t), p(\vec{r},t) \right
\rangle_{H_{\mathrm{p}}^0} -\frac{1}{2} \left \|
f_{\mathrm{p}}(\vec{r},t) \right \|_{H_{\mathrm{p}}^0}^2 -
\frac{1}{2}\left \| f_{\mathrm{q}}(\vec{r},t) \right \|_{
H_{\mathrm{q}}^0}^2
 \\ + \frac{1}{2} \int_{-\infty}^\infty \left
\|\theta(\vec{r},s,t) \right \|_{H_{\mathrm{p}}^0}^2 \mathrm{d} s -
\frac{1}{2} \int_{-\infty}^\infty
  \left \| \partial_s \varphi(s,\vec{r}) \right \|_{H_{\mathrm{p}}^0
  \oplus H_{\mathrm{q}}^0}^2
  \mathrm{d} s
   ,
\end{multline}
and
\begin{multline}
    \Phi_{i,j}(\vec{r},t) \ = \ \frac{1}{2}
    \left \langle g_{i,j} q(\vec{r},t) , p(\vec{r},t)
  \right \rangle_{V^0} \\ +
  \frac{1}{2} \int_{-\infty}^\infty \left \langle  \begin{pmatrix}
g_{i,j}^{\mathrm{p}} & 0 \\
      0 & g_{i,j}^{\mathrm{q}}
\end{pmatrix} \varphi(s,
  \vec{r},t),
  \theta(s, \vec{r},t) \right \rangle_{H_{\mathrm{p}}^0
  \oplus H_{\mathrm{q}}^0}
  \mathrm{d} s ,
\end{multline}
\begin{multline}
  \Psi_{i,j,k}(\vec{r},t) \ = \ -\frac{1}{2} \left \langle
  g_{i,j} q(\vec{r},t)  ,
  \mathbf{Y}_k^{\mathrm{T}}
  f_{\mathrm{q}}(\vec{r},t) \right \rangle_{V_0 } \\
    +\frac{1}{2}  \left \langle
  g_{j,k} q(\vec{r},t)  ,
  \mathbf{Y}_i^{\mathrm{T}}f_{\mathrm{q}}(\vec{r},t)\right \rangle_{V_0 }
    + \frac{1}{2} \left \langle
  g_{i,k} q(\vec{r},t)  ,
  \mathbf{Y}_j^{\mathrm{T}}f_{\mathrm{q}}(\vec{r},t)\right \rangle_{V_0} .
\end{multline}

 As the system is homogeneous, the total wave momentum
\begin{equation}\label{totalwavemomentum}
    \mathbf{P} \ = \ \int_{\mathbb{R}^d} \mathbf{p}(\vec{r},t) \mathrm{d}^d
    \vec{r}
\end{equation}
is conserved in the absence of a driving force, and the wave
momentum density satisfies the local conservation law
\begin{equation}\label{wavemomentumconservationlaw}
    \partial_t \mathbf{p}_i(\vec{r},t) + \partial_j \mathsf{T}_{i,j}(\vec{r},t)
    \ = \ 0 .
\end{equation}
With a driving force, \eqref{wavemomentumconservationlaw} is
modified to (see Theorem \ref{thm:drivencontinuity})
\begin{equation}\label{drivenwavemomentumconservationlaw}
    \partial_t \mathbf{p}_i(\vec{r},t) + \partial_j \mathsf{T}_{i,j}(\vec{r},t)
    \ = \ \left \langle \partial_i q(\vec{r},t), \rho(\vec{r},t) \right
    \rangle_{V_0} ,
\end{equation}
and thus
\begin{equation}
    \partial_t \mathbf{P} \ = \ \int_{\mathbb{R}^d} \left \langle \partial_i
    q(\vec{r},t), \rho(\vec{r},t) \right \rangle_{V_0}\mathrm{d}^d
    \vec{r} .
\end{equation}
Due to the term $\partial_t \Phi_{i,j}$ in the definition of the
stress tensor \eqref{momflux}, the driving force also modifies
$\mathsf{T}$. If $\mathsf{T}_{i,j}^0(\vec{r},t)$ is the stress
tensor with $\rho(\vec{r},t)=0$, then
\begin{equation}
\mathsf{T}_{i,j}(\vec{r},t) \ = \ \mathsf{T}_{i,j}^0(\vec{r},t) +
\frac{1}{2} \left \langle g_{i,j} q(\vec{r},t), \rho (\vec{r},t)
\right \rangle_{V_0} .
\end{equation}
\begin{proposition}\label{lem:stresstens}
  If the Lagrangian is homogeneous  and
  isotropic then the stress tensor \eqref{momflux} can be
  written as
  \begin{equation}\label{symmomflux}
  \mathsf{T}_{i,j}(\vec{r},t) \ = \  \frac{1}{2} \left \{ \mathsf{W}_{i,j}(\vec{r},t) +
  \mathsf{W}_{j,i}(\vec{r},t)
  \right \} +\frac{1}{2}
\left \langle g_{i,j} q(\vec{r},t), \rho (\vec{r},t) \right
\rangle_{V_0} - \delta_{i,j} \mathsf{L}(\vec{r},t) ,
  \end{equation}
  where $\rho(\vec{r},t) \in V_0$ is the external force
  and
  \begin{equation}\label{symmomflux2}
  \mathsf{W}_{i,j}(\vec{r},t) \ = \ -
    \left \langle
 \partial_{i} q(\vec{r},t) ,
 \mathbf{Y}_j^{\mathrm{T}} f_{\mathrm{q}}(\vec{r},t) \right \rangle_{V_0}
 + \partial_k \left \langle
  g_{i,k} q(\vec{r},t)  ,
  \mathbf{Y}_j^{\mathrm{T}}f_{\mathrm{q}}(\vec{r},t)\right
  \rangle_{V_0},
  \end{equation}
 with $f_{\mathrm{q}}$, $f_{\mathrm{p}}$ and $g_{i,j}$ as defined in \eqref{fq}, \eqref{fp}, and
 \emph{(\ref{qtransform}--\ref{representatives})}. In particular,
the stress tensor is symmetric if $\rho(\vec{r},t)=0$, and depends
on the state of the strings $\varphi (s,\vec{r},t)$ through the
variables $f_{\mathrm{q}}$ and $f_{\mathrm{p}}$ and the Lagrangian
density $\mathsf L$.
\end{proposition}
\begin{proof}
    By the remark preceding the theorem, it suffices to consider
    $\rho(\vec{r},t) = 0$. Using the equations of motion
    (\ref{qmotion}--\ref{thetamotion}), the definition of isotropy \ref{def:isohomo},
    and the
    skew-adjointness of the rotation generators $g_{i,j}^\sharp$, one may
    calculate that
    \begin{align}
        \partial_t \Phi_{i,j} \ &= \ \frac{1}{2}\left \langle g_{i,j} q, \partial_k
        \mathbf{Y}_k^{\mathrm{T}} f_{\mathrm{q}} \right \rangle_{V_0}
        + \frac{1}{2}\left \langle g_{i,j}^{\mathrm{q}} \mathbf{Y}_k \partial_k q,
        f_{\mathrm{q}} \right \rangle_{V_0}  \\ &= \ \notag
        \partial_k \frac{1}{2} \left \langle g_{i,j} q,
        \mathbf{Y}_k^{\mathrm{T}} f_{\mathrm{q}} \right \rangle_{V_0}
        \ - \  \left \langle  \left \{ \mathbf{Y}_k g_{i,j}-
        g_{i,j}^{\mathrm{q}} \mathbf{Y}_k \right \} \partial_k q,
        f_{\mathrm{q}} \right \rangle_{H_0^{\mathrm{q}}} \\
        &= \ \partial_k \frac{1}{2} \left \langle g_{i,j} q,
        \mathbf{Y}_k^{\mathrm{T}} f_{\mathrm{q}} \right \rangle_{V_0} \notag
         \ - \  \frac{1}{2} \left \langle
         \partial_j q ,
        \mathbf{Y}_i^{\mathrm{T}}   f_{\mathrm{q}} \right \rangle_{V_0}
        + \frac{1}{2} \left \langle
         \partial_i q,
        \mathbf{Y}_j^{\mathrm{T}}   f_{\mathrm{q}} \right \rangle_{V_0} .
    \end{align}
    Combining this expression with the definition \eqref{momflux} of $\mathsf{T}$
     gives \eqref{symmomflux}.
\end{proof}

Like the energy flux $\mathbf{S}$, the tensor field $\mathsf{W}$ at
time $t$ is a function of the fields $q(\vec{r},t)$, $p(\vec{r},t)$,
$f_{\mathrm{q}}(\vec{r},t)$, and $f_{\mathrm{p}}(\vec{r},t)$
specifying the state of the reduced TDD system. In particular, the
off-diagonal terms of the stress tensor depend only on the
instantaneous state of the reduced TDD system. The diagonal terms
depend on the Lagrangian density $\mathsf L$, requiring a more
detailed knowledge of the state of the hidden strings. However, as
with the energy density $\mathsf{H}$, we may write $\mathsf{L}$  in
terms of the history of the underlying TDD system.  To this end, we rewrite $\mathsf L$ as%
\begin{align}\label{L=pdot-H}
\mathsf{L}(\vec{r},t)\ &=\ \langle p(\vec{r},t),\partial _{t}q(\vec{r}%
,t)\rangle _{V_{0}}+\int_{-\infty }^{\infty }\left\langle \theta (s,\vec{r}%
,t),\partial _{t}\varphi (s,\vec{r},t)\right\rangle _{H_{\mathrm{p}%
}^{0}\oplus H_{\mathrm{q}}^{0}}\,\mathrm{d}s-\mathsf{H}(\vec{r},t) \\
&=\ \langle K_{\mathrm{p}}p(\vec{r},t),f_{\mathrm{p}}(\vec{r},t)\rangle _{H_{%
\mathrm{p}}^{0}}+\int_{-\infty }^{\infty }\left\Vert \theta (s,\vec{r}%
,t)\right\Vert _{H_{\mathrm{p}}^{0}\oplus H_{\mathrm{q}}^{0}}^{2}\,\mathrm{d}%
s-\mathsf{H}(\vec{r},t),  \notag
\end{align}%
using the equations of motion (\ref{qmotion}--\ref{varphimotion}).
Based on (\ref{hamdens}--\ref{systemenergy}) we introduce%
\begin{equation}
\mathsf{L}(\vec{r},t)\ =\ \mathsf{L}_{\mathrm{sys}}(\vec{r},t)+\mathsf{L}_{%
\mathrm{str}}(\vec{r},t),  \label{Ldec1}
\end{equation}
with
\begin{align}\label{Ldec2}
\mathsf{L}_{\mathrm{sys}}(\vec{r},t)\ &=\ \langle p(\vec{r},t),\partial _{t}q(%
\vec{r},t)\rangle _{V_{0}}-\mathsf{H}_{\mathrm{sys}}(\vec{r},t)
 \\
&=\ \langle K_{\mathrm{p}}p(\vec{r},t),f_{\mathrm{p}}(\vec{r},t)\rangle _{H_{%
\mathrm{p}}^{0}}-\frac{1}{2}\left\Vert
f_{\mathrm{p}}(\vec{r},t)\right\Vert
_{H_{\mathrm{p}}^{0}}^{2}-\frac{1}{2}\left\Vert f_{\mathrm{q}}(\vec{r}%
,t)\right\Vert _{H_{\mathrm{q}}^{0}}^{2},  \notag
\end{align}
\begin{align}
\mathsf{L}_{\mathrm{str}}(\vec{r},t) &=\int_{-\infty }^{\infty
}\left\langle
\theta (s,\vec{r},t),\partial _{t}\varphi (s,\vec{r},t)\right\rangle _{H_{%
\mathrm{p}}^{0}\oplus H_{\mathrm{q}}^{0}}\,\mathrm{d}s-\mathsf{H}_{\mathrm{%
str}}(\vec{r},t)=  \label{Ldec3} \\
&=\frac{1}{2}\int_{-\infty }^{\infty }\left\Vert \theta (s,\vec{r}%
,t)\right\Vert _{H_{\mathrm{p}}^{0}\oplus H_{\mathrm{q}}^{0}}^{2}\,\mathrm{d}%
s-\frac{1}{2}\int_{-\infty }^{\infty }\left\Vert \partial _{s}\varphi (s,%
\vec{r},t)\right\Vert _{H_{\mathrm{p}}^{0}\oplus H_{\mathrm{q}}^{0}}^{2}\,%
\mathrm{d}s.  \notag
\end{align}

We use the solution \eqref{solvedrivenwave} to express $\theta (s,
\vec{r},t)$ as
\begin{align}\label{solvedrivenwave2}
   \theta(s,\vec{r},t) \ &= \  \partial_t \varphi ( s, \vec{r},t)
   \\ \notag
 &= \ \frac{1}{2}\int_{-\infty}^{t} \mathrm{d} t'
 \left \{ \varsigma ( s+t-t') +
 \varsigma ( s-t+t') \right \}
 f(\vec{r},t') ,
\end{align}
where $\varsigma  =  \begin{pmatrix} \varsigma_{\mathrm{p}} &
    \varsigma_{\mathrm{q}} \end{pmatrix},$ and  $f  =
    \begin{pmatrix}
      f_{\mathrm{p}} \\
      f_{\mathrm{q}}
    \end{pmatrix} .$
Then by the definition \eqref{fb18} of $\varsigma$,  see
\eqref{fb13},
\begin{multline}\label{inttheta2} \int_{-\infty}^\infty \left \|
    \theta(s,\vec{r},t) \right \|_{H_{\mathrm{p}}^0
  \oplus H_{\mathrm{q}}^0}^2 \mathrm{d} s
  \\ = \ \frac{1}{2} \int_{-\infty}^{t} \int_{-\infty}^{t}
  \left \langle f(\vec{r},t_1),
  \left[ \partial_{\tau} \chi^{\mathrm{o}} \right ] (t_1 - t_2)
  f(\vec{r},t_2) \right \rangle_{H_{\mathrm{p}}^0
  \oplus H_{\mathrm{q}}^0} \mathrm{d} t_1 \mathrm{d} t_2\\
  + \frac{1}{2} \int_{-\infty}^{t} \int_{-\infty}^{t}
  \left \langle f(\vec{r},t_1), \left [
  \partial_{\tau} \chi^{\mathrm{o}} \right ] (2t - t_1 - t_2)
  f(\vec{r},t_2) \right \rangle_{H_{\mathrm{p}}^0
  \oplus H_{\mathrm{q}}^0} \mathrm{d} t_1 \mathrm{d} t_2 ,
\end{multline}
where $\chi^{\mathrm{o}}$ is the odd extension \eqref{chio} of the
susceptibility. Using \eqref{energydensityhistoryintegral} to
express $\mathsf{H}$ and \eqref{inttheta2} to express the
corresponding term in \eqref{L=pdot-H}, we obtain an intrinsic
definition of the Lagrangian density, and hence the stress tensor
$\mathsf{T}$, as function of the history of a TDD Hamiltonian
system. Writing
\begin{equation}\label{momdensityhistoryintegral}
    \mathbf{p}(\vec{r},t) \ = \ \int_{-\infty}^t \left \{ - \partial_j
    \mathsf{T}_{i,j}(\vec{r},t') + \langle \partial_i q(\vec{r},t'),
    \rho(\vec{r},t') \rangle \right \} \mathrm{d} t',
\end{equation}
we obtain a similar expression for the wave momentum density.

\subsection{Brillouin-type formulas for time averages}\label{sec:Brillouin}
As we have seen, to express the energy density and stress tensor of
the extended system in terms of the fields $p,q,f_{\mathrm p},
f_{\mathrm q}$ we must introduce integrals over the history, like
\eqref{energydensityhistoryintegral} and \eqref{inttheta2}. However,
it is often useful to have an approximate formula involving the
\emph{instantaneous} state of the TDD system. A well known example
is the Brillouin formula for time averaged energy density stored in
a dielectric medium (see \cite[\S 80]{LandauLif} and
\S\ref{sec:MaxBrillouin} below).

Taking inspiration from the Brillouin formula, we consider here an
evolution of the underlying TDD system which is approximately
periodic with frequency $\omega/2 \pi$.  That is, we suppose that
\begin{equation}\label{slowlyvaryingform}
    g(\vec{r},t) \ = \ \mathrm{Re} \left \{ \mathrm{e}^{-\mathrm{i}\omega t}
    g_0(\vec{r},t) \right \} ,
\end{equation}
with $g=$ $p$, $q$, $f_{\mathrm{p}}$, $f_{\mathrm{q}}$, or $\rho$.
The various functions $w_0=$ $p_0$, $q_0$, $f_{\mathrm p ;0}$,
$f_{\mathrm q ;0}$, $\rho_0$ are supposed to vary extremely slowly
over time scales of duration $1/\omega$, and may take values in the
complex Hilbert spaces $\mathbb C V_0$, $\mathbb C
H_{\mathrm{p,q}}^0$. This evolution describes a \emph{carrier wave}
of frequency $\omega/2 \pi$, which is slowly modulated in phase and
amplitude.

To quantify the notion that the functions $g_0$ vary extremely
slowly on time scale $1/\omega$, we assume the Fourier Laplace
transforms,
\begin{equation}\label{ftf0}
    \widehat g_0(\vec{r},\zeta) \ = \
    \int_{-\infty}^\infty \mathrm{e}^{\mathrm{i} \zeta t} g_0(\vec{r},t)
    \mathrm{d} t , \quad \mathrm{Im} \zeta > 0 ,
\end{equation}
for $g_0=$ $p_0$, $q_0$, $f_{\mathrm{p};0}$, $f_{\mathrm{q};0}$, or
$\rho_0$, satisfy
\begin{equation}\label{f0bound}
    \| \widehat g_0(\vec{r}, \zeta) \| \ \le \ \text{const.} \, \omega_0^{-1} \psi(|\zeta|/\omega_0) ,
    \quad \mathrm{Im} \zeta > 0 ,
\end{equation}
 with $\psi$ a
fixed rapidly decaying function and $\omega_0 << \omega$. Thus
$\delta = \omega_0/\omega$ is a dimensionless small parameter which
measures the slowness of the functions $g_0$. We are interested in
asymptotic expressions for various quantities as $\delta \rightarrow
0$ carried out to order $\delta$ and shall neglect contributions of
size $o(\delta)$.   Throughout the discussion the carrier wave
frequency $\omega$ is fixed, so $\delta \propto \omega_0$. (Recall
that $o(\delta)$ denotes any term with $o(\delta)/\delta \rightarrow
0$ as $\delta \rightarrow 0$ and $\mathcal O(\delta)$ denotes a term
bounded by $\mathrm{const.} \times \delta$.)

We use the notation $a \approx b$ to indicate that $a-b = o(\delta)$
and say that $a$ is \emph{negligible} if $a \approx 0$, i.e.,
$a=o(\delta)$. For example $\partial_t^2 g_0(\vec{r},t)$ is
negligible for each $g_0$, since
\begin{multline}\label{f0bound2} |\partial_t^2 g_0(\vec{r},t) | \ =
 \left | \int_{-\infty}^\infty \nu^2 \mathrm{e}^{-
\mathrm{i} \nu t} \widehat g_0(\vec{r}, \nu ) \right | \\ \le \
\mathrm{const.} \, \int_{-\infty}^\infty \nu^2
 \psi(|\nu|/\omega_0) \mathrm{d} \nu / \omega_0 \ = \ \mathcal{O} (\omega_0^2) \ \approx \
 0
\end{multline}
by \eqref{f0bound}. Similarly $(\partial_t g_0(\vec{r},t))^2 \approx
0$, $\partial_t^3 g_0(\vec{r},t) \approx 0$, etc.

We also write the string fields in the form
\eqref{slowlyvaryingform}, i.e.,
\begin{equation}\label{stringform}
    \theta(s,\vec{r},t) \ = \
    \mathrm{Re} \left \{ \mathrm{e}^{-\mathrm{i}\omega t}
     \theta_0(s ,\vec{r},t) \right \}   \text{ and }
    \varphi(s,\vec{r},t) \ = \
    \mathrm{Re}\left \{ \mathrm{e}^{-\mathrm{i}\omega t}
     \varphi_0(s ,\vec{r},t) \right \} .
\end{equation}
However, it is convenient to use the formulation of \S\ref{StringFT}
\begin{equation}\label{ftstringagain}
\begin{pmatrix}
    \widetilde \theta(\kappa,\vec{r},t) \\
    \widetilde \varphi(\kappa,\vec{r},t)
\end{pmatrix} \ = \  \int_{-\infty}^\infty
\mathrm{e}^{\mathrm{i} \kappa s }
\begin{pmatrix}
  \frac{1}{2 \pi} \theta(s, \vec{r},t ) \\
  \varphi(s,\vec{r},t)
\end{pmatrix} \mathrm{d} s ,
\end{equation}
involving the Fourier transform of the string variable $s$. Then
\eqref{stringform} implies
\begin{multline}\label{stringform2}
    \widetilde \theta(\kappa,\vec{r},t) \ = \
    \frac{1}{2} \left \{ \mathrm{e}^{-\mathrm{i}\omega t}
    \widetilde \theta_0(\kappa ,\vec{r},t)
    + \mathrm{e}^{\mathrm{i} \omega t}
    \widetilde \theta_0(-\kappa ,\vec{r},t)^* \right \} , \\ \text{and}
    \quad \widetilde \varphi(\kappa,\vec{r},t) \ = \
    \frac{1}{2} \left \{ \mathrm{e}^{-\mathrm{i}\omega t}
    \widetilde \varphi_0(\kappa ,\vec{r},t)
    + \mathrm{e}^{\mathrm{i} \omega t}
    \widetilde \varphi_0(-\kappa ,\vec{r},t)^* \right \} ,
\end{multline}
where $\bullet^*$ denotes complex conjugation.

The string equations of motion (\ref{thetamotion},
\ref{varphimotion}) imply the following for $\widetilde \theta_0$
and $\widetilde \varphi_0$:
\begin{align}\label{theta0}
    \partial_t \widetilde \theta_0(\kappa,\vec{r},t) -\mathrm{i}\omega
    \widetilde \theta_0(\kappa,\vec{r},t) \ &= \ -\frac{\kappa^2}{2 \pi}
    \widetilde \varphi_0(\kappa,\vec{r},t) + \frac{1}{ 2 \pi} \widehat \varsigma(\kappa)
    f_0(\vec{r},t) , \\
    \label{phi0}
    \partial_t \widetilde \varphi_0 (\kappa,\vec{r},t) -\mathrm{i}\omega
    \widetilde \varphi_0(\kappa,\vec{r},t) \ &= \ 2 \pi \widetilde \theta_0(\kappa,\vec{r},t)
\end{align}
with $\widehat \varsigma(\kappa) = \begin{pmatrix} \widehat
\varsigma_{\mathrm p} (\kappa) & \widehat \varsigma_{\mathrm q}
(\kappa) \end{pmatrix}$ and $f_0  =
\begin{pmatrix} f_{\mathrm{p} ;0}\\ f_{\mathrm{q} ; 0}
\end{pmatrix}$.
The solution to (\ref{theta0}, \ref{phi0}) with $\theta_0$ and
$\varphi_0$ vanishing as $t\rightarrow-\infty$ is expressed, by the
Fourier inversion formula,
\begin{align}\label{retthetaint}
  \widetilde \theta_0 (\kappa, \vec{r},t) \ & = \  \widehat \varsigma(
  \vec{r},\kappa)  \cdot \left [
  \frac{1}{4 \pi^2}
  \int_{-\infty}^{\infty} \mathrm{e}^{\epsilon t -\mathrm{i}
    \nu t} \frac{- \mathrm i ( \omega + \nu + \mathrm i \epsilon ) }{\kappa^2 - (\omega + \nu + \mathrm{i}
  \epsilon)^2} \widehat f_0(\vec{r}, \nu + \mathrm{i}
  \epsilon) \mathrm{d}
  \nu \right ], \\ \label{retphiint}
  \widetilde \phi_0 (\kappa, \vec{r},t) \ & = \  \widehat \varsigma(\vec{r},\kappa) \cdot \left [
  \frac{1}{2\pi}
  \int_{-\infty}^{\infty} \mathrm{e}^{\epsilon t -\mathrm{i}
    \nu t} \frac{1}{\kappa^2 - (\omega + \nu + \mathrm{i}
  \epsilon)^2} \widehat f_0(\vec{r}, \nu
   + \mathrm{i}
  \epsilon) \mathrm{d}
  \nu  \right ] ,
\end{align}
with $\epsilon > 0$ arbitrary.

The string energy density $\mathsf{H}_{\mathrm{str}}(\vec{r},t)$, as
given by \eqref{mediumenergy}, may be written
\begin{equation}\label{mediumenergyft}
\mathsf{H}_{\mathrm{str}}(\vec{r},t) \ = \ \frac{1}{2}
\int_{-\infty}^\infty
  \left [ 2\pi  \|\widetilde \theta(\kappa, \vec{r},t) \|^2
 + \frac{\kappa^2}{ 2\pi}   \| \widetilde
\varphi(\kappa,\vec{r},t) \|^2 \right ] \mathrm{d} \kappa  .
\end{equation}
Due to the dissipative dynamics of the reduced system, we expect a
steady accumulation of energy in the string degrees of freedom. That
is, $\mathsf{H}_{\mathrm{str}}(\vec{r},t)$ should grow steadily
until the work done by the external force $\rho$ is completely
dissipated. Thus $\mathsf H_{\mathrm{str}}$ should depend quite
strongly on the history of the system. Thus we consider the rate of
dissipation of energy to the strings, the \emph{power density}
$\partial_t \mathsf{H}_{\mathrm{str}}(\vec{r},t)$.

On time scales of order $1/\omega$, the power density $\partial_t
\mathsf{H}_{\mathrm{str}}(\vec{r},t)$ may fluctuate wildly. To
eliminate these fluctuations, we consider the \emph{time averaged
power density}
\begin{equation}\label{tavgHstr}
    \overline{\partial_t \mathsf{H}_{\mathrm{str}}}(\vec{r},t)
    \ = \ \frac{1}{\sigma} \int_{-\infty}^{\infty} \beta(\tau/\sigma)
     \partial_{t} \mathsf{H}_{\mathrm{str}}(\vec{r},t-\tau)
    \mathrm{d} \tau ,
\end{equation}
where $\beta $ is a fixed Schwarz class function with $\int \beta (\tau )%
\mathrm{d}\tau =1$ and $\sigma $ is a time scale much larger than
$1/\omega $ but sufficiently short that $f_{0}$ varies slowly over
intervals of length $\sigma $, i..e $1/\omega \ll \sigma \ll
1/\omega _{0}$.  To provide for that with fixed carrier frequency
$\omega $ and $\delta =\frac{\omega _{0}}{\omega } \rightarrow 0$ we
take
\begin{equation}
\sigma =\frac{1}{\omega \delta ^{\varepsilon }}=\omega ^{\varepsilon
-1}\omega _{0}^{-\varepsilon }
\end{equation}%
with $0<\varepsilon <1/2$,
readily implying%
\begin{equation}
\sigma \propto \delta ^{-\varepsilon }\propto \omega
_{0}^{-\varepsilon }\rightarrow \infty .
\end{equation}
(Recall that $1/\omega _{0}$ is the time scale for $f_{0}$ variation
and we consider the limit $\omega_0 \rightarrow 0$.)

We also assume that
\begin{equation}\label{technical}
    \int_{-\infty}^\infty \tau \beta(\tau) \mathrm{d} \tau \ = \ 0 ,
\end{equation}
as holds, for instance, if $\beta$ is symmetric about zero. Then
given a slowly varying quantity $Q(t)$, for which
\begin{equation}\label{slowlyvarying} Q(t - \tau) = Q(t) - \tau
\partial_t Q(t) + \tau^2 \mathcal O(\delta^2),
\end{equation}
we have \begin{align}\label{svapproximation}
    \overline{Q}(t) \ &= \ Q(t) - \partial_t Q(t) \frac{1}{\sigma} \int_{-\infty}^\infty
    \tau \beta(\tau/\sigma) \mathrm{d} \tau
    +  \frac{1}{\sigma} \int_{-\infty}^\infty \mathcal O(\delta^2)
    \tau^2 \beta(\tau/\sigma) \mathrm{d} \tau \\
    &= \ Q(t) + \mathcal O(\delta^2 \sigma^2) \ \approx \ Q(t)
    \notag ,
\end{align}
since $\delta^2 \sigma^2 = \delta^{2- 2 \varepsilon} \omega^{-1}  =
o(\delta)$ for $\varepsilon < 1/2$.

\begin{proposition}\label{prop:brillouin}
The time averaged power density of dissipation,
$\overline{\partial_t \mathsf H_{\mathrm{str}}}$, has the following
expression, to order $o(\delta)$,
\begin{align}\label{extBri}
  \overline{\partial_t\mathsf{H}_{\mathrm{str}}}(\vec{r},t) \
   \approx \ \frac{1}{2}
  \Biggl \{  & \left \langle f_0(\vec{r},t) ,  \omega \mathrm{Im} \widehat
  \chi(\vec{r},\omega) f_0(\vec{r},t) \right \rangle_{\mathbb C H}
  \\ \notag & + \mathrm{Im} \left \langle \partial_t f_0(\vec{r},t) ,
   \partial_\omega  \omega \mathrm{Im}\widehat
  \chi(\vec{r},\omega) f_0(\vec{r},t) \right \rangle_{\mathbb C H} \\
  \notag & \quad + \frac{1}{2}
  \partial_t  \left \langle f_0(\vec{r},t) ,
  \partial_\omega \omega
  \mathrm{Re} \widehat
  \chi(\vec{r},\omega) f_0(\vec{r},t) \right \rangle_{\mathbb C H}
  \Biggr \} .
\end{align}
\end{proposition}
\noindent \textit{Remarks}:  1.) The inner product $\langle \bullet,
\bullet \rangle_{\mathbb C H}$ denotes the \emph{complex} inner
product in $\mathbb C H$, linear in the second term and conjugate
linear in the first.  2.) In general, the last two terms on the
r.h.s.\ of \eqref{extBri} are $\mathcal O(\delta)$. However, the
first term is of order $\mathcal O(1)$ and is non-negative by the
power dissipation condition. To first order, energy is dissipated at
a steady rate governed by the size of the $\omega \mathrm{Im}
\widehat \chi(\vec{r},\omega)$:
\begin{equation}\label{dissipation}
\overline{\partial_t\mathsf{H}_{\mathrm{str}}}(\vec{r},t) \ = \
\frac{1}{2}\left \langle f_0(\vec{r},t) ,  \omega \mathrm{Im}
\widehat
  \chi(\vec{r},\omega) f_0(\vec{r},t) \right \rangle_{\mathbb C H} \ + \ \mathcal O
  (\delta) .
\end{equation}

\begin{proof}[Proof of Prop.\ \ref{prop:brillouin}]
By \eqref{mediumenergyft} and \eqref{tavgHstr}, the time averaged
power density is the sum of two terms, which may be approximated as
follows
\begin{equation}\label{thetapprox}
    \frac{\pi}{\sigma }
    \int_{-\infty}^{\infty} \beta(\tau/\sigma)
    \int_{-\infty}^\infty \partial_t
    \| \widetilde \theta(\kappa, \vec{r},t -\tau) \|^2
    \mathrm{d} \kappa \mathrm{d} \tau
    \ \approx \ \frac{\pi}{2}
    \int_{-\infty}^\infty \partial_t  \| \widetilde \theta_0(\kappa,\vec{r},t) \|^2
    \mathrm{d} \kappa  ,
\end{equation}
\begin{multline}
    \label{varphiapprox}
    \frac{1}{4 \pi \sigma }
    \int_{-\infty}^{\infty} \beta(\tau/\sigma)
    \int_{-\infty}^\infty \kappa^2 \partial_t \| \widetilde \varphi(\kappa,
    \vec{r},t -\tau) \|^2
    \mathrm{d} \kappa \mathrm{d} \tau
    \\ \approx \ \frac{1}{8 \pi}
    \int_{-\infty}^\infty   \kappa^2 \partial_t  \| \widetilde
    \varphi_0(\kappa,\vec{r},t) \|^2
    \mathrm{d} \kappa
    .
\end{multline}
On the r.h.s.'s of (\ref{thetapprox}, \ref{varphiapprox}) we have
dropped terms with the rapidly oscillating factor $\mathrm{e}^{\pm 2
\mathrm{i} \omega t}$, as their time average is smaller than any
power of $\delta$ as can be seen by repeated integration by parts.
Furthermore we have dropped time averaging from the remaining terms,
by \eqref{svapproximation}, since we will show that $\int \|
\widetilde \theta_0 \|^2 \mathrm{d} \kappa$ and $\int \kappa^2 \|
\widetilde \varphi_0 \|^2 \mathrm{d} \kappa$ are slowly varying in
the sense of \eqref{slowlyvarying}.

Let us first sketch the integration by parts argument allowing to
neglect the terms dropped.  We  focus on a single term missing from
the r.h.s.\ of \eqref{thetapprox}, namely
\begin{multline}\label{firstmisterm}
    \frac{\pi}{4 \sigma}
    \int_{-\infty}^{\infty} \beta(\tau/\sigma)
    \partial_t \int_{-\infty}^\infty  \mathrm{e}^{2 \mathrm{i} \omega (t-\tau)} \left \langle
    \widetilde \theta_0(\kappa, \vec{r},t-\tau)^* , \widetilde \theta_0(-\kappa,
    \vec{r},t-\tau)
    \right \rangle _{\mathbb C H}\mathrm{d} \kappa \mathrm{d} \tau \\
    = \  \frac{\pi}{4 \sigma^2}
    \int_{-\infty}^{\infty}\mathrm{e}^{2 \mathrm{i} \omega (t-\tau)}
     \beta'(\tau/\sigma)
    \int_{-\infty}^\infty   \left \langle
    \widetilde \theta_0(\kappa, \vec{r},t-\tau)^* , \widetilde \theta_0(-\kappa,
    \vec{r},t-\tau)
    \right \rangle_{\mathbb C H} \mathrm{d} \kappa \mathrm{d} \tau,
\end{multline}
where we  have integrated by parts once. Although we have gained a
factor of $1/\sigma$, this does not yet show this term is small,
because $\| \widetilde \theta_0 \|$ could be as large as $1/\delta
\propto \sigma^{1/\varepsilon}$ due to the large amount of energy
absorbed by the strings up to time $t$. However, using $\exp(2
\mathrm{i} \omega t) = (2 \mathrm{i} \omega)^{-n}
\partial_t^n \exp(2 \mathrm{i} \omega t)$, we may integrate by parts
as many times as we like. Thus for any $n$, the r.h.s.\ of
\eqref{firstmisterm} equals
\begin{equation}\label{firstmisterm2}
    \frac{\pi}{4 \sigma^2 (2 \mathrm{i} \omega)^{n}} \int_{-\infty}^{\infty}
    \mathrm{e}^{2 \mathrm{i} \omega (t-\tau)}
    \partial_\tau^{n} \left \{ \beta'(\tau/\sigma) \int_{-\infty}^\infty
    \left \langle
    \widetilde \theta_0(\kappa, \vec{r},t-\tau)^* , \widetilde \theta_0(-\kappa,
    \vec{r},t-\tau)
    \right \rangle_{\mathbb C H} \mathrm{d} \kappa \right \} \mathrm{d} \tau.
\end{equation}
Each $\tau$ derivative acts either on $\beta'$ or on $\langle
\widetilde \theta_0, \widetilde \theta_0^*  \rangle$.  In the first
case, we gain a factor of $1/\sigma = \omega \delta^\varepsilon$ and
in the second case a factor of $\delta$. Thus this term is $\mathcal
O(\delta^{n \varepsilon})$ and, as $n$ is arbitrary, smaller than
any power of $\delta$. The other terms missing from the r.h.s.'s of
(\ref{thetapprox}, \ref{varphiapprox}) \textemdash \ there are
 three in total \textemdash \ may be estimated similarly.

To approximate the two terms on the r.h.s.'s of (\ref{thetapprox},
\ref{varphiapprox}), we use the representations (\ref{retthetaint},
\ref{retphiint}) for $\widetilde \theta_0$ and $\widetilde
\varphi_0$. For instance by \eqref{retthetaint} we have
\begin{multline}\label{tildethetaterm}
  \frac{\pi}{2} \int_{-\infty}^\infty \partial_t \|
  \widetilde \theta_0(\kappa,\vec{r},t) \|^2 \mathrm{d} \kappa
  \\ = \ \frac{1}{32 \pi^3} \iiint
  \partial_t \mathrm{e}^{\mathrm{i}(\nu_1 - \nu_2 - 2\mathrm{i} \epsilon)
  t } \frac{(\omega + \nu_1 - \mathrm{i} \epsilon)
  (\omega + \nu_2 + \mathrm{i} \epsilon)}{\left (\kappa^2 -
  (\omega + \nu_1 - \mathrm{i} \epsilon)^2 \right ) \left (\kappa^2
  - (\omega + \nu_2 + \mathrm{i} \epsilon)^2 \right )} \\
  \times \left \langle \widehat f_0(\vec{r},\nu_1 + \mathrm{i} \epsilon),
   \widehat \varsigma(\vec{r},\kappa)^2
   \widehat f_0(\vec{r},\nu_2 + \mathrm{i} \epsilon)
   \right \rangle_{\mathbb C H}
   \mathrm{d} \nu_1 \mathrm{d} \nu_2 \mathrm{d} \kappa .
\end{multline}
Interchanging integrals to perform the $\kappa$ integration first,
we compute
\begin{multline}
  \int_{-\infty}^\infty \frac{1}{\left (\kappa^2 -
  (\omega + \nu_1 - \mathrm{i} \epsilon)^2 \right ) \left (\kappa^2
  - (\omega + \nu_2 + \mathrm{i} \epsilon)^2 \right )} \
  \widehat \varsigma(\vec r, \kappa)^2 \mathrm{d} \kappa \\
   \begin{aligned} &= \
   \frac{1}{(\omega + \nu_1 - \mathrm{i} \epsilon)^2
    - (\omega + \nu_2 + \mathrm{i} \epsilon)^2 }
    \\ & \quad \times \int_{-\infty}^\infty \left \{
    \frac{1}{\kappa^2 -
  (\omega + \nu_1 - \mathrm{i} \epsilon)^2}
  - \frac{1}{\kappa^2
  - (\omega + \nu_2 + \mathrm{i} \epsilon)^2} \right \}
  \widehat \varsigma(\vec r, \kappa)^2 \mathrm{d} \kappa \\
  & = \  \frac{2 \pi}{(2\omega
  + \nu_1 + \nu_2) (\nu_1 - \nu_2- 2\mathrm{i} \epsilon) }
  \left \{ \widehat \chi(\vec{r}, \omega +\nu_1 + \mathrm{i}
  \epsilon)^* - \widehat \chi(\vec{r}, \omega + \nu_2 + \mathrm{i}
  \epsilon) \right \} ,
  \end{aligned}
\end{multline}
by \eqref{St6fft}. Therefore, taking $\epsilon \rightarrow 0$,
\begin{multline}\label{tildethetaterm2}
  \frac{\pi}{2} \int_{-\infty}^\infty \partial_t \|
  \widetilde \theta_0(\kappa,\vec{r},t) \|^2 \mathrm{d} \kappa
  \\
    \begin{aligned}
  & = \ \frac{\mathrm{i}}{16 \pi^2} \iint
  \mathrm{e}^{+\mathrm{i}(\nu_1 - \nu_2 )
  t } \frac{(\omega + \nu_1)
  (\omega + \nu_2 )}{2\omega
  + \nu_1 + \nu_2} \\
  & \quad \times \left \langle \widehat f_0(\vec{r},\nu_1 ),
   \left \{ \widehat \chi(\vec{r}, \omega +\nu_1 )^* - \widehat \chi(\vec{r}, \omega + \nu_2
   ) \right \}
   \widehat f_0(\vec{r},\nu_2 )
   \right \rangle_{\mathbb C H}
   \mathrm{d} \nu_1 \mathrm{d} \nu_2  .
  \end{aligned}
\end{multline}
Expanding to first order around $\nu_1=\nu_2=0$ we have
\begin{multline}\label{firstorder} \mathrm i \frac{(\omega + \nu_1)
  (\omega + \nu_2 )}{2\omega
  + \nu_1 + \nu_2}  \left \{ \widehat \chi(\vec{r}, \omega +\nu_1)^* - \widehat \chi(\vec{r}, \omega
   + \nu_2  ) \right \} \\ \begin{aligned} & \approx \  \omega
  \mathrm{Im} \widehat \chi(\vec{r},\omega) + \frac{1}{2}  (\nu_1 + \nu_2 )
  \partial_\omega \omega \mathrm{Im} \widehat\chi(\vec{r}, \omega ) + \frac{1}{2} \mathrm i
  (\nu_1 - \nu_2 ) \omega \partial_\omega \mathrm{Re}
  \widehat \chi(\vec{r},\omega ) .
  \end{aligned}
\end{multline}
Thus
\begin{multline}\label{tildethetaterm3}
  \frac{\pi}{2} \int_{-\infty}^\infty \partial_t \|
  \widetilde \theta_0(\kappa,\vec{r},t) \|^2 \mathrm{d} \kappa
  \\
    \begin{aligned}
   \approx \ \frac{1}{4}
  \Biggl \{ & \left \langle f_0(\vec{r},t) ,  \omega \mathrm{Im} \widehat
  \chi(\vec{r},\omega) f_0(\vec{r},t) \right \rangle_{\mathbb C H}
  + \mathrm{Im} \left \langle \partial_t f_0(\vec{r},t) ,
   \partial_\omega  \omega \mathrm{Im}\widehat
  \chi(\vec{r},\omega) f_0(\vec{r},t) \right \rangle_{\mathbb C H} \\ &
  +
  \frac{1}{2} \partial_t  \left \langle f_0(\vec{r},t) ,  \omega
  \partial_\omega
  \mathrm{Re} \widehat
  \chi(\vec{r},\omega) f_0(\vec{r},t) \right \rangle_{\mathbb C H}
  \Biggr \} .
  \end{aligned}
\end{multline}

In a similar way, by \eqref{retphiint},
\begin{multline}\label{tildephitaterm}
  \frac{1}{8 \pi} \int_{-\infty}^\infty \kappa^2 \partial_t \|
  \widetilde \varphi_0(\kappa,\vec{r},t) \|^2 \mathrm{d} \kappa
  \\ = \ \frac{1}{32 \pi^3} \iiint
  \partial_t \mathrm{e}^{+\mathrm{i}(\nu_1 - \nu_2 - 2\mathrm{i} \epsilon)
  t } \frac{\kappa^2 }{\left (\kappa^2 -
  (\omega + \nu_1 - \mathrm{i} \epsilon)^2 \right ) \left (\kappa^2
  - (\omega + \nu_2 + \mathrm{i} \epsilon)^2 \right )} \\
  \times \left \langle \widehat f_0(\vec{r},\nu_1 + \mathrm{i} \epsilon),
  \widehat \varsigma(\vec{r},\kappa)^2
   \widehat f_0(\vec{r},\nu_2 + \mathrm{i} \epsilon)
   \right \rangle_{\mathbb C H}
   \mathrm{d} \nu_1 \mathrm{d} \nu_2 \mathrm{d} \kappa .
\end{multline}
The approximation
\begin{multline}\label{tildephiterm2} \frac{1}{8 \pi} \int_{-\infty}^\infty
\kappa^2 \partial_t \|
  \widetilde \varphi_0(\kappa,\vec{r},t) \|^2 \mathrm{d} \kappa
  \\ \begin{aligned}
   \approx \ \frac{1}{4}
  \Biggl \{ & \left \langle f_0(\vec{r},t) ,  \omega \mathrm{Im} \widehat
  \chi(\vec{r},\omega) f_0(\vec{r},t) \right \rangle_{\mathbb C H}
  + \mathrm{Im} \left \langle \partial_t f_0(\vec{r},t) ,
   \partial_\omega  \omega \mathrm{Im}\widehat
  \chi(\vec{r},\omega) f_0(\vec{r},t) \right \rangle_{\mathbb C H} \\ &
  +
  \frac{1}{2}
  \partial_t  \left \langle f_0(\vec{r},t) ,  \frac{1}{\omega}
  \partial_\omega \omega^2
  \mathrm{Re} \widehat
  \chi(\vec{r},\omega) f_0(\vec{r},t) \right \rangle_{\mathbb C H}
  \Biggr \}
  \end{aligned}
\end{multline}
follows, since
\begin{multline}
  \int_{-\infty}^\infty \frac{\kappa^2}{\left (\kappa^2 -
  (\omega + \nu_1 - \mathrm{i} \epsilon)^2 \right ) \left (\kappa^2
  - (\omega + \nu_2 + \mathrm{i} \epsilon)^2 \right )} \
  \widehat \varsigma(\vec r, \kappa)^2 \mathrm{d} \kappa \\
   \begin{aligned}
  &= \
   \frac{1}{(\omega + \nu_1 - \mathrm{i} \epsilon)^2
    - (\omega + \nu_2 + \mathrm{i} \epsilon)^2 }
    \\ & \quad \times \int_{-\infty}^\infty \left \{
    \frac{(\omega + \nu_1 - \mathrm{i} \epsilon)^2}{\kappa^2 -
  (\omega + \nu_1 - \mathrm{i} \epsilon)^2}
  - \frac{(\omega + \nu_2 + \mathrm{i} \epsilon)^2}{\kappa^2
  - (\omega + \nu_2 + \mathrm{i} \epsilon)^2} \right \}
  \widehat \varsigma(\vec r, \kappa)^2 \mathrm{d} \kappa \\
  & = \ \frac{2 \pi }{(2\omega
  + \nu_1 + \nu_2) (\nu_1 - \nu_2- 2\mathrm{i} \epsilon) } \\
  & \quad \times
  \left \{ (\omega + \nu_1 - \mathrm{i} \epsilon)^2 \widehat \chi(\vec{r}, \omega +\nu_1 + \mathrm{i}
  \epsilon)^* -
  (\omega + \nu_2 + \mathrm{i} \epsilon)^2
  \widehat \chi(\vec{r}, \omega + \nu_2 + \mathrm{i}
  \epsilon) \right \} ,
  \end{aligned}
\end{multline}
again by \eqref{St6fft}.

Combining (\ref{thetapprox}, \ref{varphiapprox},
 \ref{tildethetaterm2}, \ref{tildephiterm2}) we obtain \eqref{extBri}.
\end{proof}

If there is no dissipation at $\vec r$ at frequency $\omega $, so%
\begin{equation}
\mathrm{Im}\widehat{\chi }(\vec{r},\omega )=0\text{ (zero dissipation at $%
\omega $)},  \label{zerodis}
\end{equation}
then the string at $\vec r$ does not effectively absorb energy at
frequency $\omega$.  We expect the total dissipated energy to
fluctuate but not grow, so there should be a formula similar to
\eqref{extBri} for the time average of $\mathsf H_{\mathrm{str}}$.
In this case, only the third term contributes to the r.h.s.\ of
\eqref{extBri}. This term  is a total derivative, suggesting the
approximation
 \begin{equation}\label{zerodisenergy}
\overline{\mathsf{H}_{\mathrm{str}}}(\vec{r},t) \ = \ \frac{1}{4}
    \left \langle f_0(\vec{r},t) ,
  \partial_\omega \omega
  \mathrm{Re} \widehat
  \chi(\vec{r},\omega) f_0(\vec{r},t) \right \rangle_{\mathbb C H}  \ + \ \mathcal O( \delta)
    \quad ( \mathrm{Im} \widehat
\chi(\vec{r},\omega) =  0) .
\end{equation}
Using the methods of the proof of Prop.\ \ref{prop:brillouin} one
may verify that \eqref{zerodisenergy} is indeed correct.
By similar arguments, we find that the time average of the system energy $%
\mathsf{H}_{\mathrm{sys}}(\vec{r},t)$, defined by
(\ref{systemenergy}), satisfies
\begin{equation}
\overline{\mathsf{H}_{\mathrm{sys}}}(\vec{r},t)\ \approx \frac{1%
}{4} \left\langle
f_{0}(\vec{r},t),f_{0}(\vec{r},t)\right\rangle_{\mathbb C H},
\label{hsysr1}
\end{equation}
whether or not there is dissipation at $\omega$. Combining
(\ref{zerodisenergy}) and (\ref{hsysr1}) we obtain:
 \begin{proposition}\label{prop:trueBrillouin}
 If there is no dissipation at frequency $\omega$ at $\vec r$ then the time average of the energy
 density at $\vec r$ satisfies
 \begin{equation}
\overline{\mathsf{H}}(\vec{r},t)\ = \ \frac{1}{4}\left\langle
f_{0}(\vec{r},t),\partial _{\omega }
\left[ \omega \left( 1+%
\mathrm{Re}\widehat{\chi }(\vec{r},\omega )\right) \right] f_{0}(%
\vec{r},t)\right\rangle_{\mathbb C H} + \mathcal O (\delta) .
\label{HBril1}
\end{equation}%
 \end{proposition}
For Maxwell's equations in a TDD dielectric, the above formula
(\ref{HBril1}) reduces to the classical Brillouin formula for the
energy density,  \cite[\S 80]{LandauLif}. (See \eqref{Mbril1}
below.) Thus, \eqref{extBri} may be viewed as an extension of the
Brillouin formula to frequencies with dissipation and to arbitrary
TDD Hamiltonian systems.

Similarly, we may consider the time averaged string Lagrangian
density
\begin{equation}\label{approxL}
    \overline{\mathsf{L}_{\mathrm{str}}}(\vec{r},t) \ \approx \
\frac{1}{4}
    \int_{-\infty}^\infty  \| \theta_0(s,\vec{r},t) \|^2 \mathrm{d} s
    - \frac{1}{4}
    \int_{-\infty}^\infty \| \partial_s \varphi_0(s,\vec{r},t) \|^2 \mathrm{d} s
    ,
\end{equation}
where $\mathsf{L}_{\mathrm{str}}$ denotes the quantity
\begin{equation}\label{Lstr}
    \mathsf{L}_{\mathrm{str}}(\vec{r},t) \ = \
\frac{1}{2}
    \int_{-\infty}^\infty \| \theta(s,\vec{r},t) \|^2 \mathrm{d} s
    - \frac{1}{2}
    \int_{-\infty}^\infty   \| \partial_s \varphi(s,\vec{r},t) \|^2
    \mathrm{d} s .
\end{equation}
First note that by combining (\ref{tildethetaterm3},
\ref{tildephiterm2}) we have
\begin{equation}\label{Lapproxinit}
  \overline{\partial_t\mathsf{L}_{\mathrm{str}}}(\vec{r},t)
  \ \approx \ - \frac{1}{4}
  \partial_t  \left \langle f_0(\vec{r},t) ,
  \mathrm{Re} \widehat
  \chi(\vec{r},\omega) f_0(\vec{r},t) \right \rangle_{\mathbb C H} .
\end{equation}
This approximation holds (to order $o(\delta)$) whether or not the
dissipation $\omega \mathrm{Im} \widehat \chi(\omega)$ vanishes at
frequency $\omega$.  Both sides are total time derivatives and, in
fact, we have:
\begin{proposition}\label{prop:lagrBri}
The time averaged string Lagrangian density satisfies
\begin{equation}\label{Lapprox}
\overline{\mathsf{L}_{\mathrm{str}}}(\vec{r},t) \ = \ - \frac{1}{4}
   \left \langle f_0(\vec{r},t) ,
  \mathrm{Re} \widehat
  \chi(\vec{r},\omega) f_0(\vec{r},t) \right \rangle_{\mathbb C H} + \mathcal O(\delta) ,
\end{equation}
and the total Lagrangian density $\mathsf L$ satisfies
\begin{equation}\label{totalLapprox}
\overline{\mathsf L} (\vec{r},t) \ = \  \frac{1}{2}\mathrm{Re} \left\langle f_{\mathrm{p}%
;0},K_{\mathrm{p}}p_{0}(\vec{r},t)\right\rangle_{\mathbb C H}
-\frac{1}{4}\left\langle
f_{0}(\vec{r},t),\left( 1+\mathrm{Re}\widehat{\chi }(\vec{r}%
,\omega )\right) f_{0}(\vec{r},t)\right\rangle_{\mathbb C H}   +
\mathcal O(\delta).
\end{equation}
\end{proposition}
\begin{proof} Eq.\ \eqref{totalLapprox} follows from \eqref{Lapprox}
using the expressions (\ref{Ldec1}--\ref{Ldec3}) for $\mathsf{L}$,
the approximation \eqref{hsysr1} for the system energy density, and
further integration by parts to obtain the first term on the r.h.s.\
from the corresponding term in \eqref{Ldec2}.

To verify \eqref{Lapprox}, we follow the proof of Prop.\
\ref{prop:brillouin} to obtain
\begin{multline}
\overline{\mathsf{L}_{\mathrm{str}}}(\vec{r},t)  \\
\begin{aligned} & \approx \
  \frac{1}{16 \pi^2} \iint
  \mathrm{e}^{+\mathrm{i}(\nu_1 - \nu_2 )
  t } \frac{1}{2\omega
  + \nu_1 + \nu_2} \frac{1}{\nu_1 - \nu_2 - \mathrm i 0 } \\
&   \quad  \times \Bigl \{ \Bigl \langle \widehat f_0(\vec{r},\nu_1
),
    \left [  (\omega + \nu_1)
  (\omega + \nu_2 ) - (\omega + \nu_1)^2 \right ]
  \widehat \chi(\vec{r}, \omega +\nu_1 )^*\widehat f_0(\vec{r},\nu_2 )
   \Bigr \rangle_{\mathbb C H}  \\
   & \qquad - \Bigl \langle \widehat f_0(\vec{r},\nu_1 ),
   \left [  (\omega + \nu_1)
  (\omega + \nu_2 ) - (\omega + \nu_2)^2 \right ]  \widehat \chi(\vec{r}, \omega + \nu_2
   )   \widehat f_0(\vec{r},\nu_2 ) \Bigr \rangle_{\mathbb C H}  \Bigr \}
   \mathrm{d} \nu_1 \mathrm{d} \nu_2  \\
   & = \  -
  \frac{1}{16 \pi^2} \iint
  \mathrm{e}^{+\mathrm{i}(\nu_1 - \nu_2 )
  t } \frac{1}{2\omega
  + \nu_1 + \nu_2}  \\
&   \quad  \times \Bigl \langle \widehat f_0(\vec{r},\nu_1 ), \left
[ (\omega + \nu_1)
  \widehat \chi(\vec{r}, \omega +\nu_1 )^*
  - (\omega + \nu_2) \widehat \chi(\vec{r}, \omega + \nu_2
   )  \right ] \widehat  f_0(\vec{r},\nu_2 )
   \Bigr \rangle_{\mathbb C H}  \mathrm{d} \nu_1 \mathrm{d} \nu_2.
 \end{aligned}
\end{multline}
The key point is the cancelation of $1/(\nu_1 - \nu_2 - \mathrm i
0)$ by $\nu_1 - \nu_2$ resulting from
\begin{equation}
\begin{aligned}
(\omega + \nu_1)
  (\omega + \nu_2 ) - (\omega + \nu_2)^2 \ &= \ -(\omega + \nu_1)  (\nu_1 - \nu_2 ) , \\
  (\omega + \nu_1)
  (\omega + \nu_2 ) - (\omega + \nu_1)^2 \ & = \  (\omega + \nu_2)  (\nu_1 - \nu_2 ) .
  \end{aligned}
\end{equation}
Up to terms of order $\mathcal O(\delta)$ we may replace
\begin{equation}
\frac{1}{2\omega
  + \nu_1 + \nu_2} \left [
(\omega + \nu_1)
  \widehat \chi(\vec{r}, \omega +\nu_1 )^*
  - (\omega + \nu_2) \widehat \chi(\vec{r}, \omega + \nu_2
   )  \right ]
\end{equation}
by its value at $\nu_1 = \nu_2 =0$, which is $\mathrm{Re} \widehat
\chi(\vec{r}, \omega)$ resulting in \eqref{Lapprox}.
\end{proof}

Our main interest in \eqref{totalLapprox} is in approximating the
time averaged stress tensor.
\begin{proposition}\label{prop:stressBril} For a homogeneous and isotropic system, the time averaged stress
tensor satisfies
\begin{multline}\label{sten2}
\overline{\mathsf{T}_{i,j}}(\vec{r},t) \ = \ \frac{1}{4} \mathrm{Re} \left\{ \mathsf{W}_{i,j;0}(\vec{r},t)+\mathsf{W}_{j,i;0}(\vec{r}%
,t)\right\} +\frac{1}{4} \mathrm{Re} \left\langle g_{i,j}q_{0}(\vec{r}%
,t),\rho _{0}(\vec{r},t)\right\rangle_{\mathbb C H} \\
-\delta _{i,j} \overline{\mathsf L}(\vec{r},t)  + \mathcal
O(\delta),
\end{multline}
where $ \overline{\mathsf L}(\vec{r},t) $ is given by
\eqref{totalLapprox} and
\begin{equation}
\mathsf{W}_{i,j;0}(\vec{r},t)\ =\ -\left\langle \partial_i q_0 (\vec{r}%
,t), \mathbf{Y}_{j}^{\mathrm{T}}f_{\mathrm{q};0}(\vec{r}%
,t)\right\rangle_{\mathbb C H} +\partial _{k}\left\langle g_{i,k}q(\vec{r},t),%
\mathbf{Y}_{j}^{\mathrm{T}}f_{\mathrm{q};0}(\vec{r},t)\right\rangle_{\mathbb
C H}. \label{sten3}
\end{equation}%
\end{proposition}
\begin{proof}Using \eqref{totalLapprox} and the approximation methods of Prop.\ \ref{prop:brillouin}, this follows immediately from the definition \eqref{symmomflux} of Lemma \ref{lem:stresstens}.
\end{proof}

Observe that, unlike the Brillouin formula for the energy density
(\ref{HBril1}), the approximation  (\ref{sten2}--\ref{sten3})  for
the stress tensor does not involve the frequency differentiation of
the susceptibility but simply its value at the given frequency
$\omega $, a property discovered by L. Pitaevskii, \cite{Pi},
\cite[\S 81]{LandauLif}, for the dielectric media.

\section{Example: Maxwell's equations in an inhomogeneous TDD
medium}\label{maxwell}

In this section, we apply the general construction developed above
to the classical Maxwell equations in a material medium,
\cite[Section 1.1, Section 2.2]{Born},
\begin{gather}\label{mxx1}
\begin{cases}
\partial _{t}\mathbf{D}\left( \vec{r},t\right) =\nabla \times
\mathbf{H}\left( \vec{r},t\right) - 4 \pi \mathbf{j}_{\mathrm{ext}}(\vec{r},t) \\
\partial _{t}\mathbf{B} \left( \vec{r},t\right) =-\nabla \times
\mathbf{E}\left( \vec{r} ,t\right) ,
\end{cases} \\
\nabla \cdot \mathbf{B}\left( \vec{r},t\right) =0 \label{mxx1a}
\end{gather}
in units with the speed of light $c=1$. Here  $\mathbf{D}$,
$\mathbf{E}$, $\mathbf{B}$, $\mathbf{H}$ are the electric induction,
electric field, magnetic induction, and magnetic field respectively,
which satisfy the following material relations
\begin{equation}
\mathbf{D}\left( \vec{r},t\right) =\mathbf{E}\left( \vec{r},t\right) +4%
\mathrm{\pi }\mathbf{P}\left( \mathbf{E};\vec{r},t\right) ,\ \mathbf{B}=%
 \mathbf{H}+4\mathrm{\pi }\mathbf{M}\left(
\mathbf{H};\vec{r},t\right)   , \label{mxx2}
\end{equation}
and $\mathbf{j}_{\mathrm{ext}}$ is the external driving current. The
one remaining Maxwell equation
\begin{equation}\label{mxx1aa}
\nabla \cdot \mathbf{D}\left( \vec{r},t\right) = 4 \pi
\rho_{\mathrm{ext}}(\vec{r},t) ,
\end{equation}
with $\rho_{\mathrm{ext}}$ the external charge density, is
automatically satisfied at all times provided it holds at a given
time and that $\mathbf{j}_{\mathrm{ext}}$, $\rho_{\mathrm{ext}}$
together satisfy the equation of continuity:
\begin{equation}\label{contrho}
    \partial_t \rho_{\mathrm{ext}}(\vec{r},t) + \nabla \cdot
    \mathbf{j}_{\mathrm{ext}}(\vec{r},t) \ = \ 0 .
\end{equation}
We allow arbitrary external current $\mathbf{j}_{\mathrm{ext}}$,
taking \eqref{mxx1aa} as the definition of $\rho_{\mathrm{ext}}$.

We take the polarization $\mathbf{P}$ and magnetization $\mathbf{M}$
to be of the linear response form, \cite[Chapter IX, Section
77]{LandauLif},
\begin{eqnarray}
4 \pi \mathbf{P}\left(\vec{r},t\right) &=& \left (  \varepsilon
(\vec{r}) - \mathsf{1} \right ) \cdot
\mathbf{E}\left(\vec{r},t\right) \ + \ \int_{0}^{\infty }
\chi_{\mathrm{E}}\left( \vec{r},\tau \right) \cdot \mathbf{E}\left( \vec{r}%
,t-\tau \right) \,\mathrm{d}\tau ,  \label{mxx3} \\
4 \pi \mathbf{M}\left( \vec{r},t\right) &=& \left ( \mu(\vec{r}) -
\mathsf{1} \right ) \cdot \mathbf{H}\left(\vec{r},t\right) \ + \
\int_{0}^{\infty }
\chi_{\mathrm{H}}\left( \vec{r},\tau \right) \cdot \mathbf{H}\left( \vec{r}%
,t-\tau \right) \,\mathrm{d}\tau ,  \label{mxx3a}
\end{eqnarray}
with
\begin{itemize}\item
$\varepsilon(\vec{r})$ and $\mu(\vec{r})$ the static electric and
magnetic permeability tensors, assumed to be real symmetric and
uniformly bounded from above and below
\begin{equation}
\varepsilon_{-}\mathsf{1}\leq \varepsilon (\vec{r}) \leq
\varepsilon_{+}\mathsf{1},\quad \mu _{-}\mathsf{1}\leq \mu (\vec{r})
\leq \mu _{+}\mathsf{1}, \label{hmx5}
\end{equation}
where  $\mathsf{1}$ is the unit tensor, and $\varepsilon _{\pm },\mu
_{\pm }>0$ are constants.
\item $\chi _{\mathrm{E}}\left( \vec{r},\tau \right) $ and $\chi _{\mathrm{H}}\left(
\vec{r},\tau \right) $ the electric and magnetic susceptibility
tensors, also real symmetric and satisfying a power dissipation
condition, namely,
\begin{gather}
\operatorname{Im}\left\{ \zeta \widehat{\chi}_{\mathrm{F}}\left(
\vec{r},\zeta \right) \right\} \geq 0,\ \zeta =\omega
+\mathrm{i}\eta ,\ \eta \geq 0,
\label{mmx4} \\
\widehat{\chi}_{\mathrm{F}}\left( \vec{r},\zeta \right) \ = \ \int_{0}^{\infty }\mathrm{e}^{\mathrm{i}\zeta t}\chi _{%
\mathrm{F}}\left( \vec{r},t\right) \,\mathrm{d}t,\ \mathrm{F}=\mathrm{E%
},\mathrm{H}.  \notag
\end{gather}
\end{itemize}

\subsection{Hamiltonian structure of the field}
We parameterize the field using the electric field $\mathbf{D}$ and
the vector potential $\mathbf{A}$ as follows:
\begin{equation}\label{hs1}
  \mathbf{u}\  = \ \begin{pmatrix}\mathbf{\Pi} \\
   \mathbf{A} \end{pmatrix} \; , \quad \mathbf{\Pi} = -\mathbf{D} , \quad \mathbf{B} = \nabla \wedge \mathbf{A}
\end{equation}
in the phase space $V= L^2_{4 \pi} (\mathbb{R}^3, \mathbb{R}^3 )
\oplus L^2_{4 \pi} (\mathbb{R}^3, \mathbb{R}^3)$,  where $L^2_{4
\pi} (\mathbb{R}^3,\mathbb{R}^3)$ denotes the space of vector fields
$\mathbf{X}: \mathbb R^3 \rightarrow \mathbb R^3$, with inner
product
\begin{equation}\label{ipdefn}
\langle \mathbf{X}_1, \mathbf{X}_2 \rangle_{L^2_{4 \pi}} \ = \
\frac{1}{4 \pi}
  \int_{\mathbb{R}^3} \mathbf{X}_1(\vec{r}) \cdot \mathbf{X}_2(\vec{r}) \mathrm d^3 \vec{r}.
\end{equation}
Thus for $\mathbf u$ as in \eqref{hs1}
\begin{equation}\label{EMip}
  \left \| \mathbf{u} \right \|^2_{V} \ = \ \frac{1}{4 \pi}
  \int_{\mathbb{R}^3} \left \{
   \left | \mathbf{\Pi}(\vec{r}) \right |^2 +  \left | \mathbf{A}
  (\vec{r}) \right |^2 \right \} \mathrm{d}^3 \vec{r} .
\end{equation}
We define the symplectic operator $J$ on $V$ with the matrix
\begin{equation}\label{EMsymp}
  J \ = \ \begin{pmatrix}
    0 &  -\mathsf{1} \\
    \mathsf{1} & 0
  \end{pmatrix}.
\end{equation}
For the moment we do not impose a gauge condition on the vector
potential $\mathbf{A}$.

We take $\mathbf{\Pi} = - \mathbf{D}$ to be the canonical momentum.
The choice of sign puts the symplectic operator in the canonical
form \eqref{EMsymp}. More important than the choice of sign is the
choice of $\mathbf{D}$ as the momentum variable. This is essentially
forced on us if we wish to use the formalism of
\S\ref{sec:energyflux} and \S\ref{sec:enmom}, since we should have a
Lagrangian which does not depend on spatial derivatives of
$\partial_t Q$.  This choice is also suggested by the coupling of an
external current to the Maxwell equations and is in agreement with
the standard Lagrangian density of relativistic field theory:
$\mathsf{L} = F_{\mu \nu} F^{\mu \nu}$ with $F_{\mu \nu}=
\partial_\mu A_\nu -
\partial_\nu A_\mu$ where $A_\nu$ is the four vector potential \cite{LandauLif1}.
It is a different convention, however, from that advocated by
Sommerfeld \cite{Sommer} and adopted by us in our announcement of
these results \cite{FS3}.

In a non-dispersive medium, with $\chi_{\mathrm{E}}=
\chi_{\mathrm{H}} = 0$, the material relations are
\begin{align} \varepsilon(\vec{r}) \cdot \mathbf{E}(\vec{r}) \ &= \  \mathbf{D}
  (\vec{r}) , \label{NDmaterialb} \\ \label{NDmateriala}
\mu \left( \vec{r}\right) \cdot \mathbf{H}\left( \vec{r}\right) \ &=\mathbf{B}%
\left( \vec{r}\right) =\ \nabla \times \mathbf{A}\left(
\vec{r}\right) .
\end{align}
Identifying these equations with \eqref{hpq3} and recalling the
classical expression for the electro-magnetic field energy in a
static dielectric, i.e.,
\begin{equation}
\frac{1}{8 \pi} \int_{\mathbb R^3} \left \{   \mathbf{E}(\vec{r})
\cdot \varepsilon(\vec{r}) \cdot \mathbf{E}(\vec{r})  +
\mathbf{H}(\vec{r}) \cdot \mu(\vec{r}) \cdot \mathbf{H}(\vec{r})
\right \} \mathrm{d}^3 \vec{r} ,
\end{equation}
suggests parameterizing stress space with the vector
\begin{equation}\label{fdef}
\mathbf f \ = \ \begin{pmatrix}
\mathbf{f}_{\mathrm{E}} \\
\mathbf{f}_{\mathrm{H}}
                \end{pmatrix}
\ = \
\begin{pmatrix}
    \sqrt{\varepsilon(\vec{r})} \cdot \mathbf{E} \\
    \sqrt{\mu(\vec{r})} \cdot \mathbf{H}
                \end{pmatrix} .
\end{equation}
Thus we take the stress space $H=V =  L^2_{4 \pi} (\mathbb{R}^3,
\mathbb{R}^3 ) \oplus L^2_{4 \pi} (\mathbb{R}^3, \mathbb{R}^3)$.

The impedance $K : V \rightarrow H$ is implicitly defined by
(\ref{NDmateriala}, \ref{NDmaterialb}), since $K \mathbf u = \mathbf
f$,
\begin{equation}\label{fmax}
 K \mathbf{u} \ = \ \begin{pmatrix}
 \sqrt{\varepsilon(\vec{r})} \cdot \mathbf{E} \\
    \sqrt{\mu(\vec{r})} \cdot \mathbf{H}
                \end{pmatrix} .
 \ = \ \begin{pmatrix} - [\sqrt{\varepsilon(\vec{r})}]^{-1} \cdot \mathbf \Pi  \\
[\sqrt{\mu(\vec{r})}]^{-1} \cdot \nabla \times \mathbf{A}
                    \end{pmatrix} .
\end{equation}
Thus
\begin{equation}
  K  \ = \  \begin{pmatrix}
     \mathsf{K}_{\mathrm{E}}(\vec{r})& 0\\
    0 &   \mathsf{K}_{\mathrm{H}}(\vec{r})
    \end{pmatrix} \begin{pmatrix}
      - \mathsf{1} & 0 \\
      0 & \nabla \times
    \end{pmatrix} \ = \
\begin{pmatrix}
-\mathsf{K}_{\mathrm{E}}(\vec{r})& 0 \\
0 & \mathsf{K}_{\mathrm{H}}(\vec{r})\nabla \times%
\end{pmatrix}
\end{equation}
with
\begin{equation}
 \mathsf{K}_{\mathrm{E}}(\vec{r}) \ = \ [\sqrt{\varepsilon(\vec{r})}]^{-1} \quad \text{and} \quad \mathsf{K}_{\mathrm{H}}(\vec{r}) \ = \ [\sqrt{\mu(\vec{r})}]^{-1}
\end{equation}
well defined positive definite tensors by \eqref{hmx5}.

The Hamiltonian is therefore
\begin{equation}
  \mathrm{h} \ = \ \int_{\mathbb{R}^3} \mathsf{h}(\vec{r}) \mathrm{d}^3
  \vec{r},
\end{equation}
with
\begin{equation}\label{emndham}
  \mathsf{h}(\vec{r}) \ = \ \frac{1}{8 \pi}  \left \{ \mathbf{\Pi}(\vec{r})
  \cdot \bigl [ \varepsilon(\vec{r}) \bigr ]^{-1} \cdot \mathbf{\Pi}(\vec{r})
  + \left \{ \nabla
  \times \mathbf{A}(\vec{r}) \right \} \cdot \bigl [ \mu(\vec{r}) \bigr ]^{-1}
  \cdot \left \{ \nabla
  \times \mathbf{A}(\vec{r}) \right \} \right \} \notag .
\end{equation}
The resulting equations of motion, expressed in the form
(\ref{hpq2}, \ref{hpq3}), are
\begin{align}\label{ma3}
  \partial_t \begin{pmatrix}
    \mathbf{\Pi}(\vec{r},t) \\
    \mathbf{A}(\vec{r},t)
  \end{pmatrix} \ &= \ \begin{pmatrix} -
    \nabla \times
   \left\{ \mathsf{K}_{\mathrm{H}}(\vec{r}) \cdot \mathbf{f}_{\mathrm{H}}(\vec{r},t) \right \} \\
   - \mathsf{K}_{\mathrm{E}}(\vec{r}) \cdot \mathbf{f}_{\mathrm{E}}(\vec{r},t)
  \end{pmatrix} \\
  \begin{pmatrix}
    \mathbf{f}_{\mathrm{E}}(\vec{r},t)\\
   \mathbf{f}_{\mathrm{H}}(\vec{r},t)
  \end{pmatrix} \ &= \ \begin{pmatrix}
     - \mathsf{K}_{\mathrm{E}}(\vec{r})  \cdot \mathbf{\Pi}(\vec{r},t) \\
    \mathsf{K}_{\mathrm{H}}(\vec{r})  \cdot \left \{ \nabla \times \mathbf{A}(\vec{r},t)
    \right \}
  \end{pmatrix} \label{ma3a}\; .
\end{align}
Eq. (\ref{ma3}) implies the two dynamical Maxwell equations once we
take the curl of the second component and substitute $\mathbf{\Pi}=
-\mathbf{D} $, $\mathbf{f}_{\mathrm{E}} =
\mathsf{K}_{\mathrm{E}}(\vec{r})^{-1} \mathbf{E}$ and
$\mathbf{f}_{\mathrm{H}}= \mathsf{K}_{\mathrm{H}}(\vec{r})^{-1}
\mathbf{H}$. Similarly, the material relations (\ref{NDmaterialb},
\ref{NDmateriala}) follow from \eqref{ma3a}. The divergence
condition \eqref{mxx1a} is satisfied since $\nabla \cdot \{\nabla
\times \bullet \} \equiv 0$, which also shows that $\partial_t
\nabla \cdot \mathbf{D} =0$, using  \eqref{ma3}. Thus we may define
the time independent external charge density $\rho_{\mathrm{ext}}
(\vec{r}) =\frac{1}{4 \pi} \nabla \cdot \mathbf{D}(\vec{r},t)$,  so
that \eqref{mxx1aa} and \eqref{contrho} hold (with
$\mathbf{j}_{\mathrm{ext}}\equiv0$).

When the system is driven by an external current
$\mathbf{j}_{\mathrm{ext}}$, we replace \eqref{ma3} by
\begin{equation}\label{drivenmx}
\partial_t \begin{pmatrix}
    \mathbf{\Pi}(\vec{r},t) \\
    \mathbf{A}(\vec{r},t)
  \end{pmatrix} \ = \  \begin{pmatrix} -
    \nabla \times
   \left\{ \mathsf{K}_{\mathrm{H}}(\vec{r}) \cdot \mathbf{f}_{\mathrm{H}}(\vec{r},t) \right \} \\
   - \mathsf{K}_{\mathrm{E}}(\vec{r}) \cdot \mathbf{f}_{\mathrm{E}}(\vec{r},t)
  \end{pmatrix}
 + \begin{pmatrix}
   4 \pi \mathbf{j}_{\mathrm{ext}}(\vec{r},t)\\ 0
 \end{pmatrix}.
\end{equation}

The gauge freedom for $\mathbf{A}$ is related to the non-trivial
kernel for the impedance:
\begin{equation}\label{kerK}
\ker K \ = \ \{ 0 \} \oplus H_{\mathrm{grad}}
\end{equation}
with
\begin{equation}\label{Hgrad}
    H_{\mathrm{grad}} \ = \ \text{ closure in }L^{2}\left( \mathbb{R}^{3};\mathbb{R}%
^{3}\right) \text{ of }\left\{ \nabla \psi \left( \vec{r}%
\right) : \psi \left( \vec{r}\right) \in C^1_0\left( \mathbb{R}%
^{3} \right) \right\} .
\end{equation}
If $(\mathbf{\Pi}(\vec{r},t), \mathbf{A}(\vec{r},t))$ is a solution
to (\ref{ma3}, \ref{ma3a}) or (\ref{drivenmx}, \ref{ma3a}), so is
$(\mathbf{\Pi}(\vec{r},t)$, $\mathbf{A}(\vec{r},t) + \nabla
\psi(\vec{r}))$ for
 arbitrary (time independent) $\psi$. This is essentially the same as
 the invariance under translation of the center of mass for mechanical systems (see
 \S\ref{sec:circularstring} below), with the significant difference that only
 the magnetic field $\mathbf{B} = \nabla \times \mathbf{A}$
 is directly observable, so we cannot detect the shift.
Gauge fixing of $\mathbf{A}$ is implemented by the boundary
condition $\mathbf{A}(\vec{r}, t=-\infty) = \nabla \psi \in \ker K$
at $t = - \infty$, with the choice of $\psi$ having no effect on any
quantity expressed in terms of $\mathbf{B} = \nabla \times
\mathbf{A}$. Henceforth, we take $\psi =0$.


\subsection{Extended Hamiltonian for a TDD-Maxwell system}
To relate the TDD dielectric
medium to the general local TDD medium of \S\ref{local},  we associate $%
\mathbf{\Pi }=-\mathbf{D}$ with the momentum $p$ and $\mathbf{A}$
with the coordinate $q$. Respectively the electric field
$\mathbf{E}$ is associated with $f_{\mathrm{p}}$  and the magnetic
field $\mathbf H$ is associated with $f_{\mathrm{q}}$. In a TDD
medium, the material relations (\ref{NDmaterialb},
\ref{NDmateriala}) are replaced by
\begin{align}\varepsilon(\vec{r}) \cdot \mathbf{E}(\vec{r})
  \ + \ \int_0^\infty \mathrm{d}
 \tau  \,  \chi_{\mathrm{E}}
 (\vec{r},\tau) \cdot \mathbf{E}(\vec{r},t-\tau) \ &= \  \mathbf{D}
  (\vec{r}) \; , \label{DDmaterialb}\\
\label{DDmateriala}
 \mu(\vec{r})\cdot \mathbf{H}(\vec{r},t) \ + \ \int_0^\infty \mathrm{d}
 \tau  \,  \chi_{\mathrm{H}}
 (\vec{r},\tau) \cdot \mathbf{H}(\vec{r},t-\tau)  \ &= \  \nabla \times \mathbf{A}
 (\vec{r}) .
\end{align}
Defining
\begin{equation}\label{DDsuscept}
\chi(t) \mathbf{f}(\vec{r})\ = \ \begin{pmatrix}
\mathsf{K}_{\mathrm{E}}(\vec{r})
\cdot  \chi_{\mathrm{E}} (\vec{r},t) \cdot \mathsf{K}_{\mathrm{E}}(\vec{r}) & 0 \\
0 &  \mathsf{K}_{\mathrm{H}}(\vec{r}) \cdot
     \chi_{\mathrm{H}} (\vec{r},t) \cdot \mathsf{K}_{\mathrm{H}}(\vec{r})
\end{pmatrix} \mathbf{f}(\vec{r}) \; ,
\end{equation}
puts the system exactly in the form \eqref{driven2} considered above
\begin{equation}
  K \begin{pmatrix}
    \mathbf{\Pi} \\
    \mathbf{A}
  \end{pmatrix}(\vec{r},t) \ = \ \begin{pmatrix}
    \mathbf{f}_{\mathrm{E}} \\
    \mathbf{f}_{\mathrm{E}}
  \end{pmatrix}(\vec{r},t)  +  \int_0^\infty  \chi(\tau)\begin{pmatrix}
     \mathbf{f}_{\mathrm{E}} \\
    \mathbf{f}_{\mathrm{E}}
  \end{pmatrix}(\vec{r},t-\tau)  \mathrm{d} \tau.
\end{equation}
Note that $\chi(t)$ satisfies the power dissipation condition on $H$
by \eqref{mmx4} and \eqref{hmx5}.

The Hamiltonian for the resulting QHE  is (after a permutation of
coordinates):
\begin{equation}\label{HMx1}
    \mathrm{H}(\mathbf{U}) \ = \ \frac{1}{2} \langle \mathcal{K} \mathbf{U},
    \mathcal{K} \mathbf{U} \rangle_{\mathcal{H}} \ = \ \frac{1}{2} \langle
    \mathcal{K}_{\mathrm{E}} \mathbf{U}_{\mathrm{E}},
    \mathcal{K}_{\mathrm{E}} \mathbf{U}_{\mathrm{E}} \rangle_{\mathcal{H}_{\mathrm{E}}}
    +  \frac{1}{2} \langle
    \mathcal{K}_{\mathrm{H}} \mathbf{U}_{\mathrm{H}},
    \mathcal{K}_{\mathrm{H}} \mathbf{U}_{\mathrm{H}} \rangle_{\mathcal{H}_{\mathrm{H}}} ,
\end{equation}
with
\begin{equation}
  \mathbf{U} \ = \ \begin{pmatrix}
    \mathbf{U}_{\mathrm{E}} \\
    \mathbf{U}_{\mathrm{H}}
  \end{pmatrix} , \quad \mathbf{U}_{\mathrm{E}} = \begin{pmatrix}
    \mathbf{\Pi}(\vec{r}) \\
    \boldsymbol{\theta}_{\mathrm{E}}(\vec{r},s) \\
    \boldsymbol{\phi}_{\mathrm{E}}(\vec{r},s)
  \end{pmatrix}, \quad \mathbf{U}_{\mathrm{H}} = \begin{pmatrix}
    \mathbf{A}(\vec{r}) \\
    \boldsymbol{\theta}_{\mathrm{H}}(\vec{r},s) \\
    \boldsymbol{\phi}_{\mathrm{H}}(\vec{r},s)
  \end{pmatrix} ,
\end{equation}
and extended impedance operator
\begin{multline}
  \mathcal{K} \ = \ \begin{pmatrix}
    \mathcal{K}_{\mathrm{E}} & 0 \\
    0 &    \mathcal{K}_{\mathrm{H}}
  \end{pmatrix} , \\ \mathcal{K}_{\mathrm{E}} = \begin{pmatrix}
    - \mathsf{K}_{\mathrm{E}}(\vec{r}) & 0 & -T_{\mathrm{E}} \\
    0 & 1 & 0 \\
    0 & 0 & \partial_s
  \end{pmatrix} , \quad \mathcal{K}_{\mathrm{H}} = \begin{pmatrix}
\mathsf{K}_{\mathrm{H}}(\vec{r}) \cdot \nabla \times & 0 & -T_{\mathrm{H}} \\
    0 & 1 & 0 \\
    0 & 0 & \partial_s
  \end{pmatrix} .
\end{multline}
The extended phase space and stress space are equal
\begin{equation}
\mathcal H \ = \ \mathcal V \  = \ \mathcal V_{\mathrm{E}} \oplus
\mathcal V_{\mathrm{E}},
\end{equation}
where
\begin{align}
 \mathcal{V}_{\mathrm{E}} \ &= \
 L^2_{4\pi}(\mathbb{R}^3, \mathbb{R}^3) \oplus
  L^2(\mathbb{R}, L^2_{4\pi}(\mathbb{R}^3, \mathbb{R}^3) ) \oplus
  L^2(\mathbb{R}, L^2_{4\pi}(\mathbb{R}^3, \mathbb{R}^3) ) \\
 &= \  L^2_{4 \pi}(\mathbb{R}^3, \mathbb{R}^3) \oplus
  L^2_{4 \pi}(\mathbb{R}\times \mathbb{R}^3, \mathbb{R}^3) \oplus
  L^2_{4 \pi}(\mathbb{R}\times \mathbb{R}^3, \mathbb{R}^3) .\notag
\end{align}
The extended symplectic operator is
\begin{equation}\label{extsympl}
    \mathcal{J} \ = \ \begin{pmatrix}
      0 & 0 & 0 & -1 & 0 & 0 \\
      0 & 0 & -1 & 0 & 0 & 0 \\
      0 & 1 & 0 & 0 & 0 & 0 \\
      1 &  0 & 0 & 0 & 0 & 0 \\
      0 &  0 & 0 & 0 & 0 & -1 \\
      0 & 0 & 0 & 0 & 1 & 0
    \end{pmatrix} .
\end{equation}

The string coupling operators $T_{\mathrm{F}}$ are obtained from the
susceptibilities as follows
\begin{equation}
[T_{\mathrm{F}} \boldsymbol{\phi}_{\mathrm{F}}](\vec{r}) \ = \
\int_{-\infty}^\infty \varsigma_{\mathrm{F}}(\vec{r},s) \cdot
\boldsymbol{\phi}_{\mathrm{F}}(\vec{r},s) , \quad \mathrm{F}=
\mathrm{E, H},
\end{equation}
with coupling functions \begin{equation}
\varsigma_{\mathrm{F}}(\vec{r},s) \ = \ \frac{1}{2 \pi }
\int_{-\infty}^\infty \cos(\omega s)
\widehat{\varsigma}_{\mathrm{F}} (\vec{r}, \omega) \mathrm{d} \omega
, \quad \mathrm{F}= \mathrm{E, H},
\end{equation}
where
\begin{align}
\widehat{\varsigma}_{\mathrm{F}} (\vec{r}, \omega)\  &= \  \sqrt{ 2
\omega \mathsf{K}_{\mathrm{F}}(\vec{r}) \cdot \mathrm{Im}
\widehat{\chi}_{\mathrm{F}}(\vec{r},\omega) \cdot
\mathsf{K}_{\mathrm{F}}(\vec{r}) }, && \mathrm{F}= \mathrm{E,H}
\\&= \  \sqrt{ 2 \omega \left [ \sqrt{\varepsilon(\vec{r})} \right ]^{-1}
\mathrm{Im}
\widehat{\chi}_{\mathrm{E}}(\vec{r},\omega)   \left [ \sqrt{\varepsilon(\vec{r})} \right ]^{-1}}, && \mathrm{F} = \mathrm{E} \notag, \\
 &= \   \sqrt{ 2 \omega \left [ \sqrt{\mu(\vec{r})} \right ]^{-1}
\mathrm{Im} \widehat{\chi}_{\mathrm{E}}(\vec{r},\omega)   \left [
\sqrt{\mu(\vec{r})} \right ]^{-1}} , && \mathrm{F} = \mathrm{H}.
\notag
\end{align}

The Hamiltonian \eqref{HMx1} is conveniently expressed as
\begin{equation}\label{EMHam}\mathrm{H} \ = \
\int_{\mathbb{R}^3} \mathsf{H}(\vec{r}) \mathrm{d}^3 \vec{r},
\end{equation}
with the density
\begin{multline}\label{EMHamdens}
\mathsf{H}(\vec{r}) = \frac{1}{8 \pi} \Biggl \{ \mathbf{E}(\vec{r})
\cdot \varepsilon(\vec{r}) \cdot \mathbf{E}(\vec{r}) +
\mathbf{H}(\vec{r}) \cdot \mu(\vec{r}) \cdot \mathbf{H}(\vec{r})
\Biggr . \\ \qquad + \int_{-\infty}^\infty \Bigl [ \left |
\partial_s \boldsymbol{\phi}_{\mathrm{E}}(\vec{r}, s)\right  |^2
 + \left | \boldsymbol{\theta}_{\mathrm{E}}(\vec{r}, s) \right |^2 +
\left | \partial_s \boldsymbol{\phi}_{\mathrm{H}}(\vec{r}, s) \right
|^2
 + \left | \boldsymbol{\theta}_{\mathrm{H}}(\vec{r}, s) \right |^2 \Bigr
] \mathrm{d} s \Biggr \}  .
\end{multline}
Here $\mathbf{E}$ and $\mathbf{H}$ are related to the canonical
variables $\mathbf{\Pi}= - \mathbf{D}$, $\mathbf{A}$, and
$\boldsymbol{\varphi}_{\mathrm{E,H}}$ by
\begin{align}\label{EMHamE}
  \mathbf{E}(\vec{r}) \ &= \
  \varepsilon (\vec{r})^{-1} \cdot\mathbf{D}(\vec{r})
  -  [\sqrt{\varepsilon(\vec{r})}]^{-1} \cdot \int_{-\infty}^\infty
  \varsigma_{\mathrm{E}}(\vec{r}, s) \cdot
  \boldsymbol{\phi}_{\mathrm{E}}(\vec{r}, s) \mathrm{d} s ,\\
  \mathbf{H}(\vec{r}) \ &= \
 \mu  (\vec{r})^{-1} \cdot \left \{ \nabla \times
  \mathbf{A} \right \} (\vec{r}) -  [\sqrt{\mu(\vec{r})}]^{-1} \cdot \int_{-\infty}^\infty
  \varsigma_{\mathrm{H}}(\vec{r}, s) \cdot
  \boldsymbol{\phi}_{\mathrm{H}}(\vec{r}, s) \mathrm{d} s .
  \label{EMHamH}
\end{align}
As in \eqref{hamdens}, the total energy is a sum of terms
corresponding to the energy of the TDD system (the electromagnetic
field) and the energy of the strings (the medium), with no
interaction term. This might be puzzling, however $\mathbf{E}$ and
$\mathbf{H}$, as defined in (\ref{EMHamE}, \ref{EMHamH}),
incorporate the interaction with the strings.

The equations of motion for the extended system,
\begin{equation}
    \partial_t \mathbf{U} \ = \ \mathcal{J K^{\mathrm{T}}}\mathcal{K} \mathbf{U}
     +
  \mathbf{R}
\end{equation}
with $\mathbf{R}=\operatorname{col}(-4\pi \mathbf{j}_{\mathrm{ext}},
0 , 0 ,0 ,0 ,0)$ the external current, may be written as follows:
\begin{equation}\label{EMExtmoto}
  \left \{ \begin{aligned}
    \partial_t  \mathbf{D}(\vec{r},t)  &= \nabla \times \mathbf{H} (\vec{r},t)
    - 4 \pi \mathbf{j}_{\mathrm{ext}}(\vec{r},t)\\
   \partial_t \boldsymbol{\theta}_{\mathrm{E}}(\vec{r},s, t) &=
    \partial_s^2
  \boldsymbol{\phi}_{\mathrm{E}}(\vec{r},s,t) + \varsigma_{\mathrm{E}}(\vec{r},s)
   \cdot \sqrt{\varepsilon(\vec{r})} \cdot
  \mathbf{E}(\vec{r},t) \\
     \partial_t \boldsymbol{\phi}_{\mathrm{E}}(\vec{r},s, t) &=
     \boldsymbol{\theta}_{\mathrm{E}}(\vec{r},s,t) \\
    \partial_t \mathbf{A}(\vec{r},t) &= -\mathbf{E}(\vec{r},t) \\
     \partial_t \boldsymbol{\theta}_{\mathrm{H}}(\vec{r},s, t) &=
     \partial_s^2
  \boldsymbol{\varphi}_{\mathrm{H}}(\vec{r},s,t) + \varsigma_{\mathrm{H}}(\vec{r},s)
     \cdot \sqrt{\mu(\vec{r})} \cdot \mathbf{H}(\vec{r},t) \\
     \partial_t \boldsymbol{\phi}_{\mathrm{H}}(\vec{r},s, t) &=
     \boldsymbol{\theta}_{\mathrm{H}}(\vec{r},s,t) ,
  \end{aligned} \right .
\end{equation}
with $\mathbf{E}$ and $\mathbf{H}$ given by (\ref{EMHamE},
\ref{EMHamH}).

\subsection{Energy flux and stress tensor for the TDD Maxwell
system}\label{sec:mxenmom} Study of the stress tensor in dispersive
dielectric media has a rather long history, see \cite{JK},
\cite{Rob}, \cite{KJ},\cite{JLiu} and references therein. In
particular it is used to compute the ponderomotive and Abraham
forces, \cite{KW}, \cite{BK}, \cite[\S 75, \S 81]{LandauLif},
\cite[Section 2]{KJ}. The first formula for the stress tensor was
derived by L. Pitaevskii, \cite{Pi}, \cite[\S 81]{LandauLif},
\cite[Section 3.2]{KJ} for almost time harmonic fields in a
transparent, i.e. lossless, medium. The formula was derived by
applying thermodynamical methods and time averaging for a resonance
circuit and a capacitor filled with the dielectric. Pitaevskii's
formula is unexpectedly simple: one has to simply replace
$\varepsilon $ and $\mu $ in the expression of the stress tensor for
the case of non dispersive medium with respectively $\varepsilon
\left( \omega \right) $ and $\mu \left( \omega \right) $. This
differs dramatically from the case of the energy density where one
has to replace $\varepsilon $ and $\mu $ with nontrivial frequency
derivatives $\frac{d}{d\omega }\left[ \omega \varepsilon \left(
\omega \right) \right] $ and $\frac{d}{d\omega }\left[ \omega \mu
\left( \omega \right) \right] $.

In this section we treat the stress tensor for arbitrary  fields
\textemdash \, not necessarily almost-monochromatic \textemdash \,
in TDD dielectric media, based on the formalism of \S\ref{local}. We
recover Pitaevskii's formula in the next section using Prop.\
\ref{prop:stressBril}.

To make contact with the results of  \S\ref{local}, we take
coordinate variables $q= \mathbf{A}$, $\phi =
(\boldsymbol{\phi}_{\mathrm{E}}, \boldsymbol{\phi}_{\mathrm{H}})$
and momentum variables $p =\mathbf{\Pi}=-\mathbf{D}$, $\theta=
(\boldsymbol{\theta}_{\mathrm{E}},
\boldsymbol{\theta}_{\mathrm{H}})$, with the spaces $V_0$,
$H_{\mathrm{p,q}}^0$ all equal to $\mathbb{R}^3$ with the inner
product
\begin{equation}\label{V0Hpqip}
    \langle \mathbf{v}, \mathbf{w} \rangle_{V_0} \ = \
    \langle \mathbf{v}, \mathbf{w} \rangle_{H_{\mathrm{p}}^0} \ = \
    \langle \mathbf{v}, \mathbf{w} \rangle_{H_{\mathrm{q}}^0} \ = \
    \frac{\mathbf{v} \cdot \mathbf{w}}{4 \pi} .
\end{equation}
The map $\mathsf{K}_{\mathrm{p}}$ is
\begin{equation}
    \mathsf{K}_{\mathrm{p}}(\vec{r}) \mathbf{\Pi} \ = \ \mathsf{K}_{\mathrm{E}}
    (\vec{r}) \cdot
    \mathbf{\Pi} ,
\end{equation}
and $\mathsf{K}_{\mathrm{q}} = 0$ in (\ref{impedancedensity}),
asonly spatial derivatives of $\mathbf{A}$ appear in the Hamiltonian
\eqref{emndham}. The maps $\mathbf{Y}_i(\vec{r})$, $i=1,2,3$, are as
follows:
\begin{equation}\label{Dem}
    \mathbf{Y}_i(\vec{r}) \mathbf{A} \ = \
    \mathsf{K}_{\mathrm{H}}(\vec{r}) \cdot \left \{ \mathbf{e}_i \times \mathbf{A}
    \right \},
\end{equation}
with $\mathbf{e}_i$ the unit vector in the $i^{\mathrm{th}}$
coordinate direction, so that
\begin{equation}
    \mathbf{Y}(\vec{r}) \cdot \nabla \mathbf{A}
    \ = \ \sum_{i=1}^3 \mathbf{Y}_i(\vec{r}) \partial_{i} \mathbf{A}
    \ = \ \mathsf{K}_{\mathrm{H}}(\vec{r}) \cdot \left \{ \nabla \times \mathbf{A}
    \right \} .
\end{equation}

The general representation (\ref{lagrandens}) for the Lagrangian
density of the extended system specializes in this case to
\begin{multline}\label{EMLagrandens}
 \mathsf{L}(\vec{r},t ) \ = \ \frac{1}{8 \pi} \Biggl \{  \left |
  \mathbf{f}_{\mathrm{E}}(\vec{r},t)
  \right |^2 + 2 \mathbf{f}_{\mathrm{E}}(\vec{r},t) \cdot \left \{
  \int_{-\infty}^\infty \mathsf{\varsigma}_{\mathrm{E}}(s,\vec{r})
  \boldsymbol{\phi}_{\mathrm{E}}(s,\vec{r},t)
  \mathrm{d} s\right \}
    \\  - \left | \mathbf{f}_{\mathrm{H}}(\vec{r},t) \right |^2
  + \sum_{\mathrm{F}=\mathrm{E,H}}
  \int_{-\infty}^\infty
  \left[ \left | \partial_t \boldsymbol{\phi}_{\mathrm{F}}(s,\vec{r},t) \right
  | ^2
  \mathrm{d} s -
  \left | \partial_s \boldsymbol{\phi}_{\mathrm{F}}(s,\vec{r},t) \right
  | ^2 \right ]
  \mathrm{d} s
  \Biggr \} ,
\end{multline}
with  $\mathbf{f}_{\mathrm{p}}= \mathbf{f}_{\mathrm{E}}$ and
$\mathbf{f}_{\mathrm{q}} = \mathbf{f}_{\mathrm{H}}$ corresponding to
the material relations (\ref{fq}) and (\ref{fp}):
\begin{align}
  \mathbf{f}_{\mathrm{E}}(\vec{r},t)  \ &= \  - \mathsf{K}_{\mathrm{E}}(\vec{r})
   \cdot
    \mathbf{\Pi}(\vec{r},t)  - \int_{-\infty}^\infty \varsigma_{\mathrm{E}}(
    \vec{r},s)
    \cdot \boldsymbol{\phi}_{\mathrm{E}}(\vec{r},s,t) \mathrm{d} s \\
    \mathbf{f}_{\mathrm{H}}(\vec{r},t) \ &= \
     \mathsf{K}_{\mathrm{H}}(\vec{r}) \cdot
    \{ \nabla \times \mathbf{A} \} (\vec{r},t)
    - \int_{-\infty}^\infty \varsigma_{\mathrm{H}}(\vec{r},s)
    \cdot \boldsymbol{\phi}_{\mathrm{H}}(\vec{r},s,t) \mathrm{d} s .
\end{align}

The vector potential $\mathbf{A}$ transforms as a vector under
rotations, i.e.,
\begin{equation}
    g_{i,j} \mathbf{A}(\vec{r},t) \ = \ - \mathbf{e}_i \mathbf{A}_{j}(\vec{r},t)
    + \mathbf{e}_j \mathbf{A}_{i}(\vec{r},t) ,
\end{equation}
i.e., $g_{i,j}  =  \mathbf{e}_i \otimes \mathbf{e}_j - \mathbf{e}_j
\otimes \mathbf{e}_i$. The vectors
$\mathbf{f}_{\mathrm{E},\mathrm{H}}$ transform identically,
$g_{i,j}^{\mathrm{p},\mathrm{q}} = g_{i,j}$.
\begin{lemma}
The system is homogeneous if and only if $\mathsf{K}_{\mathrm{F}}$
and $\mathsf{\varsigma}_{\mathrm{F}}$, $\mathrm{F}=\mathrm{E,H}$,
are independent of $\vec{r}$, and is isotropic if and only if they
are scalars.
\end{lemma}
\begin{proof}
  This is obvious, except for the proof of \eqref{Yiso} for
$\mathbf{Y}$ given by \eqref{Dem} with scalar
$\mathsf{K}_{\mathrm{H}}$, which is straightforward but tedious.
\end{proof}

Although we use the formalism of \S\ref{local}, we wish to express
the resulting quantities using the usual electromagnetic field
variables. We have already written the energy density
\eqref{EMHamdens} in this form. Using (\ref{fdef}), we identify the
electric and magnetic fields and
define, as for a non-dispersive medium, the magnetic induction $
\mathbf{B}(\vec{r},t)=\nabla \times \mathbf{A}(\vec{r},t)$ Thus,
using the definition $\mathsf{L} = \langle \partial_t Q, P \rangle -
\mathsf{H}$, we may express the Lagrangian density
\eqref{EMLagrandens} as
\begin{equation}\label{EMLHrelation}
    \mathsf{L}(\vec{r},t) \ = \ \frac{1}{4 \pi} \mathbf{E}(\vec{r},t) \cdot
    \mathbf{D}(\vec{r},t)  + \frac{1}{4\pi} \sum_{\mathrm{F}=\mathrm{E,H}}
    \int_{-\infty}^\infty
    |\boldsymbol{\theta}_{\mathrm{F}}(s,\vec{r},t)|^2
     \mathrm{d} s
    - \mathsf{H}(\vec{r},t) .
\end{equation}

The following theorem follows by elementary calculations
\begin{theorem}
  The energy flux vector for the extended Maxwell system
  \eqref{EMLagrandens} is
  \begin{equation}\label{Poynting}
     \mathbf{S}(\vec{r},t) \ = \ \frac{1}{4\pi}
     \mathbf{E}(\vec{r},t) \times
     \mathbf{H}(\vec{r},t) .
  \end{equation}
  If the system is homogeneous and isotropic and perturbed
  by an external current $\mathbf{j}_{\mathrm{ext}}(\vec{r},t)$ then
  the stress tensor corresponding to \eqref{symmomflux} is
  \begin{align}
    \mathsf{T}_{i,j}(\vec{r},t)\ =& \ \frac{1}{8\pi } \left\{ \mathbf{H}_{i}%
    \mathbf{B}_{j}+\mathbf{H}_{j}\mathbf{B}_{i}
    + \mathbf{A}_i \left [ \nabla \times \mathbf{H} \right ]_j
    + \mathbf{A}_j \left [ \nabla \times \mathbf{H} \right ]_i
    \right \}(\vec{r},t) \\
    & \quad + \delta_{i,j} \left \{ \mathsf{L}(\vec{r},t) - \frac{1}{4 \pi}
    \mathbf{B} \cdot \mathbf{H} (\vec{r},t) \right \}
     + \frac{1}{2} \left \{ \mathbf{A}_i
    \mathbf{j}_{\mathrm{ext};j} - \mathbf{A}_j
    \mathbf{j}_{\mathrm{ext};i} \right \}, \notag
  \end{align}
  In view of \eqref{EMLHrelation} and the equations of motion
  \eqref{EMExtmoto}, $\mathsf{T}$  may be re-expressed as
  \begin{align}\label{Maxwellstress}
\mathsf{T}_{i,j}(\vec{r},t)\ =& \ \frac{1}{8\pi } \left\{ \mathbf{H}_{i}%
    \mathbf{B}_{j}+\mathbf{H}_{j}\mathbf{B}_{i}
    -
    2\delta
_{i,j}\mathbf{H}\cdot
\mathbf{B}\right\} (\vec{r},t) \\
& +\frac{1}{8\pi } \left\{
\mathbf{D}_{i}\mathbf{E}_{j}+\mathbf{D}_{j}\mathbf{E}_{i}-2\delta
_{i,j}\mathbf{E}\cdot \mathbf{D}\right\} (\vec{r},t)  \notag \\
& +\delta _{i,j}\left\{ \mathsf{H}(\vec{r},t)-\frac{1}{4\pi }\sum_{\mathrm{F}%
=\mathrm{E,H}}\int_{-\infty }^{\infty }|\boldsymbol{\theta }_{\mathrm{F}}(s,%
\vec{r},t)|^{2}\mathrm{d}s\right\}   \notag \\
& +\mathbf{A}_{i}\mathbf{j}_{\mathrm{ext};j}(\vec{r},t)+\frac{1}{8\pi }%
\partial _{t}\left\{ \mathbf{D}_{i}\mathbf{A}_{j}+\mathbf{D}_{j}\mathbf{A}%
_{i}\right\} (\vec{r},t),  \notag
\end{align}
where the Hamiltonian $\mathsf{H}$ is given by \eqref{EMHamdens}.

  The Hamiltonian $\mathsf{H}$ and $ \int_{-\infty}^\infty
    |\boldsymbol{\theta}_{\mathrm{F}}(s,\vec{r},t)|^2 \mathrm{d} s$, $\mathrm{F}=\mathrm{E,H}$,
  may be expressed as integrals over the history of the electro-magnetic field:
  \begin{equation}
    \mathsf{H}(\vec{r},t) \ = \ - \int_{-\infty}^t \left \{  \partial_i
    \mathbf{S}_i(\vec{r},t) + \mathbf{E}(\vec{r},t) \cdot
    \mathbf{j}_{\mathrm{ext}}(\vec{r},t)
    \right \} ,
  \end{equation}
  and
  \begin{multline}
     \int_{-\infty}^\infty
    |\boldsymbol{\theta}_{\mathrm{F}}(s,\vec{r},t)|^2 \mathrm{d} s
    \\ = \ \frac{\alpha_{\mathrm{F}}(\vec{r})}{2}  \int_{-\infty}^{t} \int_{-\infty}^{t}
   \left \{
    \partial_{\tau} \chi_{\mathrm{F}} (\vec{r}, t_1 - t_2)
    + \partial_{\tau} \chi_{\mathrm{F}}(\vec{r},2t - t_1 - t_2) \right \}
   \mathbf{F}(\vec{r},t_1) \cdot \mathbf{F}(\vec{r},t_2)
  \mathrm{d} t_1 \mathrm{d} t_2,
  \end{multline}
  with $\alpha_{\mathrm{E}}(\vec{r})=\varepsilon(\vec{r}) =
  $, $\alpha_{\mathrm{H}}(\vec{r})
  =\mu(\vec{r})$,
  and $\chi_{\mathrm{F}}(\vec{r},\tau) = -\chi_{\mathrm{F}}(\vec{r},-\tau)$, $\mathrm{F}=\mathrm{E,H}$.
\end{theorem}
\noindent \emph{Remarks}: i.) $\mathbf{S}$ is the familiar Poynting
vector for the energy flux in a dielectric. ii.) The momentum
density, by \eqref{momdens}, is
\begin{multline}\label{EMmomdens}
    \mathbf{p}_i (\vec{r},t) \ = \
    - \frac{1}{4 \pi}
    \mathbf{D} \cdot \partial_{i} \mathbf{A}(\vec{r},t)
    \\ + \frac{1}{4 \pi} \sum_{\mathrm{F}=\mathrm{E,H}}
    \int_{-\infty}^\infty \boldsymbol{\theta}_{\mathrm{F}}
    \cdot
    \partial_{i} \boldsymbol{\phi}_{\mathrm{F}}(\vec{r},s,t)
    \mathrm{d} s - \partial_j
    \Phi_{i,j}(\vec{r},t),
\end{multline}
with
\begin{multline}\label{EMPhi}
    \Phi_{i,j} (\vec{r},t) \ = \
     \frac{1}{8 \pi} \left \{
    \mathbf{D}_i \mathbf{A}_j - \mathbf{D}_j \mathbf{A}_i
    \right \}(\vec{r},t) \\
     - \frac{1}{8 \pi}\sum_{\mathrm{F}=\mathrm{E,H}}
    \int_{-\infty}^\infty \left \{ \boldsymbol{\theta}_{\mathrm{F};i}
    \boldsymbol{\phi}_{\mathrm{F};j} -
    \boldsymbol{\theta}_{\mathrm{F};j}
    \boldsymbol{\phi}_{\mathrm{F};i}\right \}(\vec{r},s,t)
    \mathrm{d} s .
\end{multline}
For a homogeneous system, the conservation law
\eqref{wavemomentumconservationlaw} holds and we can express
$\mathbf{p}$ in terms of the history of the electro-magnetic field
by \eqref{momdensityhistoryintegral}, namely
\begin{equation}
    \mathbf{p}_i(\vec{r},t) \ = \ \int_{-\infty}^t \left \{ - \partial_j
    \mathsf{T}_{i,j}(\vec{r},t') + [\partial_i\mathbf{A}]\cdot
    \mathbf{j}_{\mathrm{ext}}(\vec{r},t') \right \} \mathrm{d} t' .
\end{equation}

The last term in \eqref{Maxwellstress}, $\partial _{t}\left\{ \mathbf{D}_{i}\mathbf{A}_{j}+\mathbf{D}_{j}\mathbf{A}%
_{i}\right\} (\vec{r},t)$,  is the time derivative of a symmetric
tensor.  We may drop it from the stress tensor provided we redefine
the momentum density $\mathbf{p} \mapsto \mathbf{p} + \frac{1}{8\pi}
\partial_j    \{ \mathbf{D}_i \mathbf{A}_j + \mathbf{D}_j
\mathbf{A}_i \}(\vec{r},t)$.  Thus we may equally well take the
following  for the symmetric Maxwell stress tensor in a TDD
dielectric,
\begin{align}
\mathsf{T}_{i,j}(\vec{r},t)\ &=\ \frac{1}{8\pi }\left\{ \mathbf{H}_{i}\mathbf{%
B}_{j}+\mathbf{H}_{j}\mathbf{B}_{i}+\delta _{i,j}\left(
\mathbf{H}\cdot \mu \mathbf{H}-2\mathbf{H}\cdot \mathbf{B}\right)
\right\} (\vec{r},t)
\label{Tstress1} \\
&+\frac{1}{8\pi } \left\{
\mathbf{D}_{i}\mathbf{E}_{j}+\mathbf{D}_{j}\mathbf{E}_{i}+\delta
_{i,j}\left( \mathbf{E}\cdot \varepsilon \mathbf{E}-2\mathbf{E}\cdot \mathbf{%
D}\right) \right\} (\vec{r},t)   \notag \\
&-\delta _{i,j}\frac{1}{8\pi } \sum_{\mathrm{F}=\mathrm{E,H}}
\int_{-\infty }^{\infty }\left\{ |\boldsymbol{\theta
}_{\mathrm{F}}|^{2}- |\partial _{s}\boldsymbol{\phi }_{\mathrm{F} }
|^2  \right\}
(\vec{r},s,t)  \,\mathrm{d}s  +\mathbf{A}_{i}\mathbf{j}_{\mathrm{ext};j}(%
\vec{r},t)  ,\notag
\end{align}
where we have started with (\ref{Maxwellstress}), dropped the last
term on the r.h.s.\, and
substituted the expression (\ref{EMHamdens}) for the energy density $\mathsf{H}%
(\vec{r},t)$. The corresponding momentum density, i.e.,
$\text{r.h.s.\ of (\ref{EMmomdens})} + \frac{1}{8\pi}
\partial_j    \{
     \mathbf{D}_i \mathbf{A}_j + \mathbf{D}_j \mathbf{A}_i \}(\vec{r},t)$, is
 \begin{align}\label{EMmomdensity2}
     \mathbf{p}_i (\vec{r},t)
    &= \ \frac{1}{4 \pi} \left \{ \mathbf{D} \times \mathbf{B} \right \}_i(\vec{r},t) +
    \rho_{\mathrm{ext}}(\vec{r},t)
    \mathbf{A}_i(\vec{r},t)  \\
    \notag &+ \frac{1}{8 \pi} \sum_{\mathrm{F}=\mathrm{E,H}}
    \int_{-\infty}^\infty \left [ 2 \boldsymbol{\theta}_{\mathrm{F}}
    \cdot
    \partial_{i} \boldsymbol{\phi}_{\mathrm{F}}
    + \partial_j  \left \{ \boldsymbol{\theta}_{\mathrm{F};i}
    \boldsymbol{\phi}_{\mathrm{F};j} -
    \boldsymbol{\theta}_{\mathrm{F};j}
    \boldsymbol{\phi}_{\mathrm{F};i}\right \}   \right ]  (\vec{r},s,t)
    \mathrm{d} s ,
 \end{align}
 where  we have recalled that
     $\nabla \cdot \mathbf{D} = 4 \pi \rho_{\mathrm{ext}}$ (by definition) and used the identity
 \begin{equation}
  \mathbf{D} \cdot \partial_{i} \mathbf{A} - \mathbf{D} \cdot \nabla
    \mathbf{A}_i \ = \  \left \{ \mathbf{D} \times \mathbf{B} \right \}_i .
 \end{equation}

In the above formulas for a TDD medium, $\mathbf{D}(\vec{r}%
,t)\neq \varepsilon (\vec{r},t)\mathbf{E}(\vec{r},t)$ and $\mathbf{B}(\vec{r}%
,t)\neq \mu (\vec{r},t)\mathbf{H}(\vec{r},t)$.  However, \emph{in
the non-dissipative case} when the susceptibilities in
(\ref{DDmaterialb}, \ref{DDmateriala}) vanish, $\chi
_{\mathrm{E}}=\chi _{\mathrm{H}}=0$,
the material relations reduce to $\mathbf{D}=\varepsilon \mathbf{E%
}$ and $\mathbf{B}=\mu \mathbf{H}$ and the last term in r.h.s. of
(\ref{Tstress1}) involving $\boldsymbol{\phi }_{\mathrm{F}}$
disappears. If, furthermore, there no are external charges or
currents, $\rho_{\mathrm{ext}}=0$ and
$\mathbf{j}_{\mathrm{ext};j}=0$, then the formulas (\ref{Tstress1},
\ref{EMmomdensity2}) turn into the familiar symmetric Maxwell stress
tensor \cite[\S 33]{LandauLif1},
\begin{align}
\mathsf{T}_{i,j}(\vec{r},t) \ &= \ \frac{1}{8\pi }\left\{ \mathbf{E}_{i}\mathbf{D%
}_{j}+\mathbf{E}_{j}\mathbf{D}_{i}-\delta _{i,j}\mathbf{E}\cdot \mathbf{D}%
\right\} (\vec{r},t)  \label{Tstress2} \\
& \quad + \frac{1}{8\pi }\left\{ \mathbf{H}_{i}\mathbf{B}_{j}+\mathbf{B}_{j}\mathbf{H%
}_{i}-\delta _{i,j}\left( \mathbf{H}\cdot \mathbf{B}\right) \right\} (\vec{r}%
,t),  \notag \intertext{and momentum density} \mathbf{p} (\vec{r},t)
\ &= \ \frac{1}{4 \pi} \mathbf{D} \times \mathbf{B} (\vec{r},t) \; .
\end{align}

\subsection{Brillouin formulas for the Maxwell energy density and stress
tensor}\label{sec:MaxBrillouin} As we have discussed in
\S\ref{sec:Brillouin}, one can derive rather simple formulas for the
time averaged energy density and stress tensor produced by almost
monochromatic waves. We refer to these formulas as \emph{Brillouin
formulas}, as it was Brillouin who introduced them for the TDD dielectrics, \cite[\S 80]%
{LandauLif}. In this section we present the specific form of these
formulas for the electro-magnetic field in TDD dielectric media.

\emph{We remind the reader that the formulas are derived for almost
harmonic waves as described in \S\ref{sec:Brillouin}}.  We assume
below, without comment, that we have a solution of Maxwell's
equations with all fields in the form (\ref{slowlyvaryingform})
describing a slowly modulated carrier wave of frequency $\omega$;
i.e.,
\begin{equation}
  \mathbf{E}(\vec r, t) \ = \ \mathrm{Re} \, \left \{ \mathrm e^{-
  \mathrm i \omega t} \mathbf{E}_0(\vec{r}, t) \right \} \, , \quad
  \mathbf{H}(\vec r, t) \ = \ \mathrm{Re} \, \left \{ \mathrm e^{-
  \mathrm i \omega t} \mathbf{H}_0(\vec{r}, t) \right \} \, ,
\end{equation}
and similarly for $\mathbf{D}$, $\mathbf{B}$,  where $\mathbf{E}_0$,
$\mathbf{H}_0$, $\mathbf{D}_0$ and $\mathbf{B}_0$ denote the slowly
modulated amplitude of the wave.

We start with the energy density $\mathsf{H}$.  Let us define the
time averaged energy density with ``no losses" (even if
$\operatorname{Im}\widehat{\chi }_{\mathrm{F}} \neq 0$, $\mathrm{F}=
\mathrm{E,H}$):
\begin{multline}\label{Mbril0}
\overline{\mathsf{H}_{\mathrm{NL}}}(\vec{r},t) \\ = \ \frac{1}{16
\pi} \left \{ \mathbf{E}_0^*(\vec{r},t)  \cdot \frac{d\ }{d \omega}
[\omega \varepsilon(\vec{r},\omega) ] \cdot \mathbf{E}_0(\vec{r},t)
+ \mathbf{H}_0^* (\vec{r},t) \cdot \frac{d\ }{d \omega} [ \omega
\mu(\vec{r},\omega) ] \cdot \mathbf{H}_0(\vec{r},t) \right \} ,
\end{multline}
where $\bullet^*$ denotes complex conjugation and
\begin{equation}
\varepsilon \left( \vec{r},\omega \right)  = \varepsilon (\vec{r})+\mathrm{%
Re}\widehat{\chi }_{\mathrm{E}}(\vec{r},\omega ),\quad \mu \left(
\vec{r},\omega \right) =\mu (\vec{r})+\mathrm{Re}\widehat{\chi }%
_{\mathrm{H}}(\vec{r},\omega ). \label{Mbril2}
\end{equation}
By (\ref{HBril1}), we see that $\overline{\mathsf{H}_{\mathrm{NL}}}$
is indeed the correct first order approximation to the time averaged
energy density if the medium is lossless at $\omega$:
\begin{equation}
\overline{\mathsf{H}}(\vec{r},t) \  = \
\overline{\mathsf{H}_{\mathrm{NL}}} (\vec{r},t) + \mathcal O
(\delta) \qquad (\text{no losses}) , \label{Mbril1}
\end{equation}
where $\delta \times \frac{1}{\omega}$ is the time scale over which
the slowly varying amplitudes $\mathbf{E}_0$, $\mathbf{H}_0$,
$\mathbf{D}_0$, $\mathbf{B}_0$ change noticeably, and it is assumed
that
\begin{equation}\label{EMnodiss}
\operatorname{Im}\widehat{\chi }_{\mathrm{E}}(\vec{r},\omega ) \ = \
\mathrm{Im} \widehat \chi_{\mathrm H}(\vec r, \omega) \ = \ 0 .
\end{equation}

In general the medium is absorbing at frequency $\omega$ and
(\ref{EMnodiss}) does not hold.  As we have seen in
\S\ref{sec:Brillouin}, there is in this case no simple expression
for $\overline{\mathsf{H}}$. Instead, by (\ref{extBri}) and
(\ref{hsysr1}), we have an approximation for the time averaged power
density,
\begin{align}\label{Mbrildiss}
 \overline{\partial_t \mathsf{H}}(\vec{r},t) \
= \  \frac{1}{8 \pi} \Biggl \{ & \mathbf{E}_0^*(\vec r, t) \cdot
\omega \mathrm{Im} \widehat \chi_{\mathrm{E}} (\vec r, \omega) \cdot
\mathbf{E}_0(\vec r, t) \\ \notag & +
 \mathbf{H}_0^*(\vec r, t) \cdot \omega \mathrm{Im} \widehat \chi_{\mathrm{H}} (\vec r, \omega)
\cdot \mathbf{H}_0(\vec r, t)  \\ \notag
 & +   \mathrm{Im}   \left [ \partial_t \mathbf{E}_0^*(\vec r, t)
\right )] \cdot \frac{d}{d \omega} [ \omega \mathrm{Im} \widehat
\chi_{\mathrm{E}} (\vec r, \omega)  ] \cdot \mathbf{E}_0(\vec r, t)
\\  \notag & + \mathrm{Im}  \left [ \partial_t \mathbf{H}_0^*(\vec r, t)
\right ]  \cdot
 \frac{d}{d \omega} [  \omega \mathrm{Im} \widehat \chi_{\mathrm{H}} (\vec r, \omega) ]
\cdot \mathbf{H}_0(\vec r, t)  \Biggr \}  \\  \notag + &  \partial_t
\overline{\mathsf{H}_{\mathrm{NL}}}(\vec r, t) \ + \ o(\delta).
\end{align}
The last three terms on the r.h.s.\ involve time derivatives of the
slowly varying amplitudes and are of order $\delta$. However, the
first two terms, which are non-negative and describe steady
dissipation to the medium, are of order $1$ in general. Thus (see
\eqref{dissipation}),
\begin{multline}
\overline{\partial_t \mathsf{H}}(\vec{r},t) \ = \  \  \frac{1}{8
\pi} \bigl \{ \mathbf{E}_0^*(\vec r, t) \cdot \omega \mathrm{Im}
\widehat \chi_{\mathrm{E}} (\vec r, \omega) \cdot \mathbf{E}_0(\vec
r, t)  \\ +  \mathbf{H}_0^*(\vec r, t) \cdot \omega \mathrm{Im}
\widehat \chi_{\mathrm{H}} (\vec r, \omega) \cdot \mathbf{H}_0(\vec
r, t) \bigr \} + \mathcal O(\delta).
\end{multline}

We now turn to the stress tensor $\mathsf{T}$.  The expression
\eqref{Tstress1} was derived under the assumption of isotropy and
homogeneity, so we suppose that
\begin{equation}
\varepsilon(\vec{r}) = \varepsilon , \quad \mu(\vec r) = \mu , \quad
\widehat \chi_{\mathrm{E}}(\vec r, \omega) = \widehat
\chi_{\mathrm{E}}(\omega) , \quad \widehat \chi_{\mathrm{H}}(\vec r,
\omega) = \widehat \chi_{\mathrm{H}}(\omega)
\end{equation}
are position independent scalars. As in the general case treated in
\S\ref{sec:Brillouin} the Brillouin formula for the time averaged
Maxwell stress tensor is surprisingly
 simple.  Using Prop.\ \ref{prop:lagrBri} to express the time average of the string Lagrangian, we
 see from \eqref{Tstress1} that the time averaged stress tensor (with no external current)
 is given by
\begin{multline}\label{TBril1}
\overline{\mathsf{T}_{i,j}}(\vec{r},t) \ = \ \frac{1}{16\pi } \biggl \{ \mathrm{Re}\left [  \mathbf{E}_{0;i}^*%
\mathbf{D}_{0;j}+\mathbf{E}_{0;j}^*\mathbf{D}_{0;i} +
\mathbf{H}_{0;i}^*
\mathbf{B}_{0;j}+\mathbf{H}_{0;j}^*\mathbf{B}_{0;i}\right ](\vec{r},t) \\
 \quad +   \delta_{i,j} \left [ \varepsilon(\omega) \left | \mathbf{E}_0 \right |^2 -2 \mathbf{E}_0^*\cdot
\mathbf{D}_0 + \mu(\omega) \left | \mathbf{H}_0 \right |^2 -2
\mathbf{H}_0^*\cdot \mathbf{B}_0 \right ](\vec{r},t)  \biggr \} \ +
\ o(\delta) ,
\end{multline}
where $\varepsilon(\omega)$ and $\mu(\omega)$ are the $\vec r$
independent versions of (\ref{Mbril2}), i.e.,
\begin{equation}\label{realisreal}
\varepsilon(\omega) \ = \ \varepsilon + \mathrm{Re} \widehat
\chi_{\mathrm{E}}(\omega) , \quad \mu(\omega) \ = \ \mu +
\mathrm{Re} \widehat \chi_{\mathrm{H}}(\omega) \; .
\end{equation}
To simplify \eqref{TBril1} even further, we use an approximation for
the carrier wave amplitudes $\mathbf{D}_0$ and $\mathbf{B}_0$, which
is verified using the material relations (\ref{DDmaterialb},
\ref{DDmateriala}),
\begin{align}
\mathbf{D}_0(\vec{r},t) \ & = \ \left \{ \varepsilon + \widehat \chi_{\mathrm{E}}(\omega) \right \} \mathbf{E}_0(\vec{r},t) + \mathcal O(\delta) \\
 \mathbf{B}_0(\vec{r},t) \ & = \ \left \{ \mu+ \widehat \chi_{\mathrm{H}}(\omega) \right \} \mathbf{H}_0(\vec{r},t) + \mathcal O(\delta) .
\end{align}
Thus,
 \begin{multline}\label{TBril2}
\overline{\mathsf{T}_{i,j}}(\vec{r},t) \  = \ \frac{1}{16\pi } \Biggl \{ \varepsilon(\omega) \left [ 2 \mathrm{Re} \, \mathbf{E}_{0;i}^*%
\mathbf{E}_{0;j}- \delta_{i,j} \left | \mathbf{E}_0(\vec{r},t)
\right |^2
\right ]  (\vec{r},t) \\
 + \mu(\omega) \left [ 2 \mathrm{Re} \, \mathbf{H}_{0;i}^*
\mathbf{H}_{0;j} - \delta_{i,j} \left | \mathbf{H}_0  \right |^2
\right ](\vec{r},t) \Biggr \}  + \mathcal O(\delta),
\end{multline}

Formula (\ref{TBril2}) reproduces the Pitaevskii formula, \cite{Pi}, \cite[\S 81]{LandauLif}, \cite%
[Section 3.2]{KJ} for the Maxwell stress tensor, derived in the
references under the assumption of negligible losses at the carrier
wave frequency $\omega$.  \emph{We note, however, that
\eqref{TBril2} is valid even if there are losses at $\omega$!} The
main point of (\ref{TBril2}) is that in a TDD dielectric the Maxwell
stress tensor has the same expression as in a lossless dielectric,
with material constants incorporating the \emph{real part} of the
susceptibilities computed at the carrier wave frequency. This is in
contrast to the energy (\ref{Mbril1}) and power (\ref{Mbrildiss})
densities,  which involve frequency differentiation and, in  the
lossy case, the dissipative part of the susceptibilities.

\section{Precise formulation of the construction}\label{result} We now return
to the general problem of constructing a QHE. The construction of
Section \ref{construction} is correct, but is formal in two
respects: 1) The Fourier transform of the susceptibility function
may be an operator valued measure or distribution, in which case the
point-wise limit \eqref{pwl} does not hold.   2) We have ignored
domain questions. Specifically, we have not specified the domain of
the extended impedance $\mathcal{K}$, nor have we shown that the
dynamics of the extended system exists.

Neither of these points poses a serious technical obstacle, and both
are easily dealt with by established methods.  We shall circumvent
the first issue here by restricting ourselves to $\widehat
\chi(\omega)$ defined point-wise almost everywhere. More general
susceptibilities could be handled by replacing the spaces
$L^2(\mathbb{R}, H)$ with $L^2$ spaces with respect to an operator
valued measure, via the Naimark construction \cite{Naimark} as in
\cite{FS}. The reader familiar with the general theory can easily
fill in the details.

The second point is more essential, however, and will be dealt with
carefully below. If the operator valued string coupling function
$\varsigma(s)$ is defined pointwise and is sufficiently integrable,
then $T$ is a bounded operator and this is relatively
straightforward. However, $\varsigma(s)$ may lack integrability or
may be defined as a distribution, which might result in unbounded
$T$. Thus we need to consider the definition of $T$ and
$\mathcal{K}$ more carefully.

We rely on some standard notions and results for operators on real
Hilbert spaces, summarized in Appendix \ref{Real}. We also use some
notation defined there, in particular
\begin{align}
  \mathcal L(V,H) \ &= \ \{ \text{closed densely defined operators
  from $V \rightarrow H$} \} ,\\
  \mathcal B(V,H) \ &= \ \{\text{bounded operators from $V \rightarrow H$}\},
\end{align}
with $V$ and $H$ real Hilbert spaces. We set $\mathcal L(V) =
\mathcal L(V,V)$ and $\mathcal B(V) = \mathcal B(V,V)$.

\subsection{Hamiltonian evolution}\label{sec:hamilton}
The very first thing we require is that we can solve the evolution
equations without dissipation.  This is guaranteed by the following:
\begin{quotation}
  {\bf Hamiltonian skew-adjoint condition} (HSC):
    \emph{The symplectic operator $J \in \mathcal B(V)$ and impedance
    operator $K \in \mathcal L(V, H) $ are such that $K J K^{\mathrm{T}}$,
    defined on the domain
    \begin{equation}\label{as1}
    \mathcal D (KJ K^{\mathrm{T}}) \ = \ \left \{  f \in
    \mathcal D ( K^{\mathrm{T}} ) \, : \, J K^{\mathrm{T}} f \in
    \mathcal D (K ) \right \} \; ,
    \end{equation}
    is skew-adjoint. }
\end{quotation}
\emph{Remark}: Clearly $KJK^{\mathrm{T}}$ is anti-symmetric on the
domain \eqref{as1}. To verify skew-adjointness we need to check that
the domain is dense and the operator closed.

The Hamiltonian skew-adjoint condition gives us a one-parameter
group of orthogonal transformations $\mathrm{e}^{t
KJK^{\mathrm{T}}}$ on stress space $H$. We now show how to use this
group to solve the \emph{non-dissipative} initial value problem (see
(\ref{hpq2}, \ref{hpq3})),
\begin{equation}\label{ivp}
  \partial_t u(t) \ = \ J K^{\mathrm{T}}
  K u(t) , \quad u(0) = u_0 .
\end{equation}
Because the generator $JK^{\mathrm{T}} K$ may be unbounded, we do
not try to solve \eqref{ivp} as such, but look for a \emph{finite
energy weak solution} $u(t)$. That is, we seek a map $t \mapsto
u(t)$ with $u(0)=u_0$ such that: i.) $u(t) \in \mathcal D(K)$
(finite energy), and ii.)
\begin{equation}\label{weaksolution}
  \frac{\mathrm{d}}{\mathrm{d} t} \langle  u, J u(t) \rangle  \ = \ -
   \langle  K  u, Ku(t)
  \rangle  \; ,  \quad \text{ for any } u \in  \mathcal{D} ( K
  )\; .
\end{equation}
In particular, we require the initial value to have finite energy,
$u_0 \in \mathcal D(K)$.

To solve \eqref{weaksolution}, let $f(t) = K u(t)$ be the stress as
in \eqref{hpq3} and note that
\begin{equation}\label{fevolv}
    \partial_t \langle g, f(t) \rangle \ = \ - \langle K J K^{\mathrm{T}} g, f(t)
    \rangle ,
\end{equation}
for any $g \in \mathcal D(K J K^{\mathrm{T}})$, so
\begin{equation}\label{fsolv}
     f(t) \ = \
    \mathrm{e}^{t K J K^{\mathrm{T}}} f(0) \ = \
    \mathrm{e}^{t K J K^{\mathrm{T}}} K u_0 \; .
\end{equation}
That is, the stress is propagated by the orthogonal group $\mathrm
e^{tKJK^{\mathrm{T}}}$. The solution $u$ may be obtained by
integrating \eqref{hpq1}:
\begin{equation}\label{ivpsolve}
    u(t) \ = \ u_0 + J K^{\mathrm{T}} \int_0^t f(t') \mathrm{d} t' ,
    \quad f(t) \ = \  \mathrm{e}^{tK J K^{\mathrm{T}}} K u_0 \; .
\end{equation}
The Hamiltonian skew-adjoint condition guarantees that $\int_0^t
f(t') \mathrm{d} t' \in \mathcal D(K^{\mathrm{T}})$, since
\begin{equation}\label{welldef}
   KJK^{\mathrm{T}} \int_0^t f(t') \mathrm{d} t'
   \ = \  K J K^{\mathrm{T}} \left[ \int_0^t \mathrm{e}^{t' K J K^{\mathrm{T}}}
   \mathrm{d} t' \right ]  K u_0 \ = \ (\mathrm{e}^{t K J K^{\mathrm{T}}} - 1 )
   K u_0
\end{equation}
is bounded by $2 \| K u_0 \|$ and $\mathcal D(K J K^{\mathrm{T}})
\subset \mathcal D(K^{\mathrm{T}})$.\footnote{It is key here that we
have assumed that $K J K^{\mathrm{T}}$ is closed \emph{on the domain
specified} in the HSC. If it were only closeable, we might not have
$\mathcal D (\overline{ KJ K^{\mathrm{T}}}  ) \subset \mathcal D(
K^{\mathrm{T}}).$}

\begin{theorem}[Constant energy evolution]\label{thm:dynamics}
  Assume the Hamiltonian skew-adjoint condition
  holds for the pair $J$, $K$, and let $u_0 \in \mathcal D(K)$
  be given. Then \eqref{ivpsolve} is the unique finite energy
  weak solution $u(t) \in \mathcal D ( K)$ to the initial value problem
 \eqref{ivp} and the energy
  $\mathrm{h}(u(t)) = \frac{1}{2} \|K u(t) \|^2$ is a constant of the motion, $\mathrm{h}(u(t)) = \mathrm{h}(u_0)$ for all $t$.
\end{theorem}
\begin{proof} Taking $u(t)$ as in
    \eqref{ivpsolve}, clearly $u(0) = u_0$ and
  \begin{equation}
    \frac{d \ }{d t} \left < v, J u(t) \right > \ = \ -
    \left <  K v , f(t)
    \right >
  \end{equation}
  for $v \in \mathcal D ( K)$. Since
  \begin{equation}\label{Kutada}
    K u(t) \ = \ K u_0  + \int_0^t \partial_{t'} f(t') \mathrm{d} t'
    \ = \ f(t) \; ,
  \end{equation}
  we see that $u(t)$ is a weak solution to \eqref{ivp}.

  Conservation of energy holds for any weak
  solution by \eqref{Kutada} and the orthogonality
  of $\mathrm{e}^{t KJ K^{\mathrm{T}}}$. To show uniqueness, it
  suffices to consider $u_0 = 0$, since the equations are linear.
  By conservation of energy, the solution $u(t) \in \ker K$ at all times
  $t$, and hence by \eqref{weaksolution}
  $\partial_t \langle v, u(t) \rangle  = 0$ for $v$ in a dense set.  Thus $u(t)
  = u_0 = 0$.
\end{proof}

The general solution to the driven Hamilton equation
\eqref{hamdriven} is easily obtained by superposing solutions to the
initial value problem \eqref{ivp}, noting that \eqref{ivp} is
equivalent to \eqref{hamdriven} with $\rho(t) = u_0 \delta(t)$ and
$u(t) = 0$ for $t < 0$. Thus, if we take a driving force $\rho(t)
\in \mathcal D(K)$, the formal solution to \eqref{hamdriven} is
given by
\begin{align}\label{utjsol}
  u(t;\rho) \ &= \ u_{-\infty} + \int_{-\infty}^t  \left \{ \rho(t')  +
  J K^{\mathrm{T}} f(t';\rho) \right \} \mathrm{d} t' , \\ f(t;\rho) \ &= \
  \int_{-\infty}^t \mathrm{e}^{(t-t') K J K^{\mathrm{T}}} K\rho(t') \mathrm{d} t',
  \label{ftjsol}
\end{align}
where $\lim_{t \rightarrow -\infty} u(t) = u_{-\infty} \in \ker K$.
Some assumption on $\rho$ is necessary to guarantee that
(\ref{utjsol}, \ref{ftjsol}) make sense. We shall require that the
driving force was identically zero before some initial time,
\begin{equation}\label{rhocond0}
\rho(t) \ \equiv \ 0 \quad \text{for $t < t_0$ for some $t_0 \in
\mathbb R$},
\end{equation}
and that $\|\rho(t)\|$ and $\|K \rho(t)\|$ are locally integrable
\begin{equation}\label{rhocond}
  \int_{a}^b \left \{  \left \| \rho(t) \right \|
  + \left \| K \rho(t) \right \| \right \}  \mathrm{d} t \ <
  \ \infty \quad \text{for any $-\infty < a < b < \infty$.}
\end{equation}
In fact, \eqref{rhocond0} is overly strong as one only needs
sufficient integrability at $t=-\infty$.  However, this assumption
is convenient and not really restrictive from a physical standpoint.
In any case, the r.h.s.'s of (\ref{utjsol}, \ref{ftjsol}) are well
defined and $u(t;\rho)$ is furthermore the unique weak solution to
the driven Hamiltonian equations (\ref{hamdriven}).
\begin{theorem}[Evolution under an external force]\label{thm:dcdynamics}Assume the
Hamiltonian skew-adjoint condition and suppose given an external
force $\rho(t)$ satisfying \eqref{rhocond} and initial
  state at $t=-\infty$, $u_{-\infty} \in \ker K$.
  Then the unique weak
  solution to \emph{(\ref{hamdriven})} with $\operatorname{wk-lim}_{t \rightarrow -\infty}
  u(t) = u_{-\infty}$ is given by \emph{(\ref{utjsol}, \ref{ftjsol})}.
\end{theorem}
\noindent \textit{Remark}: A weak solution to (\ref{hamdriven}) is a
function $u(t) \in \mathcal D (K) $ satisfying
\begin{equation}\label{weakdrivenham}
\frac{\mathrm{d}}{\mathrm{d} t} \langle   u, Ju(t) \rangle  \ = \ -
   \langle  K  u, Ku(t)
  \rangle  + \langle u, J\rho(t) \rangle,
  \quad \text{ for any } u \in  \mathcal{D} ( K
  )\; .
\end{equation}

When $\ker K$ is non-trivial, as for the electromagnetic field
above, we have a \emph{gauge symmetry}: if $u(t)$ solves
\eqref{weaksolution} or \eqref{weakdrivenham} then so does $u(t) +
v_0$ with $v_0 \in \ker K$ (for suitably modified initial condition
in the case of \eqref{weaksolution}). With this symmetry comes a
conserved quantity: the component of $u$ in $\ker K J = J \ker K$.
Indeed, by \eqref{utjsol},
\begin{equation}
  \langle u, u(t) \rangle \ = \ \langle u, u_{-\infty} \rangle
  + \int_{-\infty}^t \langle u, \rho(t') \rangle \mathrm{d} t',
  \quad u \in J \ker K .
\end{equation}
Thus, $Pu(t)$, with $P$ orthogonal projection onto $J \ker K$, is
constant, unless the driving force has a component in  $J \ker K$.
As we have seen, in electromagnetism, translation by an element of
$\ker K$ corresponds to a gauge transformation of the vector
potential and projection onto $J \ker K $ singles out the
electrostatic part of the electric field.

 \subsection{Two examples} In the previous section we have presented
 a general abstract approach to quadratic Hamiltonian systems.
 Before turning to TDD systems and their extensions it may be useful
 to consider a couple of familiar examples viewed from the
 perspective of Thms.\ \ref{thm:dynamics} and \ref{thm:dcdynamics}.

\subsubsection{A String}
The vibrations of a Hilbert space valued string play a key role in
the present paper, providing the dynamics of the auxiliary fields
which give rise to the dispersion in the given TDD system.  An
$H$-valued string is also a good example of a system of the type
analyzed in the previous section. The impedance operator and
symplectic operator are
 \begin{equation}\label{str1}
   K \ = \ \begin{pmatrix}
     \mathbf{1} & 0 \\
     0 & \partial_s
   \end{pmatrix} \; , \  J \ = \ \begin{pmatrix}
     0 & -\mathbf{1} \\
     \mathbf{1} & 0
   \end{pmatrix} \; , \quad \text{ on } L^2(\mathbb{R};H)\oplus
   L^2(\mathbb{R};H) \; ,
 \end{equation}
 with $\mathcal{D}(K) = L^2(\mathbb{R};H) \oplus \mathcal{D}(\partial_s)$. This
 pair satisfies the HSC, with
 \begin{equation}\label{str2}
     KJK^{\mathrm{T}} \ = \ \begin{pmatrix}
     0 & 1 \\
     1& 0
   \end{pmatrix} \partial_s \; , \quad \text{ on } \mathcal{D}(K J
   K^{\mathrm{T}}) \ = \ \mathcal{D}(\partial_s) \oplus
   \mathcal{D}(\partial_s) \; .
 \end{equation}
 The kernel of $K$ is trivial. The orthogonal group $\mathrm{e}^{t K J
 K^{\mathrm{T}}}$ can be expressed in terms of the translation group
 $\mathrm{e}^{t \partial_s} f(s)  = f(s+t)$, i.e.,
 \begin{equation}\label{str3}
   \mathrm{e}^{t K J K^{\mathrm{T}}} \ =\  \frac{1}{2} \begin{pmatrix}
     1 & 1 \\ 1 & 1
   \end{pmatrix} \mathrm{e}^{t \partial_s} \ + \
   \frac{1}{2} \begin{pmatrix}
   1 & -1 \\
   -1 & 1
   \end{pmatrix} \mathrm{e}^{-t \partial_s} .
     \end{equation}
 Thus the solution to the initial value problem \eqref{ivp} is
 \begin{multline}\label{str4}
   \begin{pmatrix}
     \theta \\
     \phi
   \end{pmatrix}(s,t) \ = \ \begin{pmatrix}
     \theta_0 \\
     \phi_0
   \end{pmatrix}(s) \\
     +
   \frac{1}{2} \int_0^t \mathrm{d} t' \left \{ \begin{pmatrix}
     \partial_s \theta_0 +
     \partial_s^2 \phi_0 \\
     \theta_0 +
     \partial_s \phi_0
   \end{pmatrix}(s+t') + \begin{pmatrix}
     - \partial_s \theta_0 +
     \partial_s^2 \phi_0 \\ \theta_0 -
     \partial_s \phi_0
   \end{pmatrix}(s-t') \right \} \; .
 \end{multline}

 Changing the $s$-derivatives into $t$-derivatives and integrating gives
 \begin{multline}\label{str6}
     \begin{pmatrix}
     \theta \\
     \phi
   \end{pmatrix} (s,t)\ = \ \frac{1}{2}\left \{ \begin{pmatrix}
     \theta_0 + \partial_s \phi_0 \\
     \theta_0
   \end{pmatrix}(s+t) + \int_0^t \mathrm{d} t'
   \begin{pmatrix}
     0 \\
     \theta_0(s+t')
   \end{pmatrix} \right \} \\
    + \frac{1}{2} \left \{ \begin{pmatrix}
     \theta_0 - \partial_s \phi_0 \\
     \phi_0
   \end{pmatrix}(s-t) + \int_0^t \mathrm{d} t'
   \begin{pmatrix}
     0 \\
     \theta_0(s-t')
   \end{pmatrix} \right \} \; ,
 \end{multline}
 with the two terms on the right hand side giving left and right traveling
 waves, respectively. 

\subsubsection{Circular String}\label{sec:circularstring}
For the electromagnetic field, the zero modes corresponding to the
kernel $\ker K$ of the impedance describe the gauge freedom of the
magnetic potential and are not directly observable.  In other
systems these modes may be observable. An example of this type is
provided  by a circular string, with
 \begin{equation}\label{str1a}
   K \ = \ \begin{pmatrix}
     1 & 0 \\
     0 & \partial_\alpha
   \end{pmatrix} \; , \  J \ = \ \begin{pmatrix}
     0 & -1 \\
     1 & 0
   \end{pmatrix} \; , \quad \text{ on } L^2(S^1;H)\oplus L^2(S^1;H) \; ,
 \end{equation}
 with $\mathcal{D}(K) = L^2(S^1;H) \oplus \mathcal{D}(\partial_\alpha)$ and
 \begin{equation}\label{str2a}
     KJK^{\mathrm{T}} \ = \ \begin{pmatrix}
     0 & 1 \\
     1 & 0
   \end{pmatrix} \partial_\alpha \; , \quad \text{ on } \mathcal{D}(K J
   K^{\mathrm{T}}) \ = \ \mathcal{D}(\partial_\alpha) \oplus
   \mathcal{D}(\partial_\alpha) \; .
 \end{equation}
 The solution \eqref{ivpsolve} in this context is identical to \eqref{str6}:
 \begin{multline}\label{str6a}
     \begin{pmatrix}
     \theta \\
     \phi
   \end{pmatrix} (\theta,t)\ = \ \frac{1}{2}\left \{ \begin{pmatrix}
     \theta_0 + \partial_\alpha \phi_0 \\
     \phi_0
   \end{pmatrix}(\alpha+t) + \int_0^t \mathrm{d} t'
   \begin{pmatrix}
     0 \\
     \theta_0(\alpha+t')
   \end{pmatrix} \right \} \\
    + \frac{1}{2} \left \{ \begin{pmatrix}
     \theta_0 - \partial_\alpha \phi_0 \\
     \phi_0
   \end{pmatrix}(\alpha-t) + \frac{1}{2} \int_0^t \mathrm{d} t'
   \begin{pmatrix}
     0 \\
     \theta_0(\alpha-t')
   \end{pmatrix} \right \} \; ,
 \end{multline}
 The kernels of $K$ and $KJ$ are the sets \begin{align}\label{str3a} \ker K \ &=
 \ \left \{
 \begin{pmatrix}
   0 \\ \phi(\alpha)
 \end{pmatrix} \ : \ \phi(\alpha) = u = \text{ constant }
 \right \} , \\ \label{str4a} \ker K J \ &= \ \left \{
 \begin{pmatrix}
   \theta(\alpha) \\ 0
 \end{pmatrix} \ : \ \theta(\alpha) = v = \text{ constant }
 \right \} \; .
 \end{align}
 In this example, $\ker K \perp \ker K J$ and it is useful to write $
 \begin{pmatrix}\theta_0 \\
 \phi_0 \end{pmatrix}$ as
 \begin{equation}\label{str5a}
     \begin{pmatrix}
       \theta_0 \\
       \phi_0
     \end{pmatrix} \ = \ \begin{pmatrix}
       v \\ u
     \end{pmatrix} + \begin{pmatrix}
       \widetilde \theta_0 \\
       \widetilde \phi_0
     \end{pmatrix}
 \end{equation}
 with $v = \frac{1}{2\pi} \int_{S^1} \theta_0(\alpha) \mathrm{d} \alpha$, $u=
 \frac{1}{2\pi} \int_{S^1} \phi_0(\alpha) \mathrm{d} \alpha$, $\widetilde
 \theta_0(\alpha) = \theta_0(\alpha) - v$, $\widetilde \phi_0(\alpha) - u$.  In these
 coordinates, the solution \eqref{str6a} becomes the sum of clockwise and
 counterclockwise traveling waves superposed on uniform translation with
 velocity $v$:
 \begin{multline}
   \begin{pmatrix}
     \theta \\
     \phi
   \end{pmatrix} (\alpha,t)\ = \ \begin{pmatrix}
     v \\
     u + t v
   \end{pmatrix}  \\ +
   \frac{1}{2}\left \{ \begin{pmatrix}
     \widetilde \theta_0 + \partial_\alpha \widetilde \phi_0 \\
     \widetilde \phi_0 + \partial_\alpha^{-1}\widetilde \phi_0
   \end{pmatrix}(\alpha+t)
    +  \begin{pmatrix}
     \widetilde \theta_0 - \partial_\alpha \widetilde  \phi_0 \\
     \widetilde  \phi_0 - \partial_\alpha^{-1}\widetilde \theta_0
   \end{pmatrix}(\alpha-t)  \right \} \; ,
 \end{multline}
 where $\partial_\alpha^{-1}$ is a \emph{bounded} left inverse to
 $\partial_\alpha$ for functions $f$ with $\int_{S^1} f(\alpha)
 \mathrm{d} \theta = 0$, e.g., $\partial_\alpha^{-1} f(\alpha) = \int_0^\alpha
 f(\alpha') \mathrm{d} \alpha'$. The zero mode $(0,u)^{\mathrm{T}}\in \ker
 K$ is the center of mass of the string and the conserved quantity
 $(v,0)^{\mathrm{T}} \in J \ker K $ is just it's velocity!

\subsection{Construction of a Hamiltonian extension}  To construct the
Hamiltonian extension we need two conditions on the susceptibility.
The first is the
\begin{quotation}{\bf Power dissipation condition} (PDC): \emph{
The susceptibility
    $\chi(t)$ is a bounded symmetric operator on $H$
    for almost every $t$,
    \begin{equation}\label{as3}
    \mathrm{e}^{-\epsilon
    t} \| \chi(t)  \| \in L^1([0, \infty)) \; , \quad \text{for every $\epsilon >
    0$,}
\end{equation}and
$\widehat \chi(\zeta)$ \textemdash \ a well defined analytic
function of $\zeta$ in the upper half plane  by \eqref{as3}
\textemdash \ satisfies the power dissipation condition
\eqref{PDC}.}
\end{quotation}

Eq.\ \eqref{as3} allows us to define the frequency domain
susceptibility $\widehat \chi(\zeta)$, at least for $\zeta$ in the
upper half plane. It is an analytic function taking values in the
bounded operators on the complex Hilbert space $\mathbb C H$. Since
$\chi(t)$ is real symmetric for a.e.\ $t$,
\begin{equation}
\widehat \chi(\zeta)^{\mathrm{T}}  \ = \ \widehat \chi(\zeta) ,
\quad \text{and} \quad \widehat \chi(\zeta)^* \ = \ \widehat
\chi(-\zeta^*) ,
\end{equation}
where $\bullet^*$ denotes complex conjugation.  Thus $\widehat
\chi(\zeta)^\dagger = \widehat \chi(-\zeta^*) = \widehat
\chi(\zeta)^*$ and
\begin{equation}\label{Udefn}
  \Phi (\zeta) \ = \ \mathrm{Im} \, \zeta \widehat \chi(\zeta) \ = \ \frac{1}{2 \mathrm i}
  \left \{ \zeta \widehat \chi(\zeta) - \zeta^* \widehat \chi(\zeta)^* \right \}
\end{equation}
is a harmonic function of $\zeta$ in the upper half plane taking
values in the bounded operators on the \emph{real} Hilbert space
$H$. By the PDC \eqref{PDC},  $\langle f, \Phi(\zeta) f \rangle$ is
a positive harmonic function for any $f \in H$. As such, a classic
theorem of Herglotz states that there is a non-negative Borel
measure $\mu_f$ on $\mathbb{R}$ such that
\begin{equation}\label{herglotz}
    \langle f, \Phi(\zeta) f \rangle \ = \
    \frac{1}{\pi}
    \int_{\mathbb{R}} \frac{\mathrm{Im} \zeta}{|\omega - \zeta|^2} \mu_f(\mathrm{d}
    \omega) \; .
\end{equation}
The measure $\mu_f$ in \eqref{herglotz} is the weak$^*$ limit as
$\eta \rightarrow 0$ of the absolutely continuous measures $\langle
f, \Phi(\omega + \mathrm{i} \eta ) f \rangle  \mathrm{d} \omega$.
(In general there would be a linear term $a_f \mathrm{Im} \zeta$ on
the r.h.s.\ of \eqref{herglotz}, however the PDC and dominated
convergence imply that $a_f =0$ since $\|\Phi(\zeta)\| =
o(\mathrm{Im} \zeta)$ as $\mathrm{Im} \zeta \rightarrow \infty$.)

The HSC and PDC together are sufficient for the existence of a
quadratic Hamiltonian extension. However, in general, this would
require us to use the ``$L^2$ space'' with respect to the operator
valued measure $\Phi(\omega + \mathrm{i} 0) \mathrm{d} \omega$ in
place of $L^2(\mathbb{R}, H)$ in the definition of the phase and
stress spaces (\ref{eq:calV},\ref{eq:calH}). The construction of
that space is not difficult (see \cite[Appendix A]{FS} and
references therein), but is abstract. The more concrete version of
the extension, using the space $L^2(\mathbb{R}, H)$, is available
when the boundary measures are absolutely continuous. For this, we
require the
\begin{quotation}{\bf Susceptibility regularity condition} (SRC):
 \emph{The measures
$\mu_f$ appearing in \eqref{herglotz} are given by
\begin{equation}\label{as5}
   \mu_f(\mathrm{d} \omega) \ = \ \langle f, \Phi(\omega+ \mathrm{i} 0) f \rangle \mathrm{d}
   \omega \; ,
    \end{equation}
    with $\Phi(\omega + \mathrm{i} 0)$ the weak operator topology limit
\begin{equation}\label{as4}
  \Phi(\omega + \mathrm{i} 0) \ = \ \mathrm{WOT-lim}_{\eta \rightarrow 0} \Phi(\omega + \mathrm{i} \eta),
\end{equation}
assumed to  exist and be bounded for almost every $\omega \in
\mathbb{R}$.}
\end{quotation}

The power dissipation and susceptibility regularity conditions are
general enough to cover many interesting examples. For example, if
$\chi$ is of the form
\begin{equation}
  \chi(t) \ = \ \chi^\infty + h(t)
\end{equation}
with $\chi^\infty$ a bounded strictly positive operator and
$\|h(t)\| \in L^1([0,\infty))$, the conditions hold provided the
continuous function
\begin{equation}
\Phi(\omega + \mathrm{i} 0) \ = \  \chi^\infty + \omega \mathrm{Im}
\widehat h(\omega)
\end{equation}
is everywhere positive semi-definite.  The susceptibility could even
grow as $t \rightarrow \infty$: the function $\chi(t) \ = \ t^\alpha
$ for $0 \le \alpha \le 1/3$ satisfies the PDC and SRC with
\begin{equation}
\widehat{\chi}(\zeta) \ = \ \mathrm{i} \Gamma(\alpha+1)
\frac{\mathrm{e}^{\mathrm{i} \frac{\pi}{2} \alpha}}{\zeta^{\alpha +
1}} ,
\end{equation}
and
\begin{equation}
  \Phi(\omega + \mathrm{i} 0) \ = \ \Gamma(\alpha +1 ) \cos(\alpha \pi /2)
  |\omega|^{-\alpha}
     \; .
\end{equation}

The main application of the SRC is to the definition of the string
coupling operator $T : L^2(\mathbb{R}, H) \rightarrow H$
corresponding to (\ref{Excoupling2}) and \eqref{fb17}. A convenient
way to organize this is as follows. By \eqref{herglotz} and
\eqref{as5},
\begin{equation}\label{SFT}
  \widehat S f (\kappa)\ := \
  \frac{1}{1 - \mathrm{i} \kappa} \sqrt{2 \Phi(\kappa + \mathrm{i} 0)} f  \; ,
\end{equation}
defines an operator from $H \rightarrow  L^2(\mathbb{R},
  \mathbb{C} H)$, which is bounded since
  \begin{equation} \|\widehat S f \|^2 \ = \  2 \int_{-\infty}^\infty
  \frac{1}{|1- \mathrm i \kappa|^2}
  \langle f, \Phi(\kappa + \mathrm{i} 0) f \rangle \mathrm d \kappa \ = \
  2 \pi \langle f, \Phi(\mathrm{i}) f \rangle.
  \end{equation}
 Because
\begin{equation}
  [\widehat S f (\kappa)]^* \ = \ \widehat S f
  (-\kappa) \; ,
\end{equation}
the inverse Fourier transform of $\widehat S f$ is \emph{real}
(i.e., $H$-valued) almost everywhere. Thus, by the Plancherel
theorem,
\begin{equation}\label{S}
  S f (s) \ := \ \mathrm{L^2-lim}_{R \rightarrow
  \infty} \frac{1}{2 \pi} \int_{-R}^{R} \mathrm{d} \kappa \,
  \mathrm{e}^{-\mathrm{i} \kappa s} \widehat S f (\kappa)
\end{equation}
defines a bounded map $H \rightarrow L^2(\mathbb{R}, H)$. We take
the coupling operator $T$ to be
\begin{equation}\label{Tw}
  T \ := \  \, S^{\mathrm{T}}
  \left (1 - \partial_s \right ) \; ,
\end{equation}
on the domain $\mathcal D ( \partial_s) \subset L^2(\mathbb{R}, H)$,
which is the set
\begin{equation}\label{Dw}
  \mathcal D ( \partial_s) \ := \
  \left \{ \phi \in L^2(\mathbb{R}, H) \, : \,
    \kappa \widehat \phi(\kappa) \in
L^2(\mathbb{R}, \mathbb{C} H) \right \} \; .
\end{equation}
The coupling $T$ may not be closed, nor even closeable, but it is
densely defined with $T (1 - \partial_s)^{-1} = S^{\mathrm{T}}$
everywhere defined and bounded.

If the inverse Fourier transform of the coupling function
\begin{equation}
  \varsigma(s) \ = \ \frac{1 }{2 \pi}
  \int_{\mathbb{R}} \mathrm{d} \kappa \,
  \mathrm{e}^{-\mathrm{i} \kappa s} \sqrt{2 \Phi(\kappa + \mathrm{i} 0)}
\end{equation}
exists in a suitable sense, for instance if $\|\sqrt{2\Phi(\kappa +
\mathrm{i} 0)}\| \in L^2(\mathbb{R})+L^1(\mathbb{R})$, then
(\ref{Excoupling2}) holds, i.e.,
\begin{equation}\label{Tdefn}
  T \phi \ = \ \int_{-\infty}^\infty \mathrm{d} s \,
  \varsigma(s) \phi(s) \; .
\end{equation}

Even if \eqref{Tdefn} is purely formal, we may define the extended
impedance
\begin{equation}\label{calK}
\mathcal{K} \ = \ \begin{pmatrix}   K & 0 & -T \\
0 & 1 & 0 \\
0 & 0 & \partial_s
\end{pmatrix} \ = \ \begin{pmatrix}   1 & 0 &  S^{\mathrm{T}} \\
0 & 1 & 0 \\
0 & 0 & 1
\end{pmatrix} \begin{pmatrix}   K & 0 & -  S^{\mathrm{T}} \\
0 & 1 & 0 \\
0 & 0 & \partial_s
\end{pmatrix}
 ,
\end{equation}
on the domain
\begin{equation}\label{th1}
    \mathcal D(\mathcal K) = \mathcal D(K) \oplus
    L^2(\mathbb{R}, H) \oplus \mathcal D(\partial_s) \; .
    \end{equation}
\begin{theorem}[Skew-adjointness of the extended Hamiltonian]\label{thm:extension}
  Assume the Hamiltonian skew-adjoint, power dissipation, and susceptibility
  regularity conditions. Then the extended impedance
  operator $\mathcal K$, defined according to
  \emph{(\ref{calK})} on the domain \eqref{th1}, and the extended symplectic operator $\mathcal J$, defined by \eqref{eq:calJ}, together satisfy the
    the Hamiltonian skew-adjoint condition.
\end{theorem}
\begin{proof}
  That $\mathcal K$ is densely defined and closed is an easy
  consequence of \eqref{calK} and the corresponding assertion for
  $K$ and $\partial_s$ (on $L^2(\mathbb{R}, H)$). So, the main point
  of the theorem is that $\mathcal{ K J K^{\mathrm{T}}}$ is skew-adjoint. To verify this,
  consider the domain of $\mathcal D(\mathcal K^{\mathrm{T}})$.
  Since
  \begin{equation}\label{KTrans}
    \mathcal K^{\mathrm{T}} \ = \ \begin{pmatrix}
       K^{\mathrm{T}} & 0 & 0 \\
       0 & 1 & 0  \\
       -S & 0 & - \partial_s
    \end{pmatrix} \begin{pmatrix}
       1 & 0 & 0 \\
       0 & 1 & 0  \\
       S & 0 & 1
    \end{pmatrix} \; ,
  \end{equation}
  we have that
  \begin{equation}\label{DKTr}
  \mathcal D( \mathcal K^{\mathrm{T}}) \ = \
  \left \{ \begin{pmatrix}
    f \\ \theta \\ \phi
  \end{pmatrix} \ : \  \begin{pmatrix}
       1 & 0 & 0 \\
       0 & 1 & 0  \\
       S & 0 & 1
    \end{pmatrix} \begin{pmatrix}
    f \\ \theta \\ \phi
  \end{pmatrix} \in \mathcal D(K^{\mathrm{T}}) \oplus
    L^2(\mathbb{R}, H) \oplus \mathcal D(\partial_s) \right \}
  \; .
  \end{equation}
  In other words $f \in \mathcal D(K^{\mathrm{T}})$ and the linear combination
  $S f + \phi \in \mathcal D(\partial_s)$.
  Thus,
  \begin{equation}\label{calKJKT}
    \mathcal{K J K^{\mathrm{T}}}
  \ = \ \begin{pmatrix}
    1 & 0 & S^{\mathrm{T}}\\
    0 & 1 & 0 \\
    0 & 0 & 1
  \end{pmatrix} \begin{pmatrix}
    K J K^{\mathrm{T}} & -S^{\mathrm{T}} & 0 \\
    S & 0 & \partial_s \\
    0 & \partial_s & 0
  \end{pmatrix}  \begin{pmatrix}
    1 & 0 & 0 \\ 0 & 1 & 0 \\
    S & 0 & 1
  \end{pmatrix},
  \end{equation}
  on the domain
  \begin{equation}\label{DKJKT}
    \mathcal D( \mathcal{K J K}^{\mathrm{T}}) \ = \ \left \{ \begin{pmatrix}
    f \\ \theta \\ \phi
  \end{pmatrix} \ : \ f \in \mathcal D(KJK^{\mathrm{T}}) , \ \theta, \
    \phi +Sf  \in \mathcal D( \partial_s) \right \}
  \; .
  \end{equation}
  The proof of skew-adjointness is now a simple exercise (see
  \cite[Prop.\ 2.2] {FS}).
\end{proof}

\subsection{Dissipative dynamics for the open subsystem}
Combining Theorems \ref{thm:dynamics}, \ref{thm:dcdynamics}, and
\ref{thm:extension}
 we see that the Hamiltonian skew-adjoint, power dissipation, and
 susceptibility regularity conditions ensure that weak solutions to the
extended system exist for suitable initial conditions or external
forces. We now consider the relationship between these solutions and
the initial dissipative system viewed as an open subsystem of the
larger Hamiltonian system.

We are interested in the relating the solution of \eqref{driven} and
the solution of
\begin{equation}
\partial_t U(t) \ = \ \mathcal J \mathcal K^{\mathrm{T}}  \mathcal KU(t) + I_V \rho(t) ,
\end{equation}
with $\rho: \mathbb R \rightarrow V$ a driving force and $I_V$ the
isometric injection of (\ref{isoinject}). Recall that
\begin{equation}\label{isointertwine}
\mathcal{J} I_V \ = \ I_V J  \; , \text{ and } \mathcal{K} I_V \ = \
I_H K \; ,
\end{equation}
with $I_H$ as in \eqref{isoinject}. 
From the form of the impedance \eqref{calK} it is clear that
\begin{equation}
\ker \mathcal K \ = \ I_V \ker K ,
\end{equation}
which is to say that the equilibrium configurations of the extended
system correspond to equilibrium configurations of the open
subsystem.

Suppose the extended system is driven out of an equilibrium state
$I_V u_{- \infty} \in \ker \mathcal K$ at $t=-\infty$, by an
external force $I_V \rho(t)$.  The resulting trajectory is, by
\eqref{utjsol},
\begin{equation}\label{Utsjsol}
U(t) \ = \ I_V u_{-\infty} + \int_{-\infty}^t \left \{ I_V \rho(t')
+ \mathcal J \mathcal K^{\mathrm{T}} F(t') \right \} \mathrm d t' ,
\end{equation}
with the extended stress
\begin{equation}
F(t) \ = \ \int_{-\infty}^t \mathrm e^{(t-t')\mathcal K \mathcal J
\mathcal K^{\mathrm{T}}} \mathcal K I_V \rho(t') \mathrm d t' \ = \
\int_{-\infty}^t \mathrm e^{(t-t')\mathcal K \mathcal J \mathcal
K^{\mathrm{T}}}  I_H K \rho(t') \mathrm d t' .
\end{equation}
Thus the evolution of the open subsystem, described by $u(t) =
I^{\mathrm{T}}_V U(t)$, is
\begin{equation}\label{reduceu}
  u(t) \ = \  u_{\infty} + \int_{-\infty}^t \left \{ \rho(t') + J
  K^{\mathrm{T}} f(t') \right \} \mathrm{d} t'\; ,
\end{equation}
where the kinematical stress $f(t)$ is
\begin{equation}\label{reducef}
    f(t) \ = \ I_H^{\mathrm{T}}F(t)\ = \ \int_{-\infty}^t
  I^{\mathrm{T}}_H \mathrm{e}^{(t-t') \mathcal{K J K^{\mathrm{T}}}} I_H K \rho(t')
  \mathrm{d} t' \; .
\end{equation}

The key point of the extension is that the pair $(u(t), f(t))$ given
by (\ref{reduceu}, \ref{reducef}) is a solution to (\ref{driven}).
More precisely, we interpret (\ref{driven}) in the weak sense,
namely
\begin{subequations}\label{drivenweak}
\begin{align}\label{driven1weak}
\partial_t \langle v, J u(t) \rangle \
&= \ -\langle K v, f(t) \rangle + \langle v, J \rho(t) \rangle , && \text{ for any } v \in  \mathcal{D} (K) ,\\
\label{driven2weak} \langle g, K u(t) \rangle \ &= \ \langle g ,
f(t) \rangle + \int_0^\infty  \langle g, \chi(\tau) f(t - \tau)
\rangle  \mathrm{d} \tau , &&\text{ for any } g \in H \; .
\end{align}
\end{subequations}
\begin{theorem}[TDD dynamics for the open subsystem]\label{thm:dissipation}
  Assume the Hamiltonian skew-adjoint, power dissipation, and susceptibility
  regularity conditions. Let $u_\infty \in \ker K$
  and an external force $t \mapsto
  \rho(t) \in V$ satisfying \eqref{rhocond0}  and \eqref{rhocond} be given. If $u(t) = I^{\mathrm{T}}_V U(t)$
  and $f(t) = I_H^{\mathrm{T}} \mathcal K U(t)$ are as in \eqref{reduceu} and \eqref{reducef}, then $(u(t) ,f(t) )$ is the unique solution
  to \emph{(\ref{driven})} with
  $\operatorname{wk-lim}_{t \rightarrow -\infty} u(t) = u_\infty$.
\end{theorem}
\begin{proof}  Let us first prove that $(u,f)$ is a solution to
(\ref{driven}). In fact \eqref{driven1weak} follows directly from
(\ref{reduceu}), so it suffices to prove \eqref{driven2weak} or,
what is equivalent
\begin{equation}\label{FLdriven2}
\langle g, \widehat {Ku}(\zeta) \rangle \ = \ \langle g, \widehat
f(\zeta) \rangle +  \langle g, \widehat \chi(\zeta) \widehat
f(\zeta) \rangle \ , \ \text{ for any } g \in H \text{ and }
\mathrm{Im} \zeta
> 0 ,
\end{equation}
where $\widehat{Ku}$ and $\widehat f$, are the Fourier-Laplace
transforms of $Ku(t)$ and $f(t)$. Note that $\widehat{Ku}$ and
$\widehat f$ are well defined since for $t < t_0$ the external force
$\rho(t)=0$, by \eqref{rhocond0}, and therefore $Ku(t) = f(t) =0$.

To simplify calculations, it is convenient to use the formalism of
\S\ref{StringFT} in which the string coordinate is
Fourier-transformed.  This amounts to replacing $\mathcal{K}$ by the
operator appearing in (\ref{ftimpedance}),
\begin{equation}\label{calKhat}
  \widehat{\mathcal{K}} \ = \
  \begin{pmatrix} K & 0 & -\frac{1}{2 \pi} \widehat T \\
  0 & \sqrt{2 \pi} & 0 \\
  0 & 0 &  \frac{1}{\sqrt{2 \pi}} \kappa
  \end{pmatrix} \; ,
\end{equation}
with $\kappa$ multiplication by the independent variable on
$L^2(\mathbb{R}, \mathbb{C} H)$ and $\widehat T$ given by
\begin{equation}\label{That}
\widehat T \widetilde \phi \ = \ \int_{-\infty}^\infty \sqrt{2
\Phi(\kappa+ \mathrm i 0)} \, \widetilde \phi(\kappa) \mathrm d
\kappa .
\end{equation}
The domain of $\mathcal K$ is
\begin{equation}
\mathcal D(\mathcal K) \ = \ \mathcal D(K) \oplus L^2(\mathrm R, H)
\oplus  \mathcal D(\kappa) ,
\end{equation}
with
\begin{equation} \mathcal D (\kappa) \ =  \ \left \{ \widetilde \phi \ :
\ \int_{-\infty}^\infty \kappa^2 \| \widetilde \phi(\kappa) \|^2
\mathrm d \kappa \ < \ \infty \right \}.
\end{equation}
Note that $\widehat T \widetilde \phi$ is well defined for any
$\widetilde \phi \in \mathcal D(\kappa)$ by \eqref{herglotz}.

Since the transformation $\mathcal K \rightarrow \widehat {\mathcal
K}$ results from  the symplectic/orthogonal isomorphism
\eqref{eq:OSFtransfrom}, the relations (\ref{reduceu},
\ref{reducef}) hold with $\mathcal{K}$ replaced by
$\widehat{\mathcal{K}}$ on the r.h.s., and a short calculation gives
\begin{align}
\label{FLf} \widehat f(\zeta) \ &= \  \ \frac{1}{ \mathrm{i} \zeta }
\left \{ - 1 + I^{\mathrm{T}}_H \mathcal{\widehat{K} J
\widehat{K}^{\mathrm{T}}} \frac{1}{\mathcal{\widehat{K} J
\widehat{K}^{\mathrm{T}}} + \mathrm{i} \zeta} I_H  \right \}  K
\widehat \rho(\zeta),
\\
\label{FLu} \widehat{K u}(\zeta) \ &= \ \frac{1}{ \mathrm{i} \zeta }
\left \{ - 1 + K J K^{\mathrm{T}} I^{\mathrm{T}}_H
\frac{1}{\mathcal{\widehat{K} J \widehat{K}^{\mathrm{T}}} +
\mathrm{i} \zeta} I_H  \right \}  K \widehat \rho(\zeta),
\end{align}
with $\widehat \rho$ the Fourier-Laplace transform of the driving
force $\rho$. Therefore
\begin{align}\label{deltaKuf} \widehat{K u}(\zeta) - \widehat f(\zeta) \ &= \
\frac{1}{ \mathrm{i} \zeta } \left [ K J K^{\mathrm{T}}
I^{\mathrm{T}}_H - I^{\mathrm{T}}_H \mathcal{\widehat{K} J
\widehat{K}^{\mathrm{T}}} \right ] \frac{1}{\mathcal{\widehat{K} J
\widehat{K}^{\mathrm{T}}} + \mathrm{i} \zeta} I_H K \widehat \rho(\zeta) \\
\notag &=  \frac{1}{\sqrt{2 \pi} \mathrm i \zeta} \widehat T
\vartheta(\zeta) \ = \ \frac{1}{\sqrt{2 \pi} \mathrm{i} \zeta}
\int_{-\infty}^{\infty} \sqrt{2\Phi(\kappa + \mathrm{i} 0)} \,
\widehat \vartheta(\zeta;\kappa) \mathrm{d} \kappa ,
\end{align}
where we have defined
\begin{equation}\label{fthetaphi}
 \begin{pmatrix}
    \widehat f(\zeta) \\
    \widehat \vartheta(\zeta ; \kappa) \\
    \widehat \varphi(\zeta; \kappa )
  \end{pmatrix} \ = \  - \frac{1}{\mathcal{\widehat{K} J
\widehat{K}^{\mathrm{T}}} + \mathrm{i} \zeta}
  I_H K \widehat \rho(\zeta),
\end{equation}
and noted that (see \eqref{eq:ftgenerator})
\begin{equation}\label{blockmat}
  K J K^{\mathrm{T}}
  I^{\mathrm{T}}_H - I^{\mathrm{T}}_H \mathcal{\widehat{K} J
\widehat{K}^{\mathrm{T}}} \ = \
    \begin{pmatrix}
      0 & -  \frac{1}{\sqrt{2\pi}} \widehat T & 0
    \end{pmatrix} .
\end{equation}

Now, \eqref{fthetaphi} implies that
\begin{equation}
  \begin{pmatrix}
  \mathrm{i} \zeta  &  - \kappa \\
   \kappa & \mathrm{i} \zeta
\end{pmatrix} \begin{pmatrix}
  \widehat \vartheta(\zeta ; \kappa) \\
    \widehat \varphi(\zeta; \kappa )
  \end{pmatrix} \ =
  \ \frac{1}{\sqrt{2 \pi}} \begin{pmatrix} \sqrt{2\Phi(\kappa + \mathrm{i} 0 )} \widehat f(\zeta)
   \\
  0 \end{pmatrix}.
\end{equation}
Thus
\begin{equation}\label{resolve}
    \begin{pmatrix}
  \widehat \vartheta(\zeta ; \kappa) \\
    \widehat \varphi(\zeta; \kappa )
  \end{pmatrix} \ = \ \frac{1}{\sqrt{2 \pi}}\frac{1}{\kappa^2 - \zeta^2} \begin{pmatrix}
    \mathrm{i} \zeta \sqrt{\Phi(\kappa + \mathrm{i} 0 )} \widehat f(\zeta) \\
    \kappa \sqrt{\Phi(\kappa + \mathrm{i} 0 )} \widehat f(\zeta)
  \end{pmatrix} .
\end{equation}
The expressions for $\widehat \vartheta$ and $\widehat \varphi$
define square integrable functions of $\kappa$, although
$\sqrt{2\Phi(\kappa + \mathrm{i} 0 )} \widehat f(\zeta)$ may not be
(see \eqref{SFT}). Furthermore, by \eqref{deltaKuf},
\begin{align}\label{resolve2}
    \widehat{K u}(\zeta) - f(\zeta)
    \ &= \ \frac{1}{\pi} \int_{-\infty}^{\infty}
    \frac{1}{\kappa^2 - \zeta^2} \,
    \Phi(\kappa + \mathrm{i} 0 ) \, \widehat f(\zeta)  \mathrm{d}
\kappa
 \\
&= \ \frac{1}{2 \pi \zeta} \int_{-\infty}^{\infty}
    \left \{ \frac{1}{\kappa- \zeta} + \frac{1}{-\kappa - \zeta} \right \}
    \Phi(\kappa + \mathrm{i} 0) \widehat f(\zeta) \,
     \mathrm{d \kappa} \notag \\
&= \  \frac{1}{\pi \zeta} \int_{-\infty}^{\infty} \frac{1}{\kappa-
\zeta} \Phi(\kappa + \mathrm{i} 0) \widehat f(\zeta) \, \mathrm{d
\kappa} , \notag
\end{align}
since $\Phi(\kappa + \mathrm{i} 0) = \Phi(-\kappa + \mathrm{i} 0)$.
Finally, \eqref{FLdriven2} follows because
\begin{equation}\label{herglotz2}
\frac{1}{\pi}\int_{-\infty}^{\infty} \frac{1}{\kappa- \zeta} \left
\langle g, \Phi(\kappa + \mathrm{i} 0)  f \right \rangle \mathrm{d}
\kappa \ = \  \zeta \left \langle g, \widehat \chi(\zeta)f \right
\rangle ,
\end{equation}
for any $f,g \in H$.

To prove \eqref{herglotz2}, it suffices, by polarization, to
consider $g=f$. Then this identity is the well know representation
of an analytic function with positive imaginary part and sufficient
decay at infinity, see for instance \cite[Section 59]{AkhGlaz} or
\cite[Section 32.3]{Lax}. To see that it holds, note that the
imaginary parts of the right and left hand sides agree by
\eqref{herglotz}. Since the two sides are analytic functions of
$\zeta$, they can differ at most by a real constant, which must be
zero as both sides vanish at $\zeta =\infty$.

It remains to prove uniqueness of the solution to (\ref{driven}). It
suffices, by linearity, to consider $u_\infty = 0$ and $\rho \equiv
0$.  Then \eqref{driven1weak} and \eqref{FLdriven2} imply that
\begin{equation}\label{uniqueeq}
  - \langle K JK^{\mathrm{T}} g , \widehat f(\zeta) \rangle \ = \
  - \mathrm{i} \zeta \langle g, \widehat f(\zeta) \rangle - \mathrm{i}
  \zeta  \langle g, \widehat \chi(\zeta) f \rangle ,
\end{equation}
for any $g \in \mathcal D(K J K^{\mathrm{T}})$. Fix $\zeta$ and let
$g=g_\lambda = \lambda^2 | \lambda + K JK^{\mathrm{T}} |^{-2}
\widehat f(\zeta)$. Then $g_\lambda \in \mathcal D(K J
K^{\mathrm{T}})$ for $\lambda < \infty$ and the l.h.s.\ of
\eqref{uniqueeq} vanishes since
\begin{multline}
\label{uniqueeq2} \langle K JK^{\mathrm{T}} g_\lambda , \widehat
f(\zeta) \rangle \\ = \ \lambda^2 \left \langle K J K^{\mathrm{T}}
\frac{1}{ | K J K^{\mathrm{T}} + \lambda |} \widehat f(\zeta),
\frac{1}{ | K J K^{\mathrm{T}} + \lambda |} \widehat f(\zeta) \right
\rangle \ = \ 0 ,
\end{multline}
since $KJK^{\mathrm{T}}$ is skew-adjoint. Thus
\begin{equation}
  0 \ = \  \zeta \langle g_\lambda, \widehat f(\zeta) \rangle +
   \langle g_\lambda, \zeta \widehat \chi(\zeta) \widehat f(\zeta) \rangle
  \,
\end{equation}
which in the limit $\lambda \rightarrow \infty$ reduces to
\begin{equation}
  \zeta \| \widehat f(\zeta ) \|^2 \ = \ -
  \langle \widehat f(\zeta) , \zeta \widehat \chi(\zeta) \widehat f(\zeta)
  \rangle \;
  .
\end{equation}
Taking imaginary parts, we see that this violates the power
dissipation condition unless $\widehat f(\zeta) = 0$.  As $\zeta$
was arbitrary $\widehat f \equiv 0$ and so $f \equiv 0$ and $u
\equiv 0$.
\end{proof}

\appendix
\part*{Appendices}
\section{Spectral theory of operators on a real Hilbert space} \label{Real}
In this section we review some known properties of linear operators
on real Hilbert spaces. Our intention here is to set notation and to
remind the reader of the differences and similarities between the
real and complex cases by recalling the main results, without
proofs. The material presented in this section is classical and well
known to experts. Nonetheless, most of the standard textbooks focus
on the complex case, making it difficult to point to a canonical
source for the real case.

Recall that a real Hilbert space $V$ is a vector space over the
field of real numbers, complete in the norm topology induced by a
symmetric scalar
product $\left \langle \cdot ,\cdot \right \rangle _{V}$. The norm of a vector $u$ is $%
\left\| u\right\|_{V}=\sqrt{\left\langle u,u\right \rangle }$. We
drop the subscript $V$ from the norm and scalar product when it is
clear from context which space we are discussing. We assume, without
comment, that every Hilbert space considered is \emph{separable},
i.e., has a countable dense subset. We denote the space of bounded
linear maps from one real Hilbert space, $V$, to another, $H$, by
$\mathcal{B}\left( V, H\right) $, and
let $\mathcal{B}\left( V\right) =\mathcal{B}\left( V,V\right) $. A \emph{%
linear operator} $K$ from $V$ to a real Hilbert space $H$
is a linear map $K:\mathcal{D}%
\left( K\right) \rightarrow H$ defined on a linear subspace $\mathcal{D}%
\left( K\right) \subset V$. The subspace $\mathcal{D}\left( K\right)
$ is the \emph{domain of $K$}, and need
not be closed. The \emph{range of $K$} is the subspace,%
\begin{equation}\label{range}
\mathcal{R}\left( K\right) := \left\{ Ku:u\in \mathcal{D}%
\left( K\right) \right\} \subseteq H,
\end{equation}%
and also need not be closed. An operator $K$ is \emph{densely
defined} if $\mathcal{D}\left( K\right) $ is dense in $V$, and
\emph{closed} if $\mathcal{D}\left( K\right) $ is a
real Hilbert space, i.e., is complete, when endowed with the scalar product%
\begin{equation}\label{cop}
\left <  u,v\right >_{\mathcal{D}\left( K\right) }=\left <
Ku,Kv\right > _{H}+\left < u,v\right > _{V}\,.
\end{equation}
An operator $K$ is \emph{closeable} if it has a closed extension,
i.e., a closed linear operator $\widetilde{K}:\mathcal{D}(
\widetilde{K}) \rightarrow H$ with $\mathcal{D} ( \widetilde{K})
\supset
\mathcal{D}\left( K\right) $ and $\widetilde{K}u=Ku$ for $u\in \mathcal{D}%
\left( K\right) $. The closure of a closeable operator $K$, denoted
$\overline{K}$, is the minimal closed extension, i.e.,
\begin{equation}\label{DKbar} \mathcal{D}(\overline{K}) = \bigcap
\left \{\mathcal{D}( \widetilde{K}) : \widetilde{K} \text{ is a
closed extension of } K  \right \}
\end{equation} and
$\overline{K} u = \widetilde{K} u$ for any closed extension
$\widetilde K$.

The \emph{transpose }of a densely defined linear operator $K$ from
$V$ to $H$ is a linear operator $K^{\operatorname{T}}$ from $H$ to
$V$, defined as follows.  Let $\mathcal{D}\left(
K^{\operatorname{T}}\right) $ be the
set of vectors $u^{\prime }\in H$ with the property that%
\begin{equation}\label{DKT}
\left| \left < u^{\prime },Ku\right> _{H}\right| \leq C_{u^{\prime
}}\left\| u\right\| _{V},\quad u\in \mathcal{D}\left( K\right) .
\end{equation}%
Since $\mathcal{D}\left( K\right) $ is dense, the Riesz lemma
\textemdash \ valid in a real Hilbert space, by the standard proof
(e.g., see \cite[Theorem II.4]{RS1})
 \textemdash \ implies that for each $u^{\prime }\in \mathcal{D}%
\left( K^{\operatorname{T}}\right) $ there is a unique vector
$K^{\operatorname{T}}u^{\prime
}$ such that%
\begin{equation}\label{DefKT}
\left < u^{\prime },Ku\right > _{H^{\prime }}=\left <
K^{\operatorname{T}}u^{\prime },u\right > ,\quad u\in
\mathcal{D}\left( K\right) .
\end{equation}%
As it stands, $\mathcal{D}\left( K^{\operatorname{T}}\right) $ might
not be dense; indeed it might contain only the zero vector. However,
as in the complex case, if $K$ is closed, then $\mathcal{D}\left(
K^{\operatorname{T}}\right) $ is dense and
\begin{lemma}
A densely defined operator $K$ from $V$ to $H$, with $V$, $%
H$ real Hilbert spaces, is closeable if and only if
$K^{\operatorname{T}}$ is densely defined, in which case
$K^{\operatorname{T}} = \overline{K}^{\operatorname{T}}$ and
$\overline{K} = \left (K^{\operatorname{T}} \right )^{\mathrm{T}}$.
\end{lemma}

We denote the collection of closed densely defined operators from
$V$ to $H$ by $\mathcal{L}\left( V,H\right) $. Thus
$\cdot^{\operatorname{T}}:\mathcal{L}\left( V,H\right) \rightarrow
\mathcal{L}\left(
H,V\right) $ and is an involution on $\mathcal{L}\left( V\right) =%
\mathcal{L}\left( V,V\right) $. Any bounded operator is closed and
densely defined, i.e., $\mathcal{B}\left(V,H\right) \subset
\mathcal{L}\left( V,H\right) $.

A complex Hilbert space $V_{c}$ is also a real Hilbert space,
denoted here for emphasis by $ V_{c}^{\mathbb{R}}$, under the inner
product
\begin{equation}\label{reip}
\left < u,v\right >_{V_{c}^{\mathbb{R}}}=\operatorname{Re}\left
(u,v\right )_{V_{c}},
\end{equation}%
where $\left ( \cdot ,\cdot \right )_{V_{c}}$ is the complex scalar
product on $V_{c}$. Note that $\|\cdot\|_{V_{c}^{\mathbb{R}}}= \|
\cdot \|_{V_{c}}$, so these two spaces are identical as metric
spaces. We denote the space of bounded, respectively closed densely
defined, complex linear operators from one complex Hilbert
space, $V_{c}$, to another, $H_{c}$, by $\mathcal{B}_{\mathbb{C}%
}\left( V_{c},H_{c}\right) $, respectively $\mathcal{L}_{\mathbb{C}%
}\left( V_{c},H_{c}\right) $.

A complex linear operator $K \in
\mathcal{L}_{\mathbb{C}}(V_{c},H_{c})$,
considered as a map from $V_{c}^{\mathbb{R}}$ to $%
H_{c}^{\mathbb{R}}$, is obviously a real linear operator. However,
\emph{not every real linear operator is complex linear.} Indeed, a
complex Hilbert space has a canonical symplectic operator
$Ju=\mathrm{i}u$ (in the sense of \eqref{symplectic}), such that a
real linear operator $K$ is complex linear if and only if it
commutes with $J$. Conversely, given a real Hilbert space $V$ and a
symplectic operator $J\in \mathcal{B}\left( V\right) $, we may
define a complex linear structure on $V$
\begin{equation}\label{clstruct}
(a+\mathrm{i}b)\cdot u=au+bJu,
\end{equation}
and a complex inner product
\begin{equation}\label{clip}
\left ( u,v\right ) _{V_{J}}:=\left < u,v\right > _{V}-\mathrm{i}%
\left < u,Jv\right > _{V},
\end{equation}
so that $V$ becomes a complex Hilbert space, which we denote by
$V_{J}$ for emphasis.

The \emph{complexification} of a real Hilbert space $V$, denoted
$\mathbb{C}V$, is the complex Hilbert space, equal as a set to
$V\oplus V$, with multiplication by $\mathrm{i}$ given by
 \begin{equation}\label{jj2a}
    J \ = \ \begin{pmatrix} 0 & - \mathbf{1} \\
    \mathbf{1} & 0 \end{pmatrix} .
\end{equation}
An element $u_1 \oplus u_2$, $u_{j}\in V$, is denoted
$u_{1}+\mathrm{i}u_{2}$, and we have
\begin{align}\label{complexip}
\left ( u_{1}+\mathrm{i}u_{2},v_{1}+\mathrm{i}v_{2}\right )_{%
\mathbb{C}V} &:=\left < u_{1},v_{1}\right >_V-\mathrm{i}\left<
u_{2},v_{1}\right
>_V +\mathrm{i}\left < u_{1},v_{2}\right >_V +\left <
u_{2},v_{2}\right >_V  \\
\left( a+\mathrm{i}b\right) \cdot \left(
u_{1}+\mathrm{i}u_{2}\right) &:=au_{1}-bu_{2}+\mathrm{i}\left(
bu_{1}+au_{2}\right) . \label{complexmult}
\end{align}%
There is a natural operator of complex conjugation on the
complexified Hilbert space $\mathbb C V$ given by the block matrix
 \begin{equation}\label{jj2b}
    C \ = \ \begin{pmatrix}
      \mathbf{1} & 0 \\
      0 & -\mathbf{1}
    \end{pmatrix} .
\end{equation}
The operator $C$ is real linear, and, since $C J = - J C$, is
complex antilinear: $C z u = z^* C u$. We also use the notation $C u
= u^*$, so
\begin{equation}
(u_1 + \mathrm i u_2 )^* \ = \ u_1 - \mathrm i u_2, \quad u_1, u_2
\in V.
\end{equation}

Any real linear operator $A\in \mathcal{L}\left( V,H\right) $ has a
natural extension $ A_{\mathbb{C}} \in \mathcal{L}_{\mathbb{C}}
\left( \mathbb{C}V,\mathbb{C}%
H\right) $, called the complexification of $A$, namely
\begin{equation}\label{complexA}
\mathcal{D}\left( A_{\mathbb{C}}\right) =\left\{ u_{1}+\mathrm{i}%
u_{2}:u_{j}\in \mathcal{D}\left( A\right) \right\} ,\quad A_{\mathbb{C}%
}\left( u_{1}+\mathrm{i}u_{2}\right)
=Au_{1}+\mathrm{i}Au_{2}\text{.}
\end{equation}
Not only is the extension $A_{\mathbb C}$ complex linear, it also
commutes with conjugation $A_{\mathbb C} C = C A_{\mathbb C}$.
Conversely any complex linear operator $A_{\mathbb C} \in \mathcal
L_{\mathbb C}(\mathbb C V, \mathbb C H)$ which commutes with
conjugation has the block matrix form
\begin{equation}
A_{\mathbb C} \ = \ \begin{pmatrix}
A & 0 \\
0 & A
                    \end{pmatrix}
\end{equation}
for a suitable real linear operator $A \in \mathcal L(V,H)$. Thus
the category of real Hilbert spaces is equivalent to the category of
complex Hilbert spaces furnished with distinguished conjugation
operators \textemdash \ self-adjoint complex anti-linear operators
with $C^2 =1$. From this viewpoint, real linear operators are
complex linear operators which commute with conjugation.  This leads
to additional structure in the spectral theory, as we will see.

We show below, using the spectral theorem, that any symplectic operator
$J$ on $V$, has the canonical representation \eqref{jj2a} by a block
matrix for a suitable choice of real orthonormal basis on $V$.
Associated to \eqref{jj2a} is a real orthogonal decomposition
\begin{equation}
V=V_{1}\oplus V_{2}\text{ with }\dim V_{1}=\dim V_{2}, \label{jj2aa}
\end{equation}
Given such a decomposition, we may interpret $V_1$, $V_2$ as
``real'' and ``imaginary'' subspaces of $V_J$ respectively, with
complex conjugation given by the block matrix \eqref{jj2b}. (In the
context of Hamiltonian systems $V_{1}$ and $V_{2}$ can be
interpreted as the spaces of coordinate and momenta.) Note that the
spaces in the decomposition \eqref{jj2aa} are not canonical: we
cannot determine the real and imaginary subspaces uniquely from a
complex structure.

To develop a spectral theory for a real linear operator, it is
generally necessary to work with its complexification. Indeed a real
linear operator $A\in \mathcal{L}\left( V\right) $ may have empty
real spectrum. Thus we take the spectrum of $A$ to be the spectrum
of $A_{\mathbb{C}}$, denoted $\sigma \left( A\right) =\sigma \left(
A_{\mathbb{C}}\right) $, which is a non-empty closed subset of
$\mathbb{C}$
for any $A \in \mathcal L (V)$. (Recall that $\sigma \left( A_{\mathbb{C}%
}\right) = \{ \lambda \in \mathbb{C} \ \mid \ A_{\mathbb{C}}-\lambda
\text{ is not boundedly invertible }\}$.)

A curious phenomenon arises regarding the spectrum of a complex
linear operator $A\in \mathcal{L}_{\mathbb{C}%
}\left( V_c\right) $. On the one hand we can define the spectrum as
usual
\begin{equation}\label{complexspect}
\sigma \left( A\right) =\sigma _{\mathcal{B}\left( V_c\right)
}\left( A\right) =\left\{ \lambda :A-\lambda \text{ is not
invertible in }\mathcal{B}\left( V_c\right) \right\} .
\end{equation}
On the other hand, we could forget that $A$ is complex linear and
consider $V_c$ as the real Hilbert space, $V_c^{\mathbb{R}}$, which
we may then complexify to $\mathbb{C} V_c^{\mathbb{R}}$, the set
$V_c \times V_c$ with complex multiplication and inner product given
by (\ref{complexip}, \ref{complexmult}). Then the extension
$A_{\mathbb{C}}$ of $A$ has spectrum
\begin{equation}\label{realcomplexspect}
\sigma \left( A_{\mathbb{C}}\right) =\sigma _{\mathcal{B}\left(
\mathbb{C}V_c^{
\mathbb{R}}\right) }\left( A_{\mathbb{C}}\right) =\left\{ \lambda :A_{%
\mathbb{C}}-\lambda \text{ is not invertible in }\mathcal{B}\left( \mathbb{C}%
V_c^{\mathbb{R}}\right) \right\} .
\end{equation}%
The two sets (\ref{complexspect}, \ref{realcomplexspect}) are not
equal in general! Indeed, this is the case even for the map $J=$
multiplication by $\mathrm{i}$, since $\sigma_{\mathcal{B}\left( V_c
\right) }\left( J\right) =\left\{ \mathrm{i}\right\} $ while $\sigma
\left( J_{ \mathbb{C}}\right) =\left\{ \pm \mathrm{i}\right\} $.
However, this example already indicates the only difference that can
occur: in general $\sigma \left( A_{\mathbb{C}}\right) =\sigma
\left( A\right) \cup \sigma \left( A\right) ^{\ast }$.  Indeed,

\begin{lemma}[Spectral Symmetry]\label{spectsym}
Let $A\in \mathcal{L}\left( V\right) $ with $V$ a real Hilbert
space. Then the spectrum of $A$, that is $\sigma \left(
A_{\mathbb{C}}\right) $, is a nonempty closed subset of the complex
plane invariant under complex conjugation.
\end{lemma}

Using the complexification, one can carry over to real linear
operators various results from the spectral theory of complex linear
operators. To state these results, we use the terminology that an
\emph{isometry} is a linear map $T\in \mathcal{B}\left( V,H \right)
$ with $\left\| Tu\right\| _{H}=\left\|
u\right\| _{V}$ for all $u\in V$, and an \emph{orthogonal map} is an isometry $%
T$ with $\mathcal{R}\left( T\right) =H$. (We reserve the term
\emph{unitary }for isomorphisms of complex Hilbert spaces.) An
orthogonal map $T$ is invertible with $T^{-1}=T^{\operatorname{T}}$.
A \emph{partial isometry} is a linear map $T\in \mathcal{B}\left(
V,H\right) $ which is an isometry when restricted to $\left( \ker
T\right) ^{\perp }$, where
\begin{gather}\label{ker}
\ker A =\left\{ u\in \mathcal{D}\left( A\right) :Au=0\right\} ,
\intertext{for any linear operator $A$, and, for any $S \subset V$,}
S^{\perp } =\left\{ u:\left < u,v\right> _{V}=0\text{ for all }%
v\in S\right\} .
\end{gather}

In particular there is a spectral theorem for normal operators.
Recall that an operator $K \in \mathcal L(V)$ is called
\emph{normal} if $K^{\mathrm{T}} K = K K^{\mathrm{T}}$, i.e.,
$\mathcal D (K^{\mathrm{T}}K) = \mathcal D(K K^{\mathrm{T}})$ and
$K^{\mathrm{T}}K u = KK^{\mathrm{T}}u$ for any $u$ in the common
domain, and is called \emph{self-adjoint} if $K^{\operatorname{T}}=K$, i.e., $%
\mathcal{D}\left( K\right) =\mathcal{D}\left(
K^{\operatorname{T}}\right) $ and $ Ku=K^{\operatorname{T}}u$ for
all $u\in \mathcal{D}\left( K\right) $. Clearly a self-adjoint
operator is also normal.

\begin{theorem}[Spectral Theorem for Normal Operators]\label{thm:spectral}
Let $K \in \mathcal L(V)$ be a normal operator on a real Hilbert
space $V$.  Then there are i.) a $\sigma$-finite measure space $X,
\mu $, ii.) a $\mu $-measure preserving involution $\Phi : X
\rightarrow X$, i.e. $\Phi\circ \Phi(x) = x$, iii.) a
$\mu$-measurable \emph{complex} valued function $k:X\rightarrow
\mathbb{C}$, with
\begin{equation}\label{spectsymeq}
  k\circ \Phi(x) \ = \ k(x)^* \; , \ \mu \text{ a.e. } x \; ,
\end{equation}
and iv.) an isometry $T:H\rightarrow L^{2}\left( X,d\mu ; \mathbb{C}
\right) $, the complex Hilbert space of square integrable complex
valued functions on $X$, such that
\begin{equation}\label{normspth}
    \left[ TKu \right] (x)=k(x)\left[ Tu\right] (x), \ \mu \text{ a.e. } x \in X,
     \text{ for any } u\in \mathcal{D}\left( K\right) \; ,
\end{equation}
and
\begin{equation}\label{ranT}
\mathrm{ran} T \ = \ \left \{ f \in L^{2}\left( X,d\mu ; \mathbb{C}
\right)
 \ : \ f\circ \Phi(x) = f(x)^*, \ \mu \text{ a.e. } x \right \} \; .
\end{equation}

Furthermore, $\{ x : \Phi(x) = x \}$ and $\{ x: k(x)$ is real$\}$
differ only by a set of $\mu$ measure zero. In particular, $K$ is
self-adjoint if and only if we may take $\Phi$ to be the identity
and $k$ to be real valued.
\end{theorem}

Thus the spectral theorem for self adjoint real operators is
essentially the same as in the complex case. However, for a
non-self-adjoint operator $K$, the involution $\Phi$ is non-trivial
and represents additional structure compared to the complex case.
This is related to the spectral symmetry described in Lemma
\ref{spectsym}, a strong version of which is  eq.\
\eqref{spectsymeq}. Associated to this symmetry is a natural
\emph{partial symplectic operator} $J$ commuting with $K$, defined as
follows. Consider the function $\sigma(x) = \pm 1$ if $\pm
\mathrm{Im} k(x)
> 0$ and $0$ if $\mathrm{Im} k(x) =0$, and define a map $J \in \mathcal B(V)$
by
\begin{equation}\label{JPhi}
    [T J v](x) \ := \ \mathrm{i} \sigma(x) [Tv](x) \; .
\end{equation}
Because $\mathrm{i} \sigma(\Phi(x)) = - \mathrm{i} \sigma(x)$, $J$
is well defined. It is easy to see that $V_0 = \ker J$ is the
largest invariant subspace for $K$ such that the restriction of $K$
to this subspace is self adjoint. Thus $V_0^\perp=\ker J^\perp$ is
the largest invariant subspace on which $K$ is ``completely non
self-adjoint.'' Furthermore, the restriction of $J$ to $V_0^\perp$
is a symplectic operator. That is $J$ is a \emph{partial symplectic
operator}, which is a bounded operator $J$ such that $J \text{ is a
partial isometry and } J^{\mathrm{T}} = - J \; .$

A special case of a normal operator is  a symplectic operator $J$
with $J^2 =-1$ and $J^{\mathrm{T}} = - J$. The spectral
representation Theorem \ref{thm:spectral} allows us to prove easily
the canonical form \eqref{jj2a}. Indeed, a symplectic operator being
skew-adjoint has a spectral representation $T Jv(x) = j(x) T v(x)$,
with $T:V\rightarrow L^2(X,\mu, \mathbb{C})$ an isometry. Since $J^2
= -1$, the function $j$ takes values $\pm \mathrm{i}$. By
\eqref{ranT} the following subspaces define an orthogonal
decomposition of $V$:
\begin{align}
  V_1 \ &= \ \left \{ v \in V \ : \ Tv(x) \text{ is real for $\mu$ a.e. $x$} \right \} \\
  V_2 \ &= \ \left \{ v \in V \ : \ Tv(x) \text{ is imaginary for $\mu$ a.e. $x$} \right \} .
\end{align}
Clearly $J : V_1 \rightarrow V_2$ and $J: V_2 \rightarrow V_1$.
Furthermore, given a basis $\{ u_1, u_2, \ldots \}$ for $V_1$, the
sequence $\{J u_1, J u_2, \ldots \}$ is a basis for $V_2$, and with
respect to these bases the representation \eqref{jj2a} holds. We
have shown
\begin{lemma}
\label{lJ2}Let $J$ be a symplectic operator in a real Hilbert space
$V$. Then there is a real orthogonal decomposition \eqref{jj2aa} in
which the matrix of $J$ has the canonical form \eqref{jj2a}.
\end{lemma}

The functional calculus for a self adjoint operator associates to
any real valued Borel measurable function $f:\sigma(K) \rightarrow
\mathbb{R}$ a self adjoint operator $f\left( K\right) $ with
\begin{equation}\label{funccalc}
    \left[ Tf(K)u\right] (x) \ = \ f\left(
k\left( x\right) \right) \left[ Tu\right] (x) \; ,
\end{equation}
where the domain $\mathcal{D}(f(K))$ is the set of $u$ such that the
r.h.s. is square integrable. For a normal operator $K$,
\eqref{funccalc} also defines a functional calculus, with $f(K)$ a
real operator if the map $f:\sigma(K) \rightarrow \mathbb C$
satisfies $f(z^*) = f(z)^*$.

The functional calculus allows us to define the square root and
therefore the absolute value and polar decomposition, all of which
work essentially as in the complex case.  An operator $K$ is called
\emph{positive} if $K$ is self adjoint \underline{and}%
\begin{equation}\label{pos}
\left < u,Ku \right >_{V}\geq 0,\quad \text{for all }u\in
\mathcal{D}\left( K\right) .
\end{equation}%
(For an operator in a real Hilbert space self-adjointness does not
follow from \eqref{pos}, so both conditions are necessary.) Given
any $K \in \mathcal L(V, H)$  the operator $K^{\operatorname{T}}K$
on $V$
with domain $\mathcal{D}\left( K^{\operatorname{T}}K\right) =\left\{ u\in \mathcal{D}%
\left( K\right) :Tu\in \mathcal{D}\left( K^{\operatorname{T}}\right)
\right\} $ is closed, densely defined and positive. (This is proved
by the standard  argument to obtain a self-adjoint operator from a
semi-bounded quadratic form (see \cite[Theorem VIII.15]{RS1}). A
close reading of the proof shows that it works in a real Hilbert
space.) Using the functional calculus we define
\begin{equation}\label{absK}
\left| K\right| =\sqrt{K^{\operatorname{T}}K},\quad \text{with
}\left| K\right| \in
\mathcal{L}\left( V \right) ,\quad \mathcal{D}\left( \left| K\right| \right) =%
\mathcal{D}\left( K\right) .
\end{equation}%
The \emph{polar decomposition} of $K$ is the factorization%
\begin{equation}\label{poldec}
K=T\left| K\right| ,\quad \text{with }T:V\rightarrow H\text{ a
partial isometry,}
\end{equation}%
where $T$ is uniquely defined if we require $\ker T=\ker K$, in
which case
\begin{equation}\label{partiso}
T u = \lim_{\epsilon \downarrow 0} K\left( \left| K\right| +
\epsilon\right)^{-1} u \; .
\end{equation}

From the standpoint of the present work, the most important normal
operators are the \emph{skew-adjoint} operators. We call an operator
$K \in \mathcal L (V)$ \emph{skew-adjoint} if
$K=-K^{\operatorname{T}}$. This implies that the function $k$ given
by the spectral theorem is \emph{purely imaginary}, $k(x) =
\mathrm{i} \sigma(x) h(x)$, with $\sigma(x)=\pm 1, 0$  and $h(x) \ge
0$.   From the above discussion we see that the polar decomposition
for a skew-adjoint operator is
\begin{equation}
  K \ = \ J |K|
\end{equation}
with $J$ a partial symplectic operator commuting with $K$. Using $J$
to put a complex structure on $\mathrm{ran} J$, as in
(\ref{clstruct}, \ref{clip}), we see that
\begin{theorem}[Canonical decomposition for a skew-adjoint operator]\label{skadstructhm}
  Let $K$ $ \in$ $ \mathcal{L}(V)$ be skew-adjoint. Then there is a direct sum
  decomposition $V = V_0 \oplus V_0^\perp$ with a
  complex structure on $V_0^\perp$ (compatible with its real structure)
  such that $K = 0 \oplus \mathrm{i} A$, with $A$ a positive complex linear
  operator.
\end{theorem}
The skew-adjoint operators are the generators of strongly continuous
one parameter groups of orthogonal operators, by the following
analogue of the Stone-von Neumann theorem:
\begin{theorem}[Stone-von Neumann Theorem]
  \label{StoneVN} A map $t \in \mathbb{R} \rightarrow O_t \in \mathcal{B}(V)$,
  with $V$ a real Hilbert space, is a strongly continuous, one parameter group
  of orthogonal operators if and only if
  \begin{equation}
    O_t \ = \ \mathrm{e}^{t A}
  \end{equation}
  for a unique skew-adjoint operator $A \in \mathcal L(V)$.
\end{theorem}
\noindent Recall that $O_t$ is \emph{strongly continuous} if $t
\mapsto O_t v$ is a (norm) continuous map for any $v \in V$.

Combining Theorems \ref{skadstructhm} and \ref{StoneVN} we see that
any strongly continuous one parameter orthogonal group $O_t$
decomposes as $O_t = 1 \oplus U_t$ with $U_t$ a one parameter
\emph{unitary} group on a subspace.

\section{Sketch of the proof of Theorem \ref{FSthm}.}\label{Sketch}
In a remark following Theorem \ref{FSthm} above, we briefly sketched
a proof based on results from our previous work \cite{FS}.  To keep
the present work self contained, in this section we give a more
detailed sketch.

To begin, note that it suffices to show
\begin{lemma}\label{lem:frictfuncrep}
The whole line extension \eqref{ava2} of the friction function
$a_{e}(t)$ can be written in the form
\begin{equation}\label{frictfuncrep}
a_{ e}(t) \ = \ \Gamma \mathrm e^{t L_1} \Gamma^{\mathrm T},
\end{equation}
with $L_1$ a skew adjoint operator on an auxilliary Hilbert space
$H_1$ and $\Gamma : H_1 \rightarrow H$ a linear map.  Furthermore,
the minimal representation \eqref{frictfuncrep} is unique up to
isomorphism; i.e., if no proper subspace of $H_1$ containing the
range of $\Gamma^{\mathrm T}$ is invariant under $L_1$, then $L_1$
and $\Gamma$ are unique up to transformation by an orthogonal map.
 \end{lemma}

Given the unique minimal representation \eqref{frictfuncrep}, we
define  $\mathcal L$ on $H \oplus H_1$ by
\begin{equation}
\mathcal L \ = \ \begin{pmatrix}
L & \Gamma \\
- \Gamma^{\mathrm T} & L_1
                 \end{pmatrix}.
\end{equation}
One may verify that this operator gives the desired extension, i.e.,
that \eqref{solvevA1} holds.  The uniqueness of the minimal
extension follows from uniqueness in Lemma \ref{lem:frictfuncrep}.

Strictly speaking the above argument works only when the $\delta$
function contribution $2 \alpha_{\infty} \delta(t)$ to $a_{e}(t)$
vanishes, i.e., when $\chi(0) = 0$.  If this term is non-zero,
\eqref{frictfuncrep} holds with unbounded $\Gamma$, and some care
must be taken in interpreting this relation  (and also in the
definition of $\mathcal L$ as in Thm.\ \ref{thm:extension}).  It
turns out that the required $\Gamma$  is \emph{$L_1$ bounded}, i.e.,
$\Gamma : \mathcal D(L_1) \rightarrow H$ and
  \begin{equation}\label{L1bounded}
  \left \| \Gamma f \right \|_H \ \le \ a \left \| L_1 f \right \|_{H_1} + b \left \| f \right \|_{H_1}
  \end{equation}
  for suitable finite $a$, $b$ $>0$.  In place of \eqref{frictfuncrep}, we have a
  distributional limit:
  \begin{multline}\label{realfrictfuncrep}
  \int_{-\infty}^\infty \left \langle g(t), a_{\mathrm e}(t) f(t) \right \rangle \mathrm d t
  \\ =  \ \lim_{R \rightarrow \infty} \int_{-\infty}^\infty \left \langle g(t) ,
  \Gamma \mathrm e^{ tL_1} \left ( \Gamma \Phi_{R}^2 \right )^{\mathrm T} f(t) \right \rangle
  \mathrm d t , \quad \Phi_R^2 = \left ( \frac{L_1^2}{R^2} + \mathbf 1 \right )^{-1}
  \end{multline}
  for all sufficiently smooth compactly supported $H$-valued maps $f$ and $g$. The representation
  \eqref{realfrictfuncrep} and the corresponding uniqueness of the minimal representation are one direction of the following result related to an operator valued generalization of
  Bochner's Theorem \cite[Theorem 7.1]{FS}:
\begin{theorem}
Let  $H$ be a real Hilbert space and let $a_{e}(t) = 2 \alpha_\infty
\delta(t) + \alpha_{ e}(t)$, $- \infty < t < \infty$, with $\alpha_{
e}(t)$ a strongly continuous $\mathcal B(H)$-valued function and
$\alpha_{\infty}$ a non-negative bounded  operator on $H$. Then
$a_e(t)$ is representable as \eqref{realfrictfuncrep} with $L_1$ a
skew-adjoint operator on $H_1$ and $\Gamma$ an $L_1$ bounded map if
an only if $a_{ e}(t)$ satisfies the power dissipation condition
\eqref{ava4} for every continuous $H$-valued function $f(t)$ with
compact support. The operator $\Gamma$ is bounded if and only if
$\alpha_{\infty}=0$, in which case \eqref{frictfuncrep} holds.

If the space $H$ is minimal \textemdash \ in the sense that
\begin{equation}\label{minimality}
\left \{ \left [ \Gamma f(L_1)  \right ]^{\mathrm T} v \ : \  f \in
C_c(\mathbb C) \text{ with } f(z^*) = f(z)^* \text{ and } v \in H
\right \}
\end{equation}
is dense in $H_1$ \textemdash \ then $\{ H_1, L_1, \Gamma \}$ is
uniquely determined up to orthogonal transformation.
\end{theorem}
\begin{proof}[Sketch of the proof]  The proof follows quite closely the proof of
\cite[Theorem 7.1]{FS}, which deals with the complex case. We define
the Hilbert space $H_1$ to be a space of $H$ value functions with
the inner product given by
\begin{equation}\label{H1ip}
\left \langle \phi, \psi \right \rangle_{H_1} \ = \
\int_{-\infty}^\infty \int_{-\infty}^\infty \left \langle \phi(t),
a_e (t - s) \psi(s) \right \rangle_{H} \mathrm dt \mathrm d s .
\end{equation}
The r.h.s.\ of \eqref{H1ip} is non-negative by the power dissipation
condition \eqref{ava4}, but stritcly speaking only defines a
semi-inner product.  Thus we must mod out by the space of null
vectors $\psi$ with $\langle \psi, \psi \rangle_{H_1} = 0$.

The operator $L_1$ is then defined using the Stone-von Neumann
Theorem \ref{StoneVN} as the generator of the one parameter
orthogonal group of translations
\begin{equation}
\mathrm e^{s L_1}\phi(t) \ = \ \phi(t-s) .
\end{equation}
Taking
\begin{equation}
\Gamma \phi \ = \ \int_{-\infty}^\infty a_e(-t) \phi(t) \mathrm d t
,
\end{equation}
it is not difficult to show that (\ref{realfrictfuncrep}) holds.
Note that, formally,
\begin{equation}
\Gamma^{\mathrm T} f \ = \ \delta(t) f.
\end{equation}

Uniqueness is shown as follows.  Suppose we have a minimal
representation such that \eqref{minimality} is dense in $H_1$, as
can be obtained by restricting the constructed $\Gamma$ and $L_1$ to
the closure of \eqref{minimality}. Given another representation
\begin{equation}
a_{e}(t) \ = \ \widetilde \Gamma \mathrm e^{t \widetilde L_1}
\widetilde \Gamma^{\mathrm T} .
\end{equation}
with $\widetilde L_1$ and $\widetilde \Gamma_1$ defined on
$\widetilde H_1$, one defines
\begin{equation}
T \left [\Gamma f(L_1) \right ]^{\mathrm T} f \ = \ \left [
\widetilde \Gamma f(\widetilde L_1) \right ]^{\mathrm T} f ,
\end{equation}
Then $T$ extends to an isometric embedding $T: H_1 \rightarrow
\widetilde H_1$, with
\begin{equation}
T L_1  \ = \  \widetilde L T \quad \text{and} \quad \widetilde
\Gamma T \ = \ \Gamma .
\end{equation}
\end{proof}

\section{Nonlinear systems with linear friction}\label{nonlinear}
As mentioned in the introduction, the construction presented here
extends to suitable non-linear systems.  In this section we describe
a few examples of this type.

To keep the discussion simple, and to avoid the domain questions
associated with unbounded functionals, we consider a finite
dimensional mechanical system with $V=\mathbb{R}^{2n}$ for some $n$,
$J$ a $2n\times 2n$ matrix in the form \eqref{symplectic}, and
Hamiltonian
\begin{equation}\label{nonlinearham}
    \mathsf{h}(u) \ = \ \mathsf{h}_1(u) + \langle K u , K u \rangle_H
    .
\end{equation}
Here $K$ is a linear map $V \rightarrow H=\mathbb{R}^m$, for some
$m$, and $\mathsf{h}_1$ is a $C^1$ function on $V$.  Suppose the
system evolves according to
\begin{equation}\label{nonlinearmot1}
    \partial_t u(t) \ = \ J K^{\mathrm{T}} f(t) + J \nabla_{u} \mathsf{h}_1(u(t))
    + \rho(t)
\end{equation}
with dispersion in the definition of the kinematical stress $f$,
\begin{equation}\label{nonlinearmot2}
    K u(t) \ = \ f(t) + \int_0^{\infty} \chi(t-\tau) f(\tau) \mathrm{d} \tau .
\end{equation}
Here $\rho(t)$ is a driving force and $\chi(\tau)$ is an $m\times m$
matrix valued susceptibility function which satisfies the power
dissipation condition.

Now consider the extended system with Hamiltonian
\begin{equation}\label{nonlinextham}
    \mathsf{H}(u,\theta,\phi) \ = \ \frac{1}{2}\left \| K u - T \phi
    \right \|^2_H + \mathsf{h}_1(u) + \frac{1}{2} \int_{-\infty}^\infty
    \left \{ \| \theta(s) \|^2_H + \| \partial_s \phi(s) \|^2_H \right \}
    \mathrm{d} s ,
\end{equation}
where $T\phi= \int_{-\infty}^\infty \varsigma(s) \phi(s) \mathrm{d}
s$ with matrix valued string coupling function $\varsigma$ given in
terms of $\chi$ by \eqref{fb18}. The resulting equations of motion,
with driving force $\rho$, are
\begin{align}\label{nonlin1}
  \partial_t u(t) \ &= \ J K^{\mathrm{T}} f(t) + J \nabla \mathsf{h}_1(u)
  + \rho(t)\\ \label{nonlin2}
  \partial_t \phi(s,t) \ &= \ \theta(s,t) \\ \label{nonlin3}
  \partial_t \theta(s,t) \ &= \ \partial_s^2 \phi(s,t) +
  \varsigma(s)^{\mathrm{T}} f(t) \; ,
\end{align}
where
\begin{equation}\label{nonlinf}
    f(t) \ = \ Ku(t) - T \phi(t) \; .
\end{equation}

Note that given $f$, the equations for $\phi$ and $\theta$ are
identical to (\ref{St1}, \ref{St2}), so the same driven wave
equation \eqref{drivenwave} for $\phi$ results. Thus the trajectory
$\phi$, in terms of $f$, is given by \eqref{solvedrivenwave},
resulting in the same equation \eqref{fb10} relating $f$ and $Ku$.
Since $\varsigma$ is a solution to \eqref{fb11}, we see that the
`$u$'-component of any solution to (\ref{nonlin1}--\ref{nonlinf}) is
a solution to (\ref{nonlinearmot1}, \ref{nonlinearmot2}). The
function $\mathsf{h}_1$ played no roll in this argument.

Of particular interest is a point particle subject to instantaneous
friction. In this case  $u=(p,q)^{\mathrm{T}}$, with $p,q \in
\mathbb{R}^n$ and $J$ in the canonical form \eqref{symplectic}, and
\begin{equation}
  \mathsf{h}(p,q) \ = \ \mathrm{V}(q) + \frac{1}{2m} \langle p , p \rangle   \; ,
\end{equation}
with $\mathrm{V}(q)$ $C^1$. To get instantaneous friction, we set
the susceptibility
\begin{equation}
\chi(\tau) \ = \ \begin{pmatrix} \gamma & 0 \\ 0 & 0 \end{pmatrix}
\text{ for all $\tau > 0$},
\end{equation}
with $\gamma > 0$ the friction coefficient. Indeed, then
\begin{equation}\label{damper}
  \partial_t \begin{pmatrix}
    p \\
    q
  \end{pmatrix}(t) \ = \ \begin{pmatrix}
    0 & - \mathbf{1} \\
    \mathbf{1} & 0
  \end{pmatrix} \begin{pmatrix}
    \frac{1}{\sqrt{m}}f(t) \\
    \nabla \mathrm{V}(q(t))
  \end{pmatrix}
\end{equation}
with
\begin{equation}\label{dampermatrel}
  f(t) + \gamma \int_0^\infty f(t -
  \tau ) \mathrm{d} \tau \ = \ \frac{1}{\sqrt{m}} p(t)  \; .
\end{equation}
Since $f(t) = \sqrt{m} \partial_t q(t) $ by \eqref{damper}, the
``material relation'' \eqref{dampermatrel} implies
\begin{equation}
  \partial_t q(t) +  \gamma q(t) \ = \ \frac{1}{m} p(t)  \; .
\end{equation}
Thus
\begin{equation}\label{damperevol}
\partial_t^2 q(t)  \ = \ - \frac{1}{m} \nabla \mathrm{V}(q(t))  - \gamma
\partial_t q(t) ,
\end{equation}
so $q$ is subject to instantaneous linear damping.

There is a Hamiltonian extension for this system with
\begin{multline}\label{extinstlin}
    \mathrm{H}(p,\theta,q,\phi) \ = \ \frac{1}{2m} \left (p +
    \sqrt{2\gamma m}\, \phi(0) \right )^2 + \mathrm{V}(q) \\ +
    \frac{1}{2}\int_{-\infty}^\infty \left \{ ( \theta(s))^2 + ( \partial_s \phi(s) )^2
    \right \} \mathrm{d} s
\end{multline}
and associated Lagrangian
\begin{multline}\label{extinslinlagr}
  \mathrm{L}(q, \partial_t q, \phi, \partial_t \phi)
  \ = \ \frac{m}{2} ( \partial_t q)^2 + \sqrt{2\gamma m}
  \langle \partial_t q, \phi(0) \rangle + \frac{1}{2} \int_{-\infty}^\infty
  ( \partial_t \phi(s))^2 \mathrm{d} s \\
  - \mathrm{V}(q) - \frac{1}{2}
  \int_{-\infty}^\infty ( \partial_s \phi(s))^2 \mathrm{d} s \; .
\end{multline}
Indeed, the Euler-Lagrange equations associated to
\eqref{extinslinlagr} are
\begin{align}\label{damperEL1}
m \partial_t^2q(t) \ &= \ - \nabla \mathrm{V}(q(t)) - \sqrt{2\gamma m } \, \partial_t \phi(0,t) \\
\partial_t^2 \phi(s,t) \ &= \ \partial_s^2 \phi(s,t) +\sqrt{2\gamma m} \, \delta(s) \partial_t q(t) .
\label{damperEL2}
\end{align}
The solution to \eqref{damperEL2} is easily seen to be $\phi(s,t) =
\sqrt{\frac{\gamma m}{2}} q(t - |s|)$, which when inserted into
\eqref{damperEL1} gives \eqref{damperevol} for $q$.

The above ideas extend, with some care, to infinite dimensional
systems. For instance, consider the non-linear wave equation
\begin{equation}\label{nlwe}
    \partial_t^2 \psi(x,t) -  \partial_x^2\psi(x,t)  + \mathrm{V}'(\psi(x,t)) =
    \gamma \partial_t \partial_x^2 \psi(x,t) ,
\end{equation}
with dissipation proportional to $\partial_t \partial_x^2 \psi$.  By
a formal extension of \eqref{extinslinlagr} this evolution could be
seen as resulting from the reduction of the extended system
\begin{align}\label{nlwe2}
\partial_t^2 \psi(x,t) \ &= \ \partial_x^2 \psi(x,t) - \mathrm{V}'(\psi(x,t)) -
   \sqrt{ 2 \gamma } \, \partial_t \partial_x \phi(x,0,t) \\
   \partial_t^2 \phi(x,s,t) \ &= \  \partial_s^2 \phi(x,s,t) -  \sqrt{ 2 \gamma } \, \delta(s) \partial_t \partial_x \psi(x,t)
\end{align}
 with Lagrangian
\begin{multline}
\mathrm{L}(\psi, \partial_t \psi, \phi, \partial_t \phi)
  \\ = \  \int_{-\infty}^\infty \left \{ \frac{1}{2} | \partial_t \psi(x,t)|^2 +  \sqrt{2 \gamma} \,
  \partial_t \psi(x;t) \partial_x \phi(x,0;t) + \frac{1}{2} \int_{-\infty}^\infty
  | \partial_t \phi(x,s,t)|^2 \mathrm{d}s \right \}  \mathrm{d} x\\
  - \int_{-\infty}^\infty \left \{ \frac{1}{2} |\partial_x \psi(x,t)|^2 + \mathrm{V}(\psi(x,t)) +   \frac{1}{2} \int_{-\infty}^\infty | \partial_s \phi(x,s;t)|^2 \mathrm{d} s \right \}
   \mathrm{d} x\; .
\end{multline}
Of course, this begs the more difficult question of proving
existence of solutions to the non-linear equations \eqref{nlwe} and
\eqref{nlwe2}.

\section{Energy, momentum, and the stress tensor}\label{sec:stress}
Consider a Lagrangian system in $\mathbb{R}^d$, described by a
vector field $Q(\vec{r})$ taking values in a Hilbert space $V_0$ and
with Lagrangian the integral of a density:
\begin{equation}
  \mathrm{L}(Q, \partial_t Q) \ = \ \int_{\mathbb{R}^d} \mathsf{L}(Q(\vec{r}),
  \nabla Q(\vec{r}), \partial_t Q(\vec{r});\vec{r}) \mathrm{d}^d \vec{r} .
\end{equation}
Given a field configuration $Q:\mathbb{R}^d \times [t_0,t_1]
\rightarrow V_0$, the associated action is
\begin{equation}\label{action}    \mathcal{A}([Q];t_0,t_1) \ = \
\int_{t_0}^{t_1} \int_{\mathbb{R}^d} \mathsf{L}(Q(\vec{r},t),
  \nabla Q(\vec{r},t), \partial_t Q(\vec{r},t);\vec{r}) \mathrm{d}^d \vec{r}
  \mathrm{d} t  ,
\end{equation}
The physical evolution $Q(\vec{r},t)$ is a stationary point for the
action $\mathcal A$ and thus satisfies the Euler-Lagrange equations
\begin{equation}\label{spatialvariation}
    \partial_t \frac{\delta
    \mathsf{L}}{\delta \partial_t Q}(\vec{r},t)
    + \partial_i \frac{\delta
    \mathsf{L}}{\delta \partial_i Q}(\vec{r},t)
    -\frac{\delta \mathsf{L}}{\delta \mathrm{Q}}(\vec{r},t)   \ = \ 0 .
\end{equation}
(Recall the summation convention!) We use $\delta$ to indicate
partial differentiation of $\mathsf{L}$ to avoid confusion with
$\partial_t$ and $\partial_i$, and write $\mathsf{L}(\vec{r},t)$,
$\frac{\delta \mathsf{L}}{\delta Q}(\vec{r},t)$, $\ldots$ as
shorthand for $\mathsf{L}(Q(\vec{r},t),
  \nabla Q(\vec{r},t), \partial_t Q(\vec{r},t);\vec{r})$, $\frac{\delta \mathsf{L}}{\delta Q}
  (Q(\vec{r},t),
  \nabla Q(\vec{r},t), \partial_t Q(\vec{r},t);\vec{r})$, $\ldots$.

We have assumed that the Lagrangian density $\mathsf{L}$ does not
depend explicitly on the time $t$. As a result, time translation is
a symmetry of the system, and Noether's Theorem gives an expression
for the energy density $\mathsf{H}(\vec{r},t)$, which is just the
Hamiltonian, and the energy flux vector $\mathbf{S}(\vec{r})$ (see,
for example, \cite{LandauLif1, Barut}):
\begin{align}\label{Hamdens2}
  \mathsf{H}(\vec{r},t) \ &= \ \left \langle \partial_t Q(\vec{r},t), \frac{\delta
  \mathsf{L}}{\delta \partial_t Q}(\vec{r},t)  \right \rangle_{V_0} -
  \mathsf{L}(\vec{r},t), \intertext{
  and}
  \mathbf{S}_i(\vec{r},t) \ &= \ \left \langle\partial_{t} Q(\vec{r},t), \frac{ \delta
  \mathsf{L}}{\delta \partial_i Q}(\vec{r},t)  \right \rangle_{V_0}
  .
  \label{Efluxdens}
\end{align}
When evaluated ``on-shell,'' that is for $Q$ satisfying
\eqref{spatialvariation}, these two quantities satisfy a local
conservation law:
\begin{equation}\label{encontinuityequation}
\partial_t \mathsf{H}(\vec{r},t) \ + \ \partial_i
\mathbf{S}_i(\vec{r},t) \ = \ 0 ,
\end{equation}
implying that the integral of $\mathsf{H}$
\begin{equation}\label{energy}
    \mathcal{E} \ = \ \int_{\mathbb{R}^d} \mathsf{H}(\vec{r},t) \mathrm{d}^d
    \vec{r}
\end{equation}
is a \emph{conserved quantity}, which we identify with the total
energy of the system.

\begin{definition}\label{def:homogeneity}The system is
\emph{homogeneous} if $\mathsf{L}$ does not depend explicitly on
$\vec{r}$.
\end{definition}

A homogeneous system has an additional vector conserved quantity,the
total wave momentum $\mathbf{P}$. An initial expression for the wave
momentum density is
\begin{equation}\label{momedensity}
    \check{\mathbf{p}}_i(\vec{r},t) \ = \ \left \langle
     \partial_i Q(\vec{r},t), \frac{\delta \mathsf{L}}{\delta
    \partial_{t} Q}(\vec{r},t) \right \rangle_{V_0} \; ,
\end{equation}
with the associated \emph{canonical stress tensor}
\begin{equation}\label{momconservtensor}
    \check{\mathsf{T}}_{i,j}(\vec{r},t) \ = \
    \left \langle \partial_{i} Q(\vec{r},t)  , \frac{\delta \mathsf{L}}{\delta
    \partial_{j} Q}(\vec{r},t) \right \rangle_{V_0}
    - \delta_{i,j} \mathsf{L}(\vec{r},t) ,
\end{equation}
see \cite{LandauLif1,Barut,MorseFeshbach}. It is easy to show that
\begin{equation}\label{initconsvmom}
    \partial_t \check{\mathbf{p}}(\vec{r},t) + \partial_j
    \check{\mathsf{T}}_{i,j}(\vec{r},t)
    \ = \ 0 ,
\end{equation}
if the system is homogeneous, so
\begin{equation}\label{totalmomentum}
    \mathbf{P} \ = \ \int_{\mathbb{R}} \check{\mathbf{p}}(\vec{r},t) \mathrm{d}^3
    \vec{r}
\end{equation} is a conserved quantity, which we identify with the total wave
momentum.

As is well known \cite{LandauLif1,Barut}, the conservation law and
total momentum $\mathbf{P}$ are invariant under the following
``gauge transformations:''
\begin{align}\label{wavemomdensity}
    \mathbf{p}_i (\vec{r},t) \ &= \ \check{\mathbf{p}}_i(\vec{r},t) - \partial_j
    \Phi_{i,j}(\vec{r},t) \\ \label{stresstens}
    \mathsf{T}_{i,j}(\vec{r},t) \ &= \ \check{\mathsf{T}}_{i,j}(\vec{r},t)
    + \partial_t \Phi_{i,j} (\vec{r},t) + \partial_k \Psi_{i,j,k} (\vec{r},t) ,
\end{align}
with $\Phi_{i,j}$ an arbitrary two tensor and $\Psi_{i,j,k}$ a
$3$-tensor antisymmetric in the later two indices: $\Psi_{i,j,k} = -
\Psi_{i,k,j}$. Indeed, for any such $\Phi$ and $\Psi$
\begin{equation}
  \int_{\mathbf{R}^d} \mathbf{p} (\vec{r},t) \mathrm{d}^d
\vec{r}\ = \
  \int_{\mathbf{R}^d} \check{\mathbf{p}}(\vec{r},t) \mathrm{d}^d
\vec{r} ,
\end{equation}
and
\begin{equation}\label{momconserv}
    \partial_t \mathbf{p} + \partial_j \mathsf{T}_{i,j}
    \ = \ \partial_t \check{\mathbf{p}}(\vec{r},t) + \partial_j
    \check{\mathsf{T}}_{i,j} \ =  \ 0 .
\end{equation}

Thus the momentum density and stress tensor are not really uniquely
defined.
 However, there is a well known way to fix these quantities,
at least for an \emph{isotropic} system. Here isotropy (see Defn.\
\ref{def:isotropy} below) expresses the invariance of the system
under rotations and leads to another conserved current, \emph{the
angular momentum density} $\mathsf{m}_{i,j}$, an anti-symmetric
$2$-tensor, which obeys the conservation law
\begin{equation}\label{angmomconsv}
    \partial_t \mathsf{m}_{i,j}(\vec{r}) +  \partial_{k}
    \mathsf{F}_{i,j,k}(\vec{r}) \ = \ 0 ,
\end{equation}
with $\mathsf{F}$ the \emph{angular momentum flux tensor}. For
isotropic systems we demand the following relations between the wave
momentum and angular momentum densities
\begin{equation}\label{angmomdens}
    \mathsf{m}_{i,j}(\vec{r},t) \ = \ \vec{r}_i \,
    \mathbf{p}_j(\vec{r},t) - \vec{r}_j \, \mathbf{p}_i (\vec{r}_i,t) ,
\end{equation}
and between the stress tensor $\mathsf{T}$ and the angular momentum
flux tensor
\begin{equation}\label{angmomflux}
    \mathsf{F}_{i,j,k}(\vec{r},t)  \ = \ \vec{r}_i \, \mathsf{T}_{j,k}(\vec{r},t) -
    \vec{r}_j \,
    \mathsf{T}_{i,k}(\vec{r},t) .
\end{equation}
For a homogeneous and isotropic system, the conservation laws
(\ref{momconserv}, \ref{angmomconsv}) then imply that the stress
tensor is symmetric, $T_{i,j} = T_{j,i}$, since
\begin{multline}
   0 \ = \ \partial_t \mathsf{m}_{i,j} + \partial_k \mathsf{F}_{i,j,k}
   \\ = \ \vec{r}_i (\partial_t \mathbf{p}_j + \partial_k \mathsf{T}_{j,k}
   ) - \vec{r}_j (\partial_t \mathbf{p}_i + \partial_k \mathsf{T}_{i,k})
   + \mathsf{T}_{j,i} - \mathsf{T}_{i,j} \ = \ \mathsf{T}_{j,i} - \mathsf{T}_{i,j}.
\end{multline}
The canonical stress tensor $\check{\mathsf{T}}$ is not symmetric in
general, and thus is not the proper choice for a stress tensor
related to the angular momentum flux tensor via \eqref{angmomflux}.
For an isotropic system there are tensors $\Phi$ and $\Psi$ such
that \eqref{stresstens} is the symmetric stress tensor.


To proceed we must define isotropy, and to do so must specify how
the system transforms under rotations. Thus, we suppose given a
representation $\Upsilon$ of the rotation group $\mathsf{SO}(d)$ by
orthogonal operators on $V_0$. (See \cite{Chevalley,SimonGroups} for
the basics of Lie groups and representation theory.) An arbitrary
element of $\mathsf{SO}(d)$ can be expressed as
$\mathrm{e}^{\omega}$ with $\omega \in \mathsf{so}(d)$, the space of
anti-symmetric $d\times d$ matrices, which is the Lie algebra of
$\mathsf{SO}(d)$. Thus, the representation $\Upsilon:\mathsf{SO}(d)
\rightarrow \mathcal{B}(V_0)$ can be written in terms of a
corresponding representation $\upsilon$ of $\mathsf{so}(d)$
\begin{equation}
    \Upsilon (\mathrm{e}^{\omega}) \ = \ \mathrm{e}^{\upsilon(\omega)} .
\end{equation}

The matrices $\mathsf{e}_{i,j} \in \mathsf{so}(d)$,
\begin{equation}
    \mathsf{e}_{i,j ; \alpha, \beta} \ = \ \delta_{i, \alpha} \delta_{j, \beta}
    - \delta_{j, \alpha} \delta_{i,\beta},
\end{equation}
satisfy $\mathsf{e}_{i,j}  =  - \mathsf{e}_{j,i}$ and the collection
$\{ \mathsf{e}_{i,j} : 1 \le i < j \le d\}$ is a basis for
$\mathsf{so}(d)$.  An arbitrary element $\omega \in \mathsf{so}(d)$
can thus be written
\begin{equation}
    \omega \ = \ \frac{1}{2} \omega_{i,j} \mathsf{e}_{i,j} \quad
    (\text{summation convention}).
\end{equation}
Thus the representation $\upsilon$ is determined by the skew-adjoint
operators
\begin{equation}
    G_{i,j} \ = \ \upsilon(\mathsf{e}_{i,j}) ,
\end{equation}
which evidently satisfy $G_{i,j} = - G_{j,i}$ and
\begin{equation}\label{comrelation}
    \left [ G_{i,j}, G_{k,l} \right ] \ = \ \upsilon
    \left ( \left [ \mathsf{e}_{i,j}, \mathsf{e}_{k,l} \right ] \right )
\ = \
    - \delta_{i,k} G_{j,l} +
    \delta_{j,k} G_{i,l} + \delta_{i,l} G_{j,k} - \delta_{j,l} G_{i,k} .
\end{equation}
We assume that $\upsilon$ is a representation by bounded operators,
$G_{i,j} \in \mathcal{B}(V_0)$ for every $i,j$. The representative
of a generic element $\omega \in \mathsf{so}(d)$ is therefore
\begin{equation}\label{repres}
    \upsilon(\omega )\  = \ \frac{1}{2} \omega_{i,j} G_{i,j} ,
\end{equation}
and of a rotation $\mathrm{e}^{\omega} \in \mathsf{SO}(d)$
\begin{equation}
    \Upsilon(\mathrm{e}^{\omega}) \ = \ \mathrm{e}^{\frac{1}{2} \omega_{i,j} G_{i,j}} .
\end{equation}

A global rotation of the coordinate system about a given point
$\vec{r} \,^0$ involves a  transformation of $\vec{r}$
\begin{align}
  \vec{r} \ &\mapsto \  \vec{r}^{\, \omega} \ = \ \vec{r}\, ^0 + \mathrm{e}^{\omega} \cdot (\vec{r} -
  \vec{r}\, ^0), \intertext{and of the field $Q$:}
  \label{Qtransf}
  Q(\vec{r}) \ &\mapsto \ Q^{\omega}(\vec{r}) \ = \
  \mathrm{e}^{\frac{1}{2}\omega_{i,j} G_{i,j}}
  Q(\vec{r}^{\, -\omega}) .
\end{align}
Such rotations form a representation of $\mathsf{SO}(d)$ in
$L^2(\mathbb{R}^d, V_0)$ with generators
\begin{equation}\label{infgen}
    L_{i,j}^{\vec{r}\, ^0} Q(\vec{r}) \ = \ G_{i,j} Q(\vec{r}) - (\vec{r}_i - \vec{r}\,_i^0)\,
    \partial_j
    Q(\vec{r}) + (\vec{r}_j - \vec{r}\,_j^0) \partial_i Q(\vec{r}) ,
\end{equation}
so $Q^\omega$  may be written
\begin{equation}\label{Qtransf2}
    Q^\omega(\vec{r}) \ = \ [\mathrm{e}^{\frac{1}{2} \omega_{i,j} L_{i,j}^{\vec{r}\, ^0}}
    Q] (\vec{r}).
\end{equation}
\begin{definition}\label{def:isotropy}The system is \emph{isotropic at a
point $\vec{r}$} if the Lagrangian density \underline{at the given
point $\vec{r}$} is invariant under the transformations
\eqref{Qtransf2}:
\begin{multline}\label{isotropy}
    \mathsf{L}( \mathrm{e}^{\frac{1}{2}\omega_{i,j} L_{i,j}^{\vec{r}}} Q(\vec{r},t), \partial_t
    \mathrm{e}^{\frac{1}{2}\omega_{i,j} L_{i,j}^{\vec{r}}} Q(\vec{r},t),
    \nabla
    \mathrm{e}^{\frac{1}{2}\omega_{i,j} L_{i,j}^{\vec{r}}} Q(\vec{r},t))
    \\ = \ \mathsf{L}( Q(\vec{r},t), \partial_t Q(\vec{r},t), \nabla Q(\vec{r},t) ),
    \quad \text{ for every $\omega \in \mathsf{so}(d)$.}
\end{multline}
We say the system is \emph{isotropic} if it is isotropic at every
point.
\end{definition}

It is convenient to express \eqref{isotropy} in infinitesimal form
by differentiating the l.h.s.\ at $\omega =0$. To this end, note
that
\begin{equation}\label{Q}
\mathrm{e}^{\frac{1}{2}\omega_{i,j} L_{i,j}^{\vec{r}}} Q(\vec{r},t)
\ = \ \mathrm{e}^{\frac{1}{2}\omega_{i,j} G_{i,j}} Q(\vec{r},t) ,
\end{equation}
since the remaining terms vanish at the origin $\vec{r}$ of the
rotation.  Similarly,
\begin{equation}\label{partQ}
    \partial_t
    \mathrm{e}^{\frac{1}{2}\omega_{i,j} L_{i,j}^{\vec{r}}} Q(\vec{r},t)
    \ = \ \mathrm{e}^{\frac{1}{2}\omega_{i,j} G_{i,j}}
    \partial_t  Q(\vec{r},t) .
\end{equation}
However,
\begin{equation}\label{gradQ}
    \partial_k  L_{i,j}^{\vec{r}} Q(\vec{r},t)
    \ = \ G_{i,j}
    \partial_k Q(\vec{r},t) + \delta_{j,k} \partial_i Q(\vec{r},t)
    - \delta_{i,k} \partial_j Q(\vec{r},t) ,
\end{equation}
because
\begin{equation}
    \left [ \partial_k , L_{i,j}^{\vec{r}} \right ]
    \ = \ \delta_{j,k} \partial_i - \delta_{i,k} \partial_j  \ \neq \ 0 .
\end{equation}
\begin{lemma}\label{lem:infisotropy}
The system is isotropic at $\vec{r}$ if and only if for any $Q$
\begin{multline}\label{infisotropy}
    0 \ = \ \check{\mathsf{T}}_{i,j}(\vec{r},t) - \check{\mathsf{T}}_{j,i}(\vec{r},t)
   \ + \  \left \langle G_{i,j} Q(\vec{r},t) , \frac{\delta \mathsf{L}}{\delta
    Q}(\vec{r},t)  \right \rangle
    \\ +  \left \langle G_{i,j} \partial_t Q(\vec{r},t) ,
    \frac{\delta \mathsf{L}}{\delta
    \partial_t Q}(\vec{r},t)  \right \rangle
    +  \left \langle G_{i,j} \partial_k Q(\vec{r},t),
    \frac{\delta \mathsf{L}}{\delta
    \partial_k Q}(\vec{r},t)  \right \rangle ,
\end{multline}
with $\check{\mathsf{T}}$ given by \eqref{momconservtensor}. If $Q$
satisfies the Euler-Lagrange equations \eqref{spatialvariation} then
\begin{multline}\label{infisoonshell}
    0 \ = \ \check{\mathsf{T}}_{i,j}(\vec{r},t) - \check{\mathsf{T}}_{j,i}(\vec{r},t)
 \\ + \ \partial_t \left \langle G_{i,j}  Q(\vec{r},t) ,
    \frac{\delta \mathsf{L}}{\delta
    \partial_t Q}(\vec{r},t)  \right \rangle
     +  \partial_k \left \langle G_{i,j}  Q(\vec{r},t),
    \frac{\delta \mathsf{L}}{\delta
    \partial_k Q}(\vec{r},t)  \right \rangle .
\end{multline}
\end{lemma}
\begin{proof}
 Eq.\ \eqref{infisotropy} is a consequence of the definition
 \eqref{isotropy} and (\ref{Q}--\ref{gradQ}). Eq.\ \eqref{infisoonshell}
 follows from rewriting the third term on the r.h.s.\ of \eqref{infisotropy}
 using the Euler-Lagrange equations \eqref{spatialvariation} and combining terms
 with the Leibniz rule.
\end{proof}

Following \cite[Section III.4]{Barut} we define the wave momentum
density and stress tensor by gauge transformations
(\ref{wavemomdensity}, \ref{stresstens}) of $\check{\mathbf{p}}$ and
$\check{\mathsf{T}}$, with
\begin{equation}\label{Phidenf}
    \Phi_{i,j}(\vec{r},t) \ = \ \frac{1}{2} \left \langle G_{i,j} Q(\vec{r},t),
    \frac{\delta \mathsf{L}}{\delta
    \partial_{t} Q}(\vec{r},t) \right \rangle_{V_0} ,
\end{equation}
and
\begin{multline}\label{Psidefn}
    \Psi_{i,j,k}(\vec{r},t) \ = \  \frac{1}{2} \left \{
    \left \langle G_{i,j} Q(\vec{r},t)  , \frac{\delta \mathsf{L}}{\delta
    \partial_{k} Q}(\vec{r},t)\right \rangle_{V_0}  \right . \\ \left .
    - \left \langle G_{j,k} Q(\vec{r},t)  , \frac{\delta \mathsf{L}}{\delta
    \partial_{i} Q}(\vec{r},t)\right \rangle_{V_0}
    - \left \langle G_{i,k} Q(\vec{r},t)  , \frac{\delta \mathsf{L}}{\delta
    \partial_{j} Q}(\vec{r},t)\right \rangle_{V_0} \right \}.
\end{multline}
Note that $\Psi_{i,j,k}$ is anti-symmetric under interchange of $j$
and $k$.

\begin{theorem}\label{thm:stress}
    Let the stress tensor $\mathsf{T}$ be defined
    \begin{equation}\label{stresstensor}
    \mathsf{T}_{i,j}(\vec{r},t) \ := \ \check{\mathsf{T}}_{i,j}(\vec{r},t)
    + \partial_t \Phi_{i,j}(\vec{r},t) + \partial_k \Psi_{i,j,k}(\vec{r},t) ,
  \end{equation}
   with $\Phi$, $\Psi$ given by \emph{(\ref{Phidenf}, \ref{Psidefn})}. If the system
  is isotropic, then $\mathsf{T}$ is symmetric.
  If $Q$ satisfies the Euler-Lagrange equations
  and the system is homogeneous, then
  \begin{equation}\label{continuityequation}
    \partial_t \mathbf{p}_i(\vec{r},t) + \partial_j \mathsf{T}_{i,j}(\vec{r},t) =
    0 ,
  \end{equation}
  with the wave momentum density
  \begin{equation}\label{wavemomentumdensity}
    \mathbf{p}_i(\vec{r},t) \ = \ \check{\mathbf{p}}_i(\vec{r},t) - \partial_j
    \Phi_{i,j}(\vec{r},t).
  \end{equation}
  If the system is homogeneous and isotropic, then the local angular
  momentum conservation law \eqref{angmomconsv} holds
  with $\mathsf{m}$ and
  $\mathsf{F}$ defined by \eqref{angmomdens} and \eqref{angmomflux}
  respectively.
\end{theorem}
\begin{proof}
  The only point not established in the above discussion is the symmetry of
  $\mathsf{T}$.  This however follows from Lemma \ref{lem:infisotropy} since
  \begin{equation}
    \mathsf{T}_{i,j}(\vec{r},t) - \mathsf{T}_{j,i}(\vec{r},t) \ = \
    \text{r.h.s.\ of \eqref{infisoonshell}.} \qedhere
\end{equation}
\end{proof}

To close, we consider how the continuity equations
(\ref{encontinuityequation}, \ref{continuityequation}) are modified
by a driving force $R(\vec{r},t) \in V_0$ such that $Q$ satisfies
the driven Euler-Lagrange equation
\begin{equation}\label{driveneulerlagrange}
    \partial_t \frac{\delta
    \mathsf{L}}{\delta \partial_t Q}(\vec{r},t)
    + \partial_i \frac{\delta
    \mathsf{L}}{\delta \partial_i Q}(\vec{r},t)
    -\frac{\delta \mathsf{L}}{\delta Q}(\vec{r},t) \ = \  R(\vec{r},t) ,
\end{equation}
which is simply the Euler-Lagrange equation for the time dependent
Lagrangian density $\mathsf{L} + \langle Q(\vec{r},t), R(\vec{r},t)
\rangle$.  Since $R$ breaks time translation invariance and
homogeneity, energy and momentum are no longer conserved.  However,
we have
\begin{theorem}\label{thm:drivencontinuity}
  If $Q$ satisfies the driven Euler-Lagrange equations
  \eqref{driveneulerlagrange} then
  \begin{equation}\label{drivenencontinuity}
    \partial_t \mathsf{H}(\vec{r},t) + \partial_j \mathbf{S}_j(\vec{r},t)
    \ = \ \langle Q(\vec{r},t), R(\vec{r},t) \rangle_{V_0} .
  \end{equation}
  If, furthermore, the system is homogeneous then
  \begin{equation}\label{drivenmomcontinuity}
    \partial_t \mathbf{p}_i(\vec{r},t) + \partial_j \mathsf{T}_{i,j}(\vec{r},t)
    \ = \ \langle \partial_i Q(\vec{r},t), R(\vec{r},t) \rangle_{V_0} .
  \end{equation}
\end{theorem}
\begin{proof} This is a straightforward calculation. \end{proof}

\subsection*{Acknowledgments} We thank Lars Jonsson and Ilya Vitebskiy for useful
discussions. Support under AFOSR grant FA9550-04-1-0359 and an NSF
postdoctoral research fellowship (JHS) are gratefully acknowledged.

\end{document}